\documentclass{article}

\usepackage{mathtools}
\usepackage{natbib}
\usepackage{graphics}
\usepackage{graphicx}
\usepackage{setspace} 
\usepackage[portrait, margin=1in]{geometry}
\usepackage{amsmath}  
\usepackage{amsfonts}
\usepackage{indentfirst}
\usepackage{array}
\usepackage{booktabs}
\usepackage{multirow}
\usepackage{booktabs}
\usepackage{tabularx}
\usepackage{tabulary}
\usepackage{dcolumn}
\usepackage{tabularht}
\usepackage{verbatim}
\usepackage{graphicx}
\usepackage{natbib}
\usepackage{booktabs,caption}
\usepackage[flushleft]{threeparttable}
\usepackage{enumitem}   
\usepackage{placeins}
\usepackage{authblk}

\usepackage{xcolor}

\definecolor{dark-maroon}{HTML}{5D0F0D}
\definecolor{navyblue}{HTML}{0A2044}

\definecolor{purple}{HTML}{5601A4}
\definecolor{navy}{HTML}{0D3D56}
\definecolor{ruby}{HTML}{9a2515}
\definecolor{alice}{HTML}{107895}
\definecolor{daisy}{HTML}{EBC944}
\definecolor{coral}{HTML}{F26D21}
\definecolor{kelly}{HTML}{829356}
\definecolor{cranberry}{HTML}{E64173}
\definecolor{jet}{HTML}{131516}
\definecolor{asher}{HTML}{555F61}
\definecolor{slate}{HTML}{314F4F}

\usepackage{hyperref}
\hypersetup{
    colorlinks= true,
    citecolor= dark-maroon,
    linkcolor= dark-maroon,
    filecolor= dark-maroon,      
    urlcolor= dark-maroon,
}

\DeclareMathOperator*{\argmin}{arg\,min}

\newcommand{\indep}{\rotatebox[origin=c]{90}{$\models$}}
\begin{document}
\doublespacing%
\title{A Method of Moments Approach to Asymptotically Unbiased Synthetic Controls}
\author[]{Joseph Fry}  
\affil[]{\footnotesize
University of Colorado at Boulder, 256 UCB, 80309 Boulder, Colorado, United States \par\nopagebreak
  \textit{E-mail address:} \texttt{joseph.fry@colorado.edu}}
\date{}
\maketitle

\begin{abstract}
    A common approach to constructing a Synthetic Control unit is to fit on the outcome variable and covariates in pre-treatment time periods, but it has been shown by \cite{ImperfectFit} that this approach does not provide asymptotic unbiasedness when the fit is imperfect and the number of controls is fixed. Many related panel methods have a similar limitation when the number of units is fixed. I introduce and evaluate a new method in which the Synthetic Control is constructed using a General Method of Moments approach where units not being included in the Synthetic Control are used as instruments. I show that a Synthetic Control Estimator of this form will be asymptotically unbiased as the number of pre-treatment time periods goes to infinity, even when pre-treatment fit is imperfect and the number of units is fixed. Furthermore, if both the number of pre-treatment and post-treatment time periods go to infinity, then averages of treatment effects can be consistently estimated. I conduct simulations and an empirical application to compare the performance of this method with existing approaches in the literature.
\end{abstract}
\par \textbf{Keywords: }\textit{synthetic control, general method of moments, policy evaluation, factor model, instrumental variables}
\section{Introduction} \label{Intro}
The Synthetic Control Estimator (SCE) was first introduced by \cite{AandG} and \cite{Abadie2010} in the context of panel data where a single unit becomes and stays treated. The basic idea behind this approach is to construct a weighted average of the control units or a “synthetic” control (SC) unit by minimizing the difference between this SC and the treated unit on a set of pre-treatment predictor variables, with the hope that this will mean the SC is also close to the treated unit's counterfactual outcomes in the post-treatment time periods. This method has been promoted as offering several benefits over alternative approaches, including being highly transparent about how control units are being weighted to produce an estimate of the treated unit's counterfactual value. It also does not require that any specific control unit would have counterfactually had the same trend as the treated unit, as would be the case in a difference-in-differences approach. However, exactly what vector of weights $W$ are placed on the control units will depend crucially on what variables are included in the set of predictors and what relative priority is placed on having good fit for each predictor. While a significant literature has begun to emerge examining its properties and introducing variations (see \cite{LitReview} for a review of the literature), it remains an open question how to decide which predictors to include and what is the optimal way to trade-off between the goodness-of-fit that is achieved for each predictor that is included. Taking a method of moments perspective allows for a principled way to address these questions where predictors are chosen to be moment conditions that should theoretically guarantee identification and existing techniques for weighting moment conditions provide a starting point for how to weight the predictors. \cite{cherrypicking} perform simulations to show that if researchers are able to hand-pick the set of predictors themselves, then they will often times be able to cherry-pick their way to statistically significant results. Therefore, offering a theoretically justified method helps contribute to the literature by reducing the amount of ambiguity researchers face when using the Synthetic Control method, which in turn will hopefully reduce the room for specification searching in applications of this method.
\par
Much of the existing work studying the asymptotic properties of SCEs has assumed the data follow a linear factor model, so each unit's untreated potential outcomes are a linear combination of a common set of factors plus an idiosyncratic shock. Additionally, many of the asymptotic results for SCEs focus specifically on the case where the predictors are pre-treatment values of the outcome variable and covariates, and this is also a common choice in many applications of the method. In terms of asymptotic properties, there has been a particular focus on when an SCE will have its bias converge to zero.  \cite{Abadie2010} derive bounds on the bias of the SCE that go to zero as the number of pre-treatment time periods ($T_0$) goes to infinity and the number of control units ($J$) is held fixed under the assumption of perfect fit on both pre-treatment outcomes and covariates, meaning the weighted average of the controls' values of the outcome variable in every pre-treatment time period and the weighted average of the controls' values of the covariates are exactly equal to those of the treated unit. This provides some motivation for constructing the SC by fitting on these pre-treatment variables, but only when very good pre-treatment fit is achieved and there are many pre-treatment time periods. This is also why \cite{Abadie2010} only recommend using this method when those conditions are met. 
\par
\cite{ImperfectFit} analyze the case where the SC is constructed by fitting on pre-treatment outcomes but the fit is imperfect and they show that in this case the SCE will remain biased as the number of pre-treatment time periods goes to infinity when the number of controls is fixed. Intuitively, this is because better fit on the pre-treatment outcomes can be achieved not only by making the SC's loadings on the factors close to the factor loadings of the treated unit but also by spreading out the weights on the control units to reduce the variance of the SC caused by the control units' idiosyncratic shocks. This trade-off prevents the SC's factor loadings from converging to those of the treated unit, except when the contribution to the SC's variance from each of the control unit's idiosyncratic shocks is converging to zero. It is worth noting that this limitation applies to other similar SCEs that include additional predictors since \cite{cherrypicking} show other specifications will asymptotically give the same estimates provided that the values of the outcome variable in many pre-treatment time periods are being included as predictors. Additionally, \cite{IncludingAllLags} show that when common methods for weighting predictors are used, adding covariates as predictors in addition to the values of the outcome variable in all pre-treatment time periods can give the same estimate as only including the values of the outcome variable. \cite{ImperfectFit} also show that other related panel data approaches that have been studied such as \cite{Hsiao2012}, \cite{LiBell2017}, and \cite{MasimiMarcelo2021}, suffer from a similar limitation when $J$ is fixed.\footnote{They show that while some of these methods have consistency results when $J$ is fixed, these results rely on assumptions that essentially imply that there is no selection into treatment on unobservable time-varying factors.} \cite{LargeSampleProperties} analyzes a SCE that is fitted on pre-treatment outcomes as well, but they get around this trade-off by assuming that $J$ goes to infinity as $T_0$ goes to infinity and that it is possible for the SC's factor loadings to converge to the factor loadings of the treated unit while spreading out the weights on the control units more and more. Under these conditions, they show that the asymptotic unbiasedness result can be recovered.\footnote{Strictly speaking, having the SC's factor loadings be equal to those of the treated unit is neither a sufficient nor necessary condition for the estimator to be unbiased in general. Specifically, \cite{Abadie2010}, \cite{LargeSampleProperties}, as well as myself must impose additional conditions on the idiosyncratic shocks in the post-treatment time periods in order to obtain the unbiasedness results.} 
\par
Having the SC's factor loadings converge to those of the treated unit when the number of controls is fixed may seem difficult because it is known that when the set of units is fixed, it is only possible to estimate factor loadings consistently under strong assumptions on the idiosyncratic shocks (see \cite{NumberOfFactors}, \cite{Anderson1984}). Even if the factor loadings could be estimated consistently, the number of factors and the realizations of the factors in the post-treatment time periods could not be consistently estimated with $J$ fixed. This is the problem with trying to estimate the treated unit's counterfactual by estimating the latent factor structure (for example, earlier with \cite{CHAMBERLAIN1984} and \cite{LiangandZeger} and with many more recent advancements including \cite{HeckamnLeamer2001}, \cite{MoonWeidner2015}, \cite{moon_weidner_2017}, \cite{Bai2009_ife}, and \cite{Pearson2006}). \cite{IFE_with_SC} takes this approach by estimating the realizations of the factors in both the pre-treatment and post-treatment time periods using the control units and then fitting the treated unit as a linear combination of those estimated factors, but their results require both the number of controls and pre-treatment time periods to go to infinity. However, being able to consistently estimate the treated unit's factor loadings is not necessary for guaranteeing that the SC's factor loadings converges to them.  
\par
Instead, if the number of factors is $F$ so the factor loadings are $F$-dimensional vectors for each unit, then the factor loadings of the treated unit and the SC can be guaranteed to be equal by having them satisfy $F$ linearly independent moment equations. I introduce a new method that, under mild conditions on the factor structure, will cause this to be the case asymptotically by constructing the SC using a General Method of Moments (GMM) approach where other units not included in the set of controls are used as instruments. The intuition behind this approach is that if the idiosyncratic shocks are uncorrelated with the factors and across units, then the treated unit and SC will only be correlated with these instruments through the latent factors. This means the instruments will satisfy the exclusion restriction and they will be relevant as long as they are exposed to the same latent factors that the treated unit and controls are. Therefore, by choosing an SC using this GMM estimator, that SC's factor loadings must converge to the factor loadings of the treated unit. 
\par 
Two other recent papers also take a method of moments approach to estimating the weights of a SC. \cite{shi2023} use the framework of proximal causal inference. Like other work on proximal causal inference (see e.g., \cite{ProximialIntroduction}), they control for confounding caused by latent factors using moment conditions involving proxies for the confounders. In the context of SCE, untreated units not included in the SC may be thought of as proxies for the latent factors. One way their work differs is that they allow for the latent factors to be non-linearly related to the outcome variables, in which case the SC is estimated non-parametrically. This requires imposing stronger independence conditions on the idiosyncratic shocks than I do here. \cite{Powell2021} also use a GMM estimator where each moment equation is based on a control unit being uncorrelated with the difference between the treated unit and a synthetic unit that uses an estimated value of that control unit rather than its actual value. Obtaining these estimated values of the control units is aided by constructing a synthetic unit for each control unit in addition to a synthetic unit for the treated unit. Both \cite{Powell2021} and \cite{shi2023} consider the case where the number of units is held fixed, whereas I consider both cases where the number of units may be small or large relative to the number of pre-treatment time periods. The choice of moment conditions in this context to some degree parallels the choice of moment conditions in panel settings more generally. For example, \cite{Chamberlain1992} use of conditional moment equations for non-parametric estimation is similar to that of \cite{shi2023} and \cite{Holtz-EakinNewey1988} uses lagged values as instruments which is another possible choice that could be made for estimating the Synthetic Control weights. I further compare using these moment conditions and the ones I focus on in section \ref{Model} below.
 \par
 I show that when the data follow a linear factor model, an estimate of the treatment effect for the treated unit in any post-treatment time period using this SC as the counterfactual will have its bias converge to zero as $T_0$ goes to infinity and $J$ is fixed or $J$ goes to infinity, and I do so while placing comparatively weaker assumptions on the idiosyncratic shocks in the pre-treatment time periods than has previously been done in many cases. Furthermore, I show that when the number of post-treatment time periods is also large, this SC can be used to consistently estimate averages of treatment effects. Since untreated units can often times be included as either instruments or controls, in section \ref{Moment and Model Selection}, I discuss model selection methods to make this choice. I present two methods and verify that a form model selection consistency holds for each. I also supplement the formal results with simulations fitted to real data from the World Bank to compare the performance of this approach to existing SCEs. Finally, I replicate the empirical application of \cite{Abadie2015} using this new estimator to illustrate some of the practical considerations that can arise and confirm the robustness of their results. While choosing how to conduct inference is an important problem, I do not focus on promoting a particular inference method here. I instead provide a discussion of the various methods that have been proposed in section \ref{Discussion} and give an example of how formal results can be proven under stronger conditions in Appendix \hyperref[ApC]{C}.

\section{Model and Estimation Strategy} \label{Model}

Linear factor models have been a common setting to explore the properties of SCEs, beginning with \cite{Abadie2010} and continuing on in many recent papers (e.g., \cite{ImperfectFit}, \cite{covariates}, \cite{LargeSampleProperties}, \cite{SynthIntervention}, \cite{SDID}). Part of the reason for this is that it captures the intuition motivating the use of SCEs in many cases. Namely, that it is reasonable to use the control units to estimate the treated unit's counterfactual because the factors driving the trends in the treated unit's outcomes in the absence of treatment are the same factors driving the trends in the control units' outcomes. Its appeal also comes from the fact that it allows for there to be unobserved factors correlated with both the outcome variables and treatment assignment, as would be expected in the context of comparative case studies, but retains tractability by assuming that the dimension of these factors remains fixed as the sample size grows. Additionally, it has the nice property of being a generalization of a two-way fixed effect model which is often used in the difference-in-differences literature (hence why it is sometimes also called an interactive fixed-effect model), but it allows units' untreated potential outcomes to follow different trends.
\par 
 Unlike some previous work, I will be allowing for there to be multiple treated units but I assume that there is one specific unit $i = 0$ for which we are interested in estimating its treatment effects. Therefore, I will refer to this specific treated unit for which the SC is being constructed as the unit of interest. If there are multiple treated units, then this estimation procedure can be repeated for each unit to obtain unit specific treatment effect estimates, which may be averaged to estimate the average treatment effect on the treated. 
 We would ideally like to be choosing the weights of the SC so that it is as close as possible to the untreated potential outcomes of the unit of interest in the post-treatment time periods. When the units follow a linear factor model, so each unit's untreated outcomes are a linear combination of a set of common factors plus an idiosyncratic shock, we would like to do this by making the weighted average of the control units' loadings on these factors equal to the loadings of the unit of interest. I index the units $\{-K, -K+1, ...., -1, 0, 1,...,J\}$. I let $\mathcal{J} = \{1,...,J\}$ denote the indices for the units that will compose the Synthetic Control, which I refer to as the control units, and let $\mathcal{K} = \{-K, -K+1,..., -1\}$ denote the indices for a disjoint set of units that will only be used as instruments. For the method to work, it is key that the control units are never treated whereas for the instrument units it only matters that they are untreated in the pre-treatment time periods (i.e., time periods that are prior to the treatment of the $0$-th unit). This means that units besides the $0$-th unit which are untreated in the pre-treatment time periods but become treated at some point in the post-treatment time periods can only be included in $\mathcal{K}$, but for units who are never treated, the researcher has a choice of whether to include them in $\mathcal{J}$ or $\mathcal{K}$. For the formal results in this section, I will assume that this choice has been fixed, but in section \ref{Moment and Model Selection} I discuss methods for making this choice. I denote the sets of indices for the pre-treatment time periods and post-treatment time periods as $\mathcal{T}_0$ with $T_0 = |\mathcal{T}_0|$ and $\mathcal{T}_1$ with $T_1 = |\mathcal{T}_1|$ respectively. In all of my formal results, I treat the factor loadings and treatment assignment as fixed but the latent factors, dynamic treatment effects, and idiosyncratic shocks as stochastic. 

\textbf{Assumption 1} \phantomsection\label{A1} (Linear Factor Model) For all units $i \in \{ -K,...,J\}$ and time periods $t \in \mathcal{T}_0\bigcup \mathcal{T}_1$, outcomes follow a linear factor model with $F$ factors so that
\begin{equation*}
    Y_{it} = \alpha_{it} D_{it} + \lambda_t \mu_i  + \epsilon_{it} 
\end{equation*}

where $D_{it}$ is an indicator function equal to 1 if and only if the $i$-th unit is treated in the $t$-th time period. 
\par
I let $\mu$ denote the $F \times (K+J+1)$ matrix of factor loadings with $\mu_i$ being its $i$-th column, $\lambda$ denote the $(T_0+T_1) \times F$ matrix of realizations of the factors with $\lambda_t$ being is $t$-th row, and $\epsilon$ denote the $(K+J+1) \times (T_0 + T_1)$ matrix of idiosyncratic shocks. Additionally, I use the subscripts $\mathcal{J}$ and $\mathcal{K}$ to denote sub-matrices for only the units $j \in \mathcal{J}$ and $k \in \mathcal{K}$ respectively and the superscripts $pre$ and $post$ to denote the sub-matrices for only pre-treatment and post-treatment values respectively. I allow for treatment effects to vary by both time and unit by letting $\alpha_{it}$ denote the effect of treatment on the $i$-th unit in the $t$-th time period. It is worth noting that in Assumption \hyperref[A1]{1}, by making $Y_{it}$ only a function of treatment for the $i$-th unit in the $t$-th time period, this implicitly rules out any spillover effects or anticipatory effects of treatment as is common in this literature. If anticipatory effects are a concern, it may be best to include a small number of time periods immediately prior to treatment for which anticipatory effects are most plausible in $\mathcal{T}_1$ instead of $\mathcal{T}_0$.\footnote{For example, the time periods in which a law has been passed but not yet enacted, or a program has been announced but not yet implemented.} Similarly, if spillover effects are a concern, then units neighboring the unit of interest should likely be included in $\mathcal{K}$ instead on $\mathcal{J}$.
\par
We would like to make the factor loadings of the Synthetic Control $\mu_{\mathcal{J}} W$ equal to the factor loadings of the unit of interest $\mu_0$ and in this case we would have $$Y_{0t} = \lambda_t \mu_0 + \epsilon_{0t} = \lambda_t \mu_{\mathcal{J}}W + \epsilon_{0t}.$$
So if we observed the latent factor structure, we could estimate the control weights by an OLS regression of $Y_{0t}$ on $\lambda_t \mu_{\mathcal{J}}$. However, because $\lambda_t$ and $\mu_{\mathcal{J}}$ are not (fully) observed, we could instead perform a regression on $Y_{\mathcal{J},t}$, which can be thought of as a noisy estimate of $\lambda_t \mu_{\mathcal{J}}$. I refer to this method as OLS-SCE. This gives us a regression based on
\begin{equation*}
    Y_{0t} = (Y_{\mathcal{J},t} - \epsilon_{\mathcal{J},t})W + \epsilon_{0t} = Y_{\mathcal{J},t}W + (\epsilon_{0t} - \epsilon_{\mathcal{J},t}W).
\end{equation*}
Here we can see that $Y_{\mathcal{J},t}$ is endogenous because it is correlated with $\epsilon_{\mathcal{J},t}$. Based on this, it is shown formally by \cite{ImperfectFit} that the OLS-SCE will experience bias analogous to the attenuation bias caused by measurement error in other context, where the weights are pulled towards zero based on the variances of the idiosyncratic shocks of the control units. One solution used to deal with this bias is to find valid instruments and estimate the parameters using GMM. In the context of the linear factor model, other units can be a natural choice. This approach is analogous to the standard strategy for dealing with classical measurement error in linear models when multiple measurements are available.
\par
For each $k \in \mathcal{K}$, there will be a moment condition based on $\sum_{t \in \mathcal{T}_0} E[ Y_{kt}(Y_{0t} -  Y_{\mathcal{J},t} W)] = 0$.
Using a unit $k \in \mathcal{K}$ as an instrument this way, it will clearly be relevant (i.e., correlated with $Y_{\mathcal{J},t}$) when it is exposed to the same latent factors as the control units. Furthermore, if the idiosyncratic shocks are uncorrelated across units and uncorrelated with the factors, then $\sum_{t \in \mathcal{T}_0} E[Y_{kt}(\epsilon_{0t} - \epsilon_{\mathcal{J},t}W)] = 0$ so it will satisfy the exclusion restriction. In addition to having $K$ moment conditions based on second moments, I also have a moment condition based on the $0$-th unit and the SC having the same first moment. That is, I will have a moment condition based on $\sum_{t \in \mathcal{T}_0} E[Y_{0t} - Y_{\mathcal{J},t} W] = 0$. Typically for GMM, the number of moment equations needs to be at least as large as the number of parameters being estimated to achieve identification, which in this case would mean $K+1 \geq J$. However, if we are actually interested only in making the factor loadings of the SC $\mu_{\mathcal{J}} W$ equal to the $0$-th unit's factor loadings $\mu_0$ and not in finding the “true” weights of the SC, then identification of $\mu_0$ is all that is required. Since $\mu_0$ is an F-dimensional vector, it will instead only be necessary to have $F$ linearly independent moment equations. I allow for the distributions of the outcome variables to be different in each time period, so $E[ Y_{kt}(Y_{0t} - Y_{\mathcal{J},t} W)]$ and $E[Y_{0t} - Y_{\mathcal{J},t} W]$ may vary with $t$. Since the sample average over pre-treatment time periods of these moment conditions is what is actually used in the estimation of the control weights, it is not necessary for these expectations to be exactly equal to zero in each pre-treatment time period to have $\mu_{\mathcal{J}}W = \mu_0$ (e.g., $E[\epsilon_{0t}]$ or $E[\epsilon_{kt}\epsilon_{0t}]$ may be non-zero for a vanishingly small fraction of $t \in \mathcal{T}_0$). Instead, it will be sufficient for $\mu_0$ to be asymptotically identified that $plim_{T_0 \rightarrow \infty} \frac{1}{T_0} \sum_{t \in \mathcal{T}_0}Y_{kt}(Y_{0t} - Y_{\mathcal{J},t}W) = 0 \; \text{for all } k \in \mathcal{K}$ and $plim_{T_0 \rightarrow \infty} \frac{1}{T_0} \sum_{t \in \mathcal{T}_0} (Y_{0t} - Y_{\mathcal{J},t} W) = 0$ if and only if $\mu_{\mathcal{J}} W = \mu_0$. Using these moment conditions, we can estimate the control weights as
\begin{equation} \label{eq: Time-Series Averaged SC}
    \hat{W} \in \argmin_{W \in \Delta^{J}} (\frac{1}{T_0} \Tilde{Y}_{\mathcal{K}}^{pre}
     (Y_0^{pre'}
     - Y_{\mathcal{J}}^{pre'} W))'A_{T_0}(\frac{1}{T_0} {\Tilde{Y}_{\mathcal{K}}}^{pre} 
     ({Y_{0}}^{pre'} - {Y_{\mathcal{J}}^{pre}}' W))
\end{equation}
where $\Delta^J = \{ W \geq 0_J : \sum_{j \in \mathcal{J}} W_j = 1 \}$ denotes the unit $J-1$-simplex, $A_{T_0}$ is some $(K+1 )\times (K+1)$ matrix used to weight the moment conditions, and $\Tilde{Y}_{\mathcal{K}}^{pre} = \begin{pmatrix}     \textbf{1}_{T_0} \\     Y_{\mathcal{K}}^{pre}     \end{pmatrix}$ where $\textbf{1}_{T_0}$ denotes a $1 \times T_0$ vector of ones.

\par
I refer to Synthetic Control Estimators that estimate the control weights using GMM as in equation \eqref{eq: Time-Series Averaged SC} as a GMM-SCE. As shown below in Proposition \hyperref[P1]{1}, a variety of choices for $A_{T_0}$ will work to achieve asymptotically unbiased estimates of the treatment effect with $\hat{\alpha}_{0t} \coloneqq Y_{0t} - \sum_{j \in \mathcal{J}} \hat{W}_j Y_{jt} \; \text{for all } t \in \mathcal{T}_1$, but they will differ in terms of the asymptotic variance of $\hat{W}$. A simple choice would be to make it equal to the identity matrix, $A_{T_0} = I_{K+1}$, or to do a two-step feasible approach where $A_{T_0} = I_{K+1}$ in the first step and in the second step $A_{T_0} = \hat{\mathcal{V}}(\hat{W})^{-1}$ where $\hat{\mathcal{V}}(\hat{W})$ is an estimate of the long-run variance matrix of the sample moment conditions using the first step estimate of the control weights. The two-step estimator is a common choice is many settings, however, it may fail to be optimal for several reasons. First, long-run variance estimators can perform poorly in small samples and \cite{HWANG2018} show that the two-step GMM estimator may even be inferior to the one-step estimator in such cases. Furthermore, if $W$ is only partially identified, then $\hat{\mathcal{V}}(\hat{W})$ will have a stochastic limit. Lastly, since $W$ is not the parameter of interest, rather than choosing $A_{T_0}$ to minimize the asymptotic variance of $\hat{W}$, it would be more reasonable to choose it to minimize the mean squared error (MSE) for the estimate of interest (most likely the average treatment effect on the treated). As shown in Appendix \hyperref[ApC]{C}, I need to impose strong conditions in order to characterize the asymptotic distribution of the estimated average treatment effect and as a result characterizing its asymptotic MSE is beyond the scope of this paper. However, if an expression for its asymptotic MSE can be found under more plausible conditions, then choosing $A_{T_0}$ to minimize that expression would be reasonable.
\par 
 Relatedly, it is possible for the solution to equation \eqref{eq: Time-Series Averaged SC} to not be unique. However, the argument made by \cite{LitReview} can be extended to this estimator to show that the simplex constraints induced sparsity, and as a result $\hat{W}$ is almost always unique. More specifically, assume prefect fit is not feasible (i.e., $A_{T_0}^{1/2}\Tilde{Y}_{\mathcal{K}}^{pre}Y_0^{pre'}$ lies outside the convex hull of the columns of $A_{T_0}^{1/2}\Tilde{Y}_{\mathcal{K}}^{pre}Y_{\mathcal{J}}^{pre'}$ so the sample moment conditions cannot all be made exactly equal to zero). This is often the case due to the curse of dimensionality. This means $A_{T_0}^{1/2}\Tilde{Y}_{\mathcal{K}}^{pre}Y_0^{pre'}$ will be projected onto a face of the convex hull of the columns of $A_{T_0}^{1/2}\Tilde{Y}_{\mathcal{K}}^{pre}Y_{\mathcal{J}}^{pre'}$ and only a small number of control units $S$ whose vectors lie on that face will receive positive weight in any solution to equation \eqref{eq: Time-Series Averaged SC}. Therefore, if there is no set of $m$ columns of $A_{T_0}^{1/2}\Tilde{Y}_{\mathcal{K}}^{pre}Y_{\mathcal{J}}^{pre'}$, with $2 \leq m \leq S + 1$, that fall into an ($m-2$)-dimensional hyperplane, then $\hat{W}$ is unique and sparse with the number of non-zero weights bounded by $S$. This means that in practice the non-uniqueness of $\hat{W}$ is not a significant concern, but when $\hat{W}$ is not unique, the formal results I present below will hold regardless of which optimal solution is chosen.
\par 
In order to analyze this method's formal properties, I impose additional conditions on the factor model.
\textbf{Assumption 2} \phantomsection\label{A2} As $T_0 \rightarrow \infty$ while $K$ is fixed and either $J$ is fixed or $J \rightarrow \infty$, 
\begin{enumerate}
    \item $\max_{k <0, i \geq 0} \{ | \frac{1}{T_0} \sum_{t \in \mathcal{T}_0} \epsilon_{it}\epsilon_{kt} | \} \overset{p}{\rightarrow} 0$.
    \item $(\lambda^{pre'}\lambda^{pre}/T_0) \overset{p}{\rightarrow} \Omega_0$, $\frac{1}{T_0}\sum_{t \in \mathcal{T}_0} \lambda_t \overset{p}{\rightarrow} \Bar{\lambda}_0$ for some non-stochastic $\Omega_0$ and $\Bar{\lambda}_0$, and $\sup_{t \in \mathcal{T}_1} E[||\lambda_t ||_2^2] < \infty$
    \item $rank(\Lambda_{\mathcal{K}} ) = F$ for the non-stochastic $K+1 \times F$ matrix $\Lambda_{\mathcal{K}} = \begin{pmatrix}     \Bar{\lambda}_0 \\ \mu_{\mathcal{K}}' \Omega_0   \end{pmatrix}$ and the sequence of factor loadings $\mu_i$ is uniformly bounded.
    \item $\max_{-K \leq i \leq J} \{ | \frac{1}{T_0} \sum_{t \in \mathcal{T}_0} \lambda_{tf} \epsilon_{it} | \} \overset{p}{\rightarrow} 0 \; \text{for all } f \in \{1,...,F\}$.
    \item $E[\epsilon_{it}] = 0,  \; 
\sup_{i \in \{0,...,J\}} E[\epsilon_{it}^2] < \infty \; \text{for all } t \in \mathcal{T}_1$ and $\max_{0 \leq i \leq J} \{| \frac{1}{T_0} \sum_{t \in \mathcal{T}_0} \epsilon_{it} | \} \overset{p}{\rightarrow} 0$.
\end{enumerate}
\par 
Assumption \hyperref[A2]{2.1} and Assumption \hyperref[A2]{2.4} help ensure the exclusion restriction holds by limiting the degree of correlation of the idiosyncratic shocks across units and with the factors so that any of the covariance of the outcomes across units that remains asymptotically must be attributable to the factors. The first part of Assumption \hyperref[A2]{2.3} helps to guarantee that there are at least $F$ linearly independent moment conditions to make $\mu_0$ identified. It implies that $K + 1 \geq F$ and also rules out the possibility of any new factors “turning on” in the post-treatment time periods. The second part of Assumption \hyperref[A2]{2.3} is imposed following  \cite{LargeSampleProperties} to ensure that the set of factor loadings that can be chosen for the SC is compact in the case where $J$ goes to infinity, but note that it is trivially satisfied when the set of control units is fixed. It is also important to highlight that Assumption \hyperref[A2]{2.5} only applies to the unit of interest and the control units. This means that $\epsilon_{kt}$ for $k \in \mathcal{K}$ may not be mean-zero and can have long-run trends. Intuitively, because only the control units are used in the counterfactual of the unit of interest, it is only necessary for their trends to be driven by the same factors driving the trends of the unit of interest. Whereas for the instrument units, it is really only necessary that the set of factors they and the unit of interest share exposure to is the same set of factors they and the control units share exposure to. Also, note that since the analysis is being done conditional on treatment assignment, having the idiosyncratic shocks for both the controls and unit of interest be mean-zero means that if we were treating $D_{it}$ as stochastic, we would want it to be uncorrelated with the idiosyncratic shocks. This captures the fact that the Synthetic Control method is able to adjust for confounding variables that affect the outcome across units but not confounding variables that are idiosyncratic to a particular unit.
\par 
In the case where $J \rightarrow \infty$, the reason it is sufficient to have a maximum of averages converges to zero in Assumptions \hyperlink{A2}{2.1}, \hyperref[A2]{2.4}, and \hyperref[A2]{2.5} is because $W$ is constrained to have a bounded L1 norm. For example, for convergence of the sample moment conditions to the population moment conditions uniformly in $W$, we want $\sup_{W \in \Delta^J} |\sum_{j \in \mathcal{J}} W_j \frac{1}{T_0} \sum_{t \in \mathcal{T}_0} \epsilon_{jt}| \overset{p}{\rightarrow} 0$, but $\sup_{W \in \Delta^J} |\sum_{j \in \mathcal{J}} W_j \frac{1}{T_0} \sum_{t \in \mathcal{T}_0} \epsilon_{jt}| \leq \max_{j \in \mathcal{J}} | \frac{1}{T_0} \sum_{t \in \mathcal{T}_0} \epsilon_{jt}|$ because $||W||_1 = 1$. Only requiring that the maximum of these averages converge to zero allows for cases where $J$ is growing faster than $T_0$, which illustrates how the constraints function as a form of regularization when $J$ is large. Suppose that $E[\epsilon_{it}] = 0$, $E[\lambda_t] = \Bar{\lambda}_0$, $E[\epsilon_{it}\epsilon_{ht}] = 0$, $E[\lambda_t'\lambda_t] = \Omega_0$, and $E[\lambda_t \epsilon_{it}] = 0_F$ for all $t,h,$ and $i$ and there exists $\delta > 0$ such that $J/T_0^d \rightarrow 0$. I show in Appendix \hyperref[ApB]{B}, that Assumptions \hyperref[A2]{2.1}, \hyperref[A2]{2.2}, \hyperref[A2]{2.4}, and \hyperref[A2]{2.5} would be satisfied if $(\lambda_t, \epsilon_t ')$ is $\alpha$-mixing with exponential speed, has uniformly bounded fourth moments, and has exponential bounds on the tails of their distributions. Note that this would allow the distributions of $\lambda_t$ and $\epsilon_t$ to change over time so the factors and idiosyncratic shocks would not be stationary, but Assumption \hyperref[A2]{2} would still hold. Imposing that the factors are mean-invariant may still seem undesirable but it is possible to loosen this condition as well. For example, the approach of \cite{t-test2022} could be taken where there is a factor $\phi_t$ that is allowed to be diverging but all units have the same exposure to it so $Y_{it} = \phi_t + \lambda_t \mu_i + \epsilon_{it}$. Because $W \in \Delta^J$, $Y_{0t} - Y_{\mathcal{J},t}W = \epsilon_0 - \epsilon_{\mathcal{J},t}W + \lambda_t(\mu_0 - \mu_{\mathcal{J}}W)$. Then if the differences between units in $\mathcal{K}$ are used as instruments, the diverging factor has no effect on the estimation of $\hat{W}$. This allows for any cointegration system whose nonstationarity is driven by $\phi_t$. Furthermore, Assumption \hyperref[A2]{2.2} may be altered so that $\sum_{t \in \mathcal{T}_0} \lambda^{pre'}\lambda^{pre}/T_0^\gamma \overset{p}{\rightarrow} \Omega_0$ and $\sum_{t \in \mathcal{T}_0} \lambda_t /T_0^\gamma \overset{p}{\rightarrow} \Bar{\lambda}_0$ for some $\gamma \geq 1$ which can allow for more general forms of nonstationarity in the outcome variables of a cointegration-type. See Appendix \hyperref[ApB]{B} for more details and examples.
\par
The fact that Assumption \hyperref[A2]{2} places restrictions related to the first and second moments of the factors and idiosyncratic shocks is what allows me to use moment conditions based on the first and second moments of the outcome variables. If we were to place additional restrictions on the factors and idiosyncratic shocks, then additional moment conditions could be used. For example, if the factors and idiosyncratic shocks were pair-wise independent and $E[g(Y_{kt})^2] < \infty$ for some measurable function $g$, then the moment equations $E[g(Y_{kt})(Y_{0t} - Y_{\mathcal{J},t}W)] = 0 \; \text{for all } k \in \mathcal{K}$ would hold when $\mu_0 = \mu_{\mathcal{J}}W$ as well. Such stronger independence conditions are also necessary for the conditional moment equations used by \cite{shi2023}. Using $Y_{kt}$ as one of their proxy variables, we would want to non-parametrically estimate a function $g$ satisfying $E[Y_{0t} - g(Y_{\mathcal{J},t})|Y_{kt}] = 0$. \cite{Powell2021} uses a moment equation for each control unit $j \in \mathcal{J}$ of $E[Y_{jt}(Y_{0t} - \sum_{1\leq h \ne j \leq J} W_h Y_{ht} - W_j \hat{Y}_{jt})]$ where $\hat{Y}_{jt}$ is estimated by constructing synthetic units for all the control units using similar moment conditions. Because the synthetic units affect the moment conditions and vice versa, they must both be iteratively re-estimated (see section \ref{Simulations} for more details). This approach bypasses the need for two separate groups $\mathcal{J}$ and $\mathcal{K}$, potentially making it superior when the number of units is small and avoiding the question of how to split units into $\mathcal{J}$ and $\mathcal{K}$. On the other hand, the results only hold for a fixed number of controls and because synthetic units are estimated for all units, there must exist good weights for some of the controls' synthetic units and not just the synthetic unit for the $0$-th unit.\footnote{They require for each unit $i \in \{0,...,J\}$ that either its factor loadings can be expressed as a convex combination of the other units factor loadings or there exists another unit $j \in \{0,...,J\}$ whose factor loadings can be expressed as a convex combination of the other units factor loadings where positive weight is placed on the $i$-th unit.} If the idiosyncratic shocks are also strongly exogenous in the sense that $E[\lambda_t \epsilon_{is}] = 0_F$ and $E[\epsilon_{it}\epsilon_{hs}] = 0$ for all $i,h$ and for all $t, s \in \mathcal{T}_0$, then lagged values of the units' outcomes may also be used as instruments (e.g., $E[Y_{0,t-1}(Y_{0t} - Y_{\mathcal{J},t}W)] = 0$). \cite{sunBen-MichaelFeller2023} consider the setting where there are multiple outcome variables that are exposed to the same underlying factors. In this context, it could be reasonable to use values for some outcomes as instruments when estimating treatment effects for other outcomes if the idiosyncratic shocks for different outcomes are uncorrelated.
\par
Additionally, if the factor loadings for some factors are observable, this information could be used to create additional moment equations. For example, if the $f$-th factor loading corresponded to an observable covariate, the moment equation $\mu_{f0} = \sum_{j \in \mathcal{J}} \mu_{fj} W$ could be used. Similarly, knowledge of the structure could be used to alter the estimating equation. For example, if one of the factors is a unit fixed effect or a linear time trend, then an intercept term $W_0$ or a linear function of time $t W_0$ could be added. If some factor loadings are estimated directly, then it is not required that it's possible to give the SC the same loadings on those factors as the unit of interest. The resulting estimator can be handled using the same GMM approach.\footnote{One potential complication is if these additional parameters are unconstrained, then the parameter space would no longer be compact, where compactness plays a role in the proof of Lemma \hyperref[L1]{1} as described below. However, it may be possible to deal with this using the convexity of the objective function (see, for example, Theorem 2.7 of \cite{Newey-McFadden}).} I focus on presenting results for the version of the estimator that doesn't directly use knowledge of the factor structure, because the results for this estimator will be valid regardless of whether some factors or factor loadings are observable.
\par 
 Focusing on the version of the GMM-SCE given in equation \eqref{eq: Time-Series Averaged SC}, I show that this Synthetic Control's factor loadings $\mu_{\mathcal{J}} \hat{W}$ will converge in probability to $\mu_0$ if it is feasible. For the case where $J$ is fixed, by feasible I mean that a set of control weights exists that will give the SC the same factor loadings as the unit of interest, so $\mu_0 \in \mathcal{M}_{\mathcal{J}} \coloneqq \{\mu : \mu = \mu_{\mathcal{J}} W \text{ for some } W \in \Delta^J \}$. When $J \rightarrow \infty$, there only needs to be a sequence of feasible factor loadings which converge to the factor loadings of the $0$-th unit, so $\mu_0 \in \mathcal{M} \coloneqq cl(\cup_{J \in \mathbb{N}} \mathcal{M}_{\mathcal{J}})$. The plausibility of this assumption must be evaluated based on the relative size of $J$ and $F$ as well as specific knowledge about the factor structure. If additional assumptions are made about the factor structure, the assumption that $\mu_0 \in \mathcal{M}_{\mathcal{J}}$ can become much simpler to analyze. For example, when the data follow a two-way fixed effect model, there are only two factors: one for time fixed effects and one for unit fixed effects. Since all units have identical loadings on the time fixed effect factor, all that is needed for $\mu_0 \in \mathcal{M}_{\mathcal{J}}$ is for the unit fixed effect of the $0$-th unit to be in between the minimum and maximum unit fixed effects of the control units. If the idiosyncratic shocks have mean zero, this means that $\mu_0 \in \mathcal{M}_{\mathcal{J}}$ exactly when $\min_{j\in \mathcal{J}} E[Y_{jt}] \leq E[Y_{0t}] \leq \max_{j \in \mathcal{J}} E[Y_{jt}]$. For the more general case, I show in the simulation results that when $J$ is much larger than $F$ it is very likely that $\mu_0 \in \mathcal{M}_{\mathcal{J}}$ even when no restrictions are placed on the factor model.

\textbf{Lemma 1} \phantomsection\label{L1} Suppose that Assumptions \hyperref[A1]{1} and \hyperref[A2]{2} hold. Additionally, $A_{T_0} \overset{p}{\rightarrow} A$ where $A$ is positive definite, as $T_0 \rightarrow \infty$ and $K$ is fixed while either
\begin{enumerate}[label=(\roman*)]
    \item $J$ is fixed and $\mu_0 \in \mathcal{M}_{\mathcal{J}}$ or
    \item $J \rightarrow \infty$ and $\mu_0 \in \mathcal{M}$.
\end{enumerate}
Then $\hat{\mu}_0 \coloneqq \mu_{\mathcal{J}}\hat{W} \overset{p}{\rightarrow} \mu_0$. 
\par 
 There are some additional complications in the proof in the case where $J \rightarrow \infty$ because the dimension of $\hat{W}$ is going to infinity, but this is dealt with using the technique of \cite{LargeSampleProperties} where the optimization problem in equation \eqref{eq: Time-Series Averaged SC} is reformulated as choosing the implied factor loadings of the SC $\hat{\mu}_0$ rather than the control weights $\hat{W}$. The idea behind this reformulation is that choosing the control weights is equivalent to first choosing the factor loadings for the SC from the set $\mathcal{M}_{\mathcal{J}}$ and then choosing the control weights among those consistent with these factor loadings. In other words, for any $\hat{W}$ satisfying equation \eqref{eq: Time-Series Averaged SC},
\begin{equation*}
    \mu_{\mathcal{J}}\hat{W} = \min_{\Tilde{\mu} \in \mathcal{M}_{\mathcal{J}}} \min_{W \in \Delta^J: \mu_{\mathcal{J}} W = \Tilde{\mu}} \{  (\frac{1}{T_0} \Tilde{Y}_{\mathcal{K}}^{pre}(Y_{0}^{pre'} - Y_{\mathcal{J}}^{pre'}W))'A_{T_0}(\frac{1}{T_0} \Tilde{Y}_{\mathcal{K}}^{pre}(Y_{0}^{pre'} - Y_{\mathcal{J}}^{pre'}W))  \}.
\end{equation*}
Additionally, to deal with the fact that the feasible set $\mathcal{M}_{\mathcal{J}}$ is expanding as $J$ grows, one can instead have the feasible set be $\mathcal{M}$ and add a penalty term to the objective function to ensure that at each $J$, $\hat{\mu}_0 \in \mathcal{M}_{\mathcal{J}}$. This can be done by making the objective function equal to 
 \begin{equation*}
    H_J(\mu) =  \min_{\Tilde{\mu} \in \mathcal{M}_{\mathcal{J}}} \{ \min_{W \in \Delta^J: \mu_{\mathcal{J}} W = \Tilde{\mu}} \{  (\frac{1}{T_0} \Tilde{Y}_{\mathcal{K}}^{pre}(Y_{0}^{pre'} - Y_{\mathcal{J}}^{pre'}W))'A_{T_0}(\frac{1}{T_0} \Tilde{Y}_{\mathcal{K}}^{pre}(Y_{0}^{pre'} - Y_{\mathcal{J}}^{pre'}W))  \} + \eta||\mu - \Tilde{\mu}||_2 \}.
\end{equation*}
The exact value of $\eta$ is given in the proof of Lemma \hyperref[L1]{1} but for $\mu_{\mathcal{J}}\hat{W} = \argmin_{\mu \in \mathcal{M}} H_J(\mu)$, it is sufficient that $\eta > 0$. Note that since the factor loadings of the control units are bounded by Assumption \hyperref[A2]{2.3} and $W \in \Delta^J$, the set of asymptotically feasible factor loadings $\mathcal{M}$ will be compact. This is also part of the reason why it is convenient to treat the factor loadings as fixed, since otherwise the feasible set $\mathcal{M}$ would be stochastic. The constraints $W \in \Delta^J$ play a key role in the proof of Lemma \hyperref[L1]{1}, because the consistency of $\hat{\mu}_0$ is shown using the compactness of $\mathcal{M}$ along with the fact that $H_J(\mu)$ is getting close to a continuous function which is uniquely minimized at $\mu_0$. Using this result, I now examine the formal properties of the estimated treatment effects $\hat{\alpha}_{0t} \coloneqq Y_{0t} - \sum_{j \in \mathcal{J}} \hat{W}_j Y_{jt} \; \text{for all } t \in \mathcal{T}_1$. I show that the bias of $\hat{\alpha}_{0t}$ will converge to zero for both fixed $J$ and when $J \rightarrow \infty$.

\textbf{Proposition 1} \phantomsection\label{P1} Suppose that Assumptions \hyperref[A1]{1} and \hyperref[A2]{2} are satisfied and $A_{T_0} \overset{p}{\rightarrow} A$ where $A$ is positive definite, as $T_0 \rightarrow \infty$ and $K$ is fixed while either 
\begin{enumerate}[label=(\roman*)]
    \item $J$ is fixed and $\mu_0 \in \mathcal{M}_{\mathcal{J}}$ or
    \item $J \rightarrow \infty$, $\mu_0 \in \mathcal{M}$, and $\epsilon_{\mathcal{J}}^{post} \indep Y^{pre}$.
\end{enumerate}
Then  $\text{for all } t \in \mathcal{T}_1, \; \lim_{T_0 \rightarrow \infty}E[\hat{\alpha}_{0t} - \alpha_{0t}] = 0$.
\par
Here, the constraints placed on $\hat{W}$ help to guarantee the existence of $E[\hat{\alpha}_{0t} - \alpha_{0t}]$.\footnote{Note that $E[\hat{\alpha}_{0t} - \alpha_{0t}]$ exists because  $|E[\hat{\alpha}_{0t} - \alpha_{0t}]| \leq E[||\lambda_t||_2^2]^{\frac{1}{2}}E[||\mu_0 - \hat{\mu}_0||_2^2]^{\frac{1}{2}} + |E[\epsilon_{0t}]| + E[||\hat{W}||_2^2]^{\frac{1}{2}}E[||\epsilon_{\mathcal{J},t}||_2^2]^{\frac{1}{2}}$ where $E[\epsilon_{0t}]$, $E[||\lambda_t||_2^2]$, and $E[||\epsilon_{\mathcal{J},t}||_2^2]$ exist due to Assumption \hyperref[A2]{2}, $E[||\hat{W}||_2^2] \leq 1$, and $E[||\mu_0 - \hat{\mu}_0||_2^2]^{\frac{1}{2}}$ exists because $\hat{\mu}_0$ is bounded by $||\hat{W}||_1 = 1$ and Assumption \hyperref[A2]{2.3}.} The additional assumption on the independence of the control units' idiosyncratic shocks in the post-treatment time periods and the pre-treatment outcomes in the case where $J \rightarrow \infty$ is imposed because, even if $\hat{\mu}_0 = \mu_0$, $\hat{\alpha}_{0t}$ may still be biased if $ E[\sum_{j \in \mathcal{J}} \hat{W}_j \epsilon_{jt}] \ne 0$. Therefore, by the same reasoning as \cite{LargeSampleProperties}, since $\hat{W}$ is estimated using only pre-treatment outcomes it can be made uncorrelated with these post-treatment idiosyncratic shocks if they are independent of the pre-treatment outcomes. For the case where $J$ is fixed, this independence assumption is not necessary because $\hat{W}$ will converge to a fixed set of control weights that only depend on the factor loadings and as a result $\hat{W}$ can be shown to be asymptotically uncorrelated with the idiosyncratic shocks.
\par
Unfortunately, it is impossible to consistently estimate the treatment effect for the unit of interest in a single post-treatment time period $\alpha_{0t}$ due to the idiosyncratic shocks that unit incurs in that time period $\epsilon_{0t}$. However, if we are instead interested in the average effect of treatment for this unit $\frac{1}{T_1} \sum_{t \in \mathcal{T}_1}  \alpha_{0t}$, which is the case in many policy applications, this can be consistently estimated if the number of post-treatment time periods is also large so that the idiosyncratic shocks of the $0$-th unit as well as the idiosyncratic shocks of the control units in the post-treatment time periods may be averaged out. More generally, if we are interested in estimating some weighted average of the treatment effects in the post-treatment time periods, $\sum_{t \in \mathcal{T}_1} v_t \alpha_{0t}$\footnote{Here, the dependence of the vector $v$ on $T_1$ is suppressed for notational simplicity.}, our estimate of this $\sum_{t \in \mathcal{T}_1} v_t \hat{\alpha}_{0t}$ will be consistent as long as the sequence of weights $v$ are being spread out over more and more post-treatment time periods to be able to average out the idiosyncratic shocks of the control units and the unit of interest.

\textbf{Assumption 3} \phantomsection\label{A3}
Suppose that as $T_1 \rightarrow \infty$ while either $J$ is fixed or $J \rightarrow \infty$, $\max_{0 \leq i \leq J} \{| \sum_{t \in \mathcal{T}_1} v_t \epsilon_{it} | \} \overset{p}{\rightarrow} 0$.
\par
What conditions this places on the weights $v$ will depend on the dependence across the idiosyncratic shocks and how fast $J$ grows relative to $T_1$. For example, if all idiosyncratic shocks were independent across time with bounded fourth moments, then 
$\max_{0 \leq i \leq J} \{| \sum_{t \in \mathcal{T}_1} v_t \epsilon_{it} | \} \overset{p}{\rightarrow} 0$ as $T_1 \rightarrow \infty$ if $J^{\frac{1}{4}} ||v||_2 \rightarrow 0$ as $T_1 \rightarrow \infty$ (see Appendix \hyperref[ApB]{B} for more details). 
We will often be particularly interested in the average effect of treatment so $v = (\frac{1}{T_1},...,\frac{1}{T_1})$. For this case, Assumption \hyperref[A3]{3.2} is essentially the same as Assumption \hyperref[A2]{2.5} but imposed in the post-treatment time periods so it will hold under analogous conditions. While equal weights is the most common choice, there are also cases where we would expect an intervention's effect to be delayed. In such cases, it could be reasonable to not place weight on the time periods at the start of the post-treatment block.  

\textbf{Proposition 2}\phantomsection\label{P2} Suppose that Assumptions \hyperref[A1]{1}, \hyperref[A2]{2}, and \hyperref[A3]{3} are satisfied and $A_{T_0} \overset{p}{\rightarrow} A$ where $A$ is positive definite, as $T_0 , T_1 \rightarrow \infty$ and $K$ is fixed and while either 
\begin{enumerate}[label=(\roman*)]
    \item $J$ is fixed and $\mu_0 \in \mathcal{M}_{\mathcal{J}}$ or
    \item $J \rightarrow \infty$ and $\mu_0 \in \mathcal{M}$.
\end{enumerate}
Then $ \sum_{t \in \mathcal{T}_1} v_t (\hat{\alpha}_{0t} - \alpha_{0t}) \overset{p}{\rightarrow} 0$.
\par
Note that no restrictions are placed on $\{ \alpha_{0t} \}_{t \in \mathcal{T}_1}$. In cases where the weighted average of treatment effects is converging to some long-run average (e.g., $\frac{1}{T_0}\sum_{t \in \mathcal{T}_1}\alpha_{0t} \overset{p}{\rightarrow} \Bar{\alpha}$ for some $\Bar{\alpha}$), the weighted average of estimated treatment effects will consistently estimate this. When this is not the case, say if $\sum_{t \in \mathcal{T}_1}v_t \alpha_{0t}$ is diverging, we can still be confident that the difference between $\sum_{t \in \mathcal{T}_1}v_t \hat{\alpha}_{0t}$ and $\sum_{t \in \mathcal{T}_1} v_t \alpha_{0t}$ is small when $T_0$ and $T_1$ are large. When dealing with long panels, this provides motivation for using this estimator even if having low bias isn't a priority relative to having low variance. However, there are two reasonable concerns one can have with regard to large-$T_1$ asymptotic results of this form. First, the effect of some interventions may be expected to be transitory (e.g., only be non-zero for several years after the intervention occurred/started). In this case, including time periods that are too far into the future may cause us to incorrectly conclude that there was no effect of the intervention. Second, in many cases, the longer a time period is from the start of treatment, the less plausible the SC will be as a counterfactual. This is because of the accumulating risk of non-transitory shocks besides the treatment effect, which weren't relevant in the pre-treatment time periods. As mentioned earlier, Assumption \hyperref[A2]{2.3} implicitly imposes that no new latent factors turn on in the post-treatment time periods, but this condition is less realistic when $T_1$ is very large. Because of this, Proposition \hyperref[A2]{2} and other large-$T_1$ asymptotic results in the literature are likely most relevant in cases where the number of time periods is large due to the post-treatment being a moderately sized length of time and the frequency of data collection being relatively high. For example, \cite{Carl2021} use monthly crime rate data for nine years following their intervention. Other recent papers have also started using SCEs in settings with monthly or weekly data such as \cite{GibsonSun2023}, \cite{ClarkeEtAl2023}, and \cite{Li2020}. 
\par 
In the cases where $T_1$ is small, the consistency result of Proposition \hyperref[P2]{2} is not applicable, but the asymptotic unbiasedness result of Proposition \hyperref[P1]{1} will still hold for any single time period and Lemma \hyperref[L1]{1} also implies that for each $t \in \mathcal{T}_1$, $\lambda_t(\mu_0 - \mu_{\mathcal{J}}\hat{W}) \overset{p}{\rightarrow} 0$ as $T_0 \rightarrow \infty$. This may still prove useful for conducting inference with methods that work with a fixed $T_1$, such as the conformal inference method of \cite{ConformalInference} and the end-of-sample instability test of \cite{Andrews2003}. For example, $\sum_{t \in \mathcal{T}_0}(\lambda_t(\mu_0 - \mu_{\mathcal{J}}\hat{W}))^2/T_0 \overset{p}{\rightarrow} 0$ and $\lambda_t(\mu_0 - \mu_{\mathcal{J}}\hat{W}) \overset{p}{\rightarrow} 0$ for each $t \in \mathcal{T}_1$ imply Assumption 3 of \cite{ConformalInference}. I discuss these methods further in section \ref{Discussion}.

\section{Moment and Model Selection} \label{Moment and Model Selection}

I now discuss the choice of assigning units to the sets $\mathcal{J}$ and $\mathcal{K}$. As mentioned before, the instrument units need to be untreated in the pre-treatment time periods, but because the instruments' post-treatment data is not used, there are no restrictions that need to be imposed on their behavior during the post-treatment time periods. Therefore, there will be cases in which some units satisfy the conditions to be instruments but not to be control units. One important case is when there are units besides the unit of interest that become treated at the same time or a later time. Additionally, if something else happens to a unit in the post-treatment time periods that makes it no longer satisfy the assumptions above, such as receiving some different intervention, then it would only be reasonable to use it as an instrument unit. In the empirical application in section \ref{Empirical Application}, I provide an example of this. Another way this could happen is when spillover effects are a concern. In such cases, units which neighbor the unit of interest should be excluded from the set of controls, but could still be included as instruments. Lastly, if post-treatment data is missing for certain units, it may be natural to still use them as instruments. However, assuming a never-treated unit satisfies the assumptions of the model in both the pre-treatment and post-treatment time periods, there will be a trade-off between increasing $J$ by including it as a control and increasing $K$ by including it as an instrument. It is important to have a procedure to address this choice, both to improve the performance of the GMM-SCE and to limit the room for specification searching by researchers.\footnote{The assumptions in section \ref{Model} are done given the units' assignment to $\mathcal{J}$ and $\mathcal{K}$. This means that taking a data-driven approach to assigning units to $\mathcal{J}$ and $\mathcal{K}$ may cause these assumptions to be violated. Further work is needed to examine if these results continue to hold after different model selection procedures are used.} Here, I present two methods for choosing $\mathcal{J}$ and $\mathcal{K}$.
\par
First, I discuss the method of \cite{MomentSelection}. They provide a method of consistent model and moment selection for GMM estimators based on the over-identification test statistic. I interpret their consistency results in this context and show that their conditions are satisfied for the GMM-SCE under conditions similar to the assumptions in section \ref{Model}. I now consider the case where there are a total of $N_0$ never-treated units indexed $\mathcal{N}_0 = \{1,...,N_0\}$, the $0$-th unit which is the unit of interest, and $N_1$ units indexed $\mathcal{N}_1 = \{-N_1,...,-1\}$. The units in $\mathcal{N}_1$ don't necessarily need to be treated. They just need to be the set of units for which we have strong theoretical grounds for only using as instruments, whereas the units in $\mathcal{N}_0$ may be included in $\mathcal{J}$ or $\mathcal{K}$. Because the results of \cite{MomentSelection} require that the number of models being consider is fixed, I analyze their method in the case where the number of units is fixed. Since it is also common in applications to have the number of units be a similar size to the number of pre-treatment time periods, I also present an alternative model selection method and compare their performance in the simulations. Let $\mathcal{L} = \{(\mathcal{J},\mathcal{K}) : \mathcal{K} \cup \mathcal{J} = \mathcal{N}_0 \cup \mathcal{N}_1,\;\mathcal{J}\cap\mathcal{N}_1= \emptyset, \text{ and } \mathcal{K} \cap \mathcal{J} = \emptyset \}$ denote the set of possible partitions of units into controls and instruments. I now impose the following conditions, similar to Assumptions \hyperref[A1]{1} and \hyperref[A2]{2}.
\par 
\textbf{Assumption 1*} \phantomsection\label{A1*} For all units $i \in \{-N_1,...,N_0\}$ and time periods $t \in \mathcal{T}_0\bigcup \mathcal{T}_1$, outcomes follow a linear factor model with $F$ factors so that
\begin{equation*}
    Y_{it} = \alpha_{it} D_{it} + \lambda_t \mu_i  + \epsilon_{it} 
\end{equation*}

where $D_{it} = 0$ if $t \in \mathcal{T}_0$ or if $i \in \mathcal{N}_0$. 

\textbf{Assumption 2*} \phantomsection\label{A2*}
\begin{enumerate}
    \item $\max_{-N_1 \leq h \ne i \leq N_0} \{ | \frac{1}{T_0} \sum_{t \in \mathcal{T}_0} \epsilon_{it}\epsilon_{ht} | \} = O_p(T_0^{-\frac{1}{2}})$ and for each $i \in \mathcal{N}_0$, $\frac{1}{T_0} \sum_{t \in \mathcal{T}_0} \epsilon_i^2 = \sigma_i^2 + O_p(T_0^{-\frac{1}{2}})$.
    \item $(\lambda^{pre'}\lambda^{pre}/T_0) = \Omega_0 + O_p(T_0^{-\frac{1}{2}})$ and $\frac{1}{T_0}\sum_{t \in \mathcal{T}_0} \lambda_t = \Bar{\lambda}_0 + O_p(T_0^{-\frac{1}{2}})$.
    \item $rank(\Omega_0) = F$.
    \item $\max_{-N_1 \leq i \leq N_0} \{ || \frac{1}{T_0} \sum_{t \in \mathcal{T}_0} \lambda_{t} \epsilon_{it} ||_1 \} = O_p(T_0^{-\frac{1}{2}})$.
    \item $\max_{-N_1 \leq i \leq N_0} \{| \frac{1}{T_0} \sum_{t \in \mathcal{T}_0} \epsilon_{it} | \} = O_p(T_0^{-\frac{1}{2}})$.
    \item For each, $(\mathcal{\mathcal{J}},\mathcal{K}) \in \mathcal{L}$, there exists a positive definite matrix $A_{\mathcal{J},\mathcal{K}}$ such that $A_{T_0,\mathcal{J},\mathcal{K}} \overset{p}{\rightarrow} A_{\mathcal{J},\mathcal{K}}$ as $T_0 \rightarrow \infty$.
\end{enumerate}
Here, $A_{T_0,\mathcal{J},\mathcal{K}}$ denotes the weighting matrix used for a specific choice of $\mathcal{J}$ and $\mathcal{K}$. Under Assumptions \hyperref[A1*]{1*} and \hyperref[A2*]{2*}, if $\mathcal{N}_0$ was randomly partitioned into $\mathcal{J}$ and part of $\mathcal{K}$, all the conditions needed for Lemma \hyperref[L1]{1} would be guaranteed if for that partition it also holds that $rank(\Lambda_{\mathcal{K}}) = F$ (where $\Lambda_{\mathcal{K}} = \begin{pmatrix}
    \Bar{\lambda}_0 \\
    \mu_{\mathcal{K}}\Omega_0
\end{pmatrix}$) and $\mu_0 \in \mathcal{M}_{\mathcal{J}}$.\footnote{
While the rate of convergence of $O_p(T_0^{-\frac{1}{2}})$ is stronger than the convergence conditions in Assumption \hyperref[A2]{2}, many Law of Large Numbers results can provide this rate of convergence. Furthermore, if a different rate of convergence holds, the method of \cite{MomentSelection} can easily be adjusted by changing what function of $T_0$ the test statistic is multiplied by and how the critical value used grows with $T_0$} Therefore, in order to select a choice of parameters and moment conditions that will result in the SC reconstructing the factor loadings of the treated unit, we want a choice of $\mathcal{J}$ and $\mathcal{K}$ such that $rank(\Lambda_{\mathcal{K}}) = F$ and $\mu_0 \in \mathcal{M}_{\mathcal{J}}$. Let $\mathcal{L}^0 = \{ (\mathcal{J},\mathcal{K}) \in \mathcal{L} : rank(\Lambda_{\mathcal{K}}) = F,\; \mu_0 = \mu_{\mathcal{J}} W \text{ for some } W \in \Delta^J \}$ denote the set of "correct" partitions and $\mathcal{B}\mathcal{C} \subseteq \mathcal{L}$ denote the set of partitions to be chosen from. \cite{MomentSelection} provide a downward testing selection procedure which asymptotically will choose the model in $\mathcal{BC}$ that provides the most over-identifying restrictions while still having all chosen population moment conditions be satisfied by a choice of the parameter values in the feasible set. In this context, this will mean that as long as $\mathcal{L}^0 \cap \mathcal{BC}$ is non-empty, with probability approaching one we can select a model in $\mathcal{L}^0 \cap \mathcal{BC}$ where $K$ is as large as possible relative to $J$. Let $\mathcal{MBCL}^0 = \{ (\mathcal{J},\mathcal{K}) \in \mathcal{L}^0\cap \mathcal{BC} : |\mathcal{J}| \leq |\mathcal{J}^*| \text{ for all } (\mathcal{J}^*,\mathcal{K}^*) \in \mathcal{L}^0\cap \mathcal{BC}\}$ denote this set. This procedure is done by starting with the partitions in $\mathcal{BC}$ where $J$ is smallest and estimating the Sargan–Hansen statistic
\begin{equation*}
    SH_{T_0}(\mathcal{J},\mathcal{K}) = \min_{W \in \Delta^J} T_0g_{T_0,\mathcal{J},\mathcal{K}}(W)'A_{T_0,\mathcal{J},\mathcal{K}}g_{T_0,\mathcal{J},\mathcal{K}}(W),
\end{equation*}
where $g_{T_0,\mathcal{J},\mathcal{K}}(W) = \begin{pmatrix}
    \textbf{1}_{T_0} \\
    Y_{\mathcal{K}}^{pre}
\end{pmatrix}(Y_0^{pre'} - Y_{\mathcal{J}}^{pre'}W)/T_0$. This is then tested against some threshold $\gamma_{T_0,J,K}$. Based on \cite{MomentSelection}, we would use $\gamma_{T_0,J,K} = \chi^2_{K+1-J}(\alpha_{T_0})$ so it is the critical value at the $\alpha_{T_0}$ level of significance of a chi-squared distribution with degrees of freedom equal to the number of over-identifying restrictions. To allow for the fact that $J$ may be larger than $K+1$, I use $\gamma_{T_0,J,K} = \chi^2_{\max \{1, K+1-J\}}(\alpha_{T_0})$. This doesn't affect their consistency results, as the exact distribution that the critical values are calculated from is not used at all. Instead, all that is required is that the critical values diverge at a sufficiently slow rate as $T_0$ grows.  

\textbf{Assumption 3*}\phantomsection\label{A3*} $\gamma_{T_0,J,K} \rightarrow \infty$ and $\gamma_{T_0,J,K} = o(T_0)$ as $T_0 \rightarrow \infty$ for all $J,K$. 

By Theorem 5.8 of \cite{Potscher1983}, this will hold when $\alpha_{T_0} \rightarrow 0$ as $T_0 \rightarrow \infty$ with $\ln (\alpha_{T_0}) = o(T_0)$. This procedure is repeated with each partition in $\mathcal{B}\mathcal{C}$ starting from those with the smallest $J$ moving to those with the largest $J$ until a partition is found for which $SH_{T_0}(\mathcal{J},\mathcal{K}) < \gamma_{T_0,J,K}$. I denote this selected partition as $(\hat{\mathcal{J}},\hat{\mathcal{K}})$. Intuitively, the consistency comes from the fact that under Assumptions \hyperref[A1*]{1*} and \hyperref[A2*]{2*}, $SH_{T_0}(\mathcal{J},\mathcal{K})$ should diverge if and only if $E[g_{T_0,\mathcal{J},\mathcal{K}}(W)] \ne 0_{K+1}$ for all $W \in \Delta^J$. Note that there will be partitions where $E[g_{T_0,\mathcal{J},\mathcal{K}}(W)] = 0_{K+1}$ for some $W \in \Delta^J$ but there are not enough instruments to identify $\mu_0$. However, because of the downward nature of the testing procedure where we test models with more moments first, it is easier to choose $\mathcal{BC}$ such that we'll test a "correct” model in $\mathcal{L}^0$ before testing a model where $\mu_0$ is unidentified. Then using Theorem 2 of \cite{MomentSelection}, we will achieve a form of consistent model selection.

\textbf{Proposition 3}\phantomsection\label{P3} Suppose Assumptions \hyperref[A1*]{1*}, \hyperref[A2*]{2*}, and \hyperref[A3*]{3*} hold and $\mathcal{BC} \cap \mathcal{L}^0 \ne \emptyset$. Furthermore, suppose that for all $(\mathcal{J},\mathcal{K}) \in \mathcal{BC}$, if $|\mathcal{K}| \geq |\mathcal{K}^*|$ for some $(\mathcal{J}^*,\mathcal{K}^*) \in \mathcal{MBCL}^0$, then $rank(\Lambda_{\mathcal{K}}) = F$. Then $P((\hat{\mathcal{J}},\hat{\mathcal{K}}) \in \mathcal{MBCL}^0) \rightarrow 1$ as $T_0 \rightarrow \infty$ and $N_0$ and $N_1$ are fixed. If additionally, $\mathcal{MBCL}^0$ contains a single element $(\mathcal{J}^0,\mathcal{K}^0)$ and there is a unique $W^0 \in \Delta^{|\mathcal{J}^0|}$ such that $\mu_0 = \mu_{\mathcal{J}^0} W^0$, then $\hat{W} \overset{p}{\rightarrow} W^0$. 
\par
The condition that if $|\mathcal{K}| \geq |\mathcal{K}^*|$ for some $(\mathcal{J}^*,\mathcal{K}^*) \in \mathcal{MBCL}^0$, then $rank(\Lambda_{\mathcal{K}}) = F$, helps prevent models where $\mu_0$ is unidentified from being considered before a "correct” model is considered. One simple way to help guarantee this holds is to have the models in $\mathcal{BC}$ simply be a set of $N_0$ models where the units in $\mathcal{N}_0$ are sequentially moved from $\mathcal{K}$ to $\mathcal{J}$. Another benefit of this approach is that even though the size of $\mathcal{L}$ will be growing exponentially in $N_0$, the size of $\mathcal{BC}$ will then only be growing linearly in $N_0$. This both makes the model selection problem computationally more tractable and makes the consistency result more applicable than it would be if $|\mathcal{BC}| >> T_0$. One choice for how to pick the order in which units are switched from instruments to controls is to start by using the units that are most similar to the unit of interest in the pre-treatment time periods. Specifically, calculate the pre-treatment mean squared difference for each unit $i \in\mathcal{N}_0$ as $MSE(i) = \sum_{t \in \mathcal{T}_0}(Y_{0t} - Y_{it})^2/T_0$. Then sort them from smallest to largest and let $I_{MSE(i)}$ denote the place of $MSE(i)$ in that ordering. Then the models to be considered are $\hat{\mathcal{BC}} = \{(\mathcal{J}_1,\mathcal{K}_1), ... ,(\mathcal{J}_{N_0},\mathcal{K}_{N_0})\}$ where $\mathcal{J}_n = \{i \in \mathcal{N}_0 : I_{MSE(i)} \leq n\}$ and $\mathcal{K}_n = \mathcal{N}_1 \cup \{i \in \mathcal{N}_0 : I_{MSE(i)} > n\}$. Since $\sum_{t \in \mathcal{T}_0} (Y_{0t} - Y_{it})^2/T_0 \overset{p}{\rightarrow} (\mu_0 - \mu_i)'\Omega_0(\mu_0 - \mu_i) + \sigma_0^2 + \sigma_i^2$ under Assumptions \hyperref[A1*]{1*} and \hyperref[A2*]{2*}, this procedure will tend to include units as controls whose factor loadings are more similar to $\mu_0$ and who have less variation in their idiosyncratic shocks.

\textbf{Remark 1} Let $\mathcal{BC}^* = \{(\mathcal{J},\mathcal{K}) \in \mathcal{L}: \text{for each } j \in \mathcal{J} \text{ and } k \in \mathcal{K},\text{ either } k \in \mathcal{N}_1 \text{ or } 
(\mu_0 - \mu_j)' \Omega_0(\mu_0 - \mu_j) + \sigma_j^2 < (\mu_0 - \mu_k)' \Omega_0(\mu_0 - \mu_k) + \sigma_k^2\}$. Since $\hat{\mathcal{BC}} = \mathcal{BC}^*$ with probability approaching one under Assumptions \hyperref[A1*]{1*} and \hyperref[A2*]{2*}, we want it to be the case that there exists a partition in $\mathcal{BC}^* \cap \mathcal{L}^0$. When  $rank(\Lambda_{\mathcal{N}_1}) = F$ and $\mu_0 \in \mathcal{M}_{\mathcal{N}_0}$, this trivially holds with $\mathcal{J} = \mathcal{N}_0$ and $\mathcal{K} = \mathcal{N}_1$. In the case where $N_1 = 0$, if the idiosyncratic shocks are homogeneous across units, then the condition $(\mu_0 - \mu_j)'\Omega_0(\mu_0 - \mu_j) + \sigma_j^2 < (\mu_0 - \mu_k)' \Omega_0(\mu_0 - \mu_k) + \sigma_k^2$ becomes $(\mu_0 - \mu_j)'\Omega_0(\mu_0 - \mu_j) < (\mu_0 - \mu_k)' \Omega_0(\mu_0 - \mu_k)$. So if $\mu_0$ is in the convex hull of the $N_0 - F$ closest factor loadings of units in $\mathcal{N}_0$ (and the remaining $F$ factor loadings are linearly independent), then $\mathcal{BC}^* \cap \mathcal{L}^0 \neq \emptyset$. In this case, $\hat{\mathcal{BC}} \cap \mathcal{L}^0 \neq \emptyset$ with probability approaching one.
\par 
Since the consistency results of \cite{MomentSelection} imposes that the number of models is fixed, they may not be applicable when $N_0$ is not small relative to $T_0$. More recently, there have been GMM estimators proposed that perform parameter and/or moment selection using a regularization penalty, where consistency is achieved while allowing the number of moments or parameters to go to infinity. \cite{ChengLiao2015} introduce a GMM estimator that uses LASSO to select moments when there are many potentially valid moment conditions. Similarly, \cite{ElasticNetGMM} provide a GMM estimator that uses an Elastic Net penalty to simultaneously choose which moment conditions are included and which parameters will be non-zero. However, one limitation is that their methods require that a subset of the moment conditions be known initially to be valid. In this context, their condition would require that identification can be achieved with only the units in $\mathcal{N}_1$ so $rank(\Lambda_{\mathcal{N}_1}) = F$.\footnote{An additional subtlety in applying these approaches to this context is that, unlike in their papers, there is a dependence between whether a moment condition is valid and whether a parameter is non-zero. Namely, whether a unit is a valid instrument depends on whether it is receiving positive weights in the SC.} Like the LASSO penalty, the simplex constraints also induce sparsity. I therefore consider a second approach to model selection that takes advantage of this and can achieve a form of consistency under a condition similar to the one needed by \cite{ChengLiao2015} and \cite{ElasticNetGMM}. If the researcher is in the scenario where there are several units in $\mathcal{N}_1$, a two-step procedure can be performed. In the first step, $\hat{W}$ is estimated with $\mathcal{J} = \mathcal{N}_0$ and $\mathcal{K} = \mathcal{N}_1$, and in the second step all control units who received zero weight in $\hat{W}$ are transferred from $\mathcal{J}$ to $\mathcal{K}$. For the consistency of this two-step approach, if we impose that $rank(\Lambda_{\mathcal{N}_1}) = F$, then $rank(\Lambda_{\hat{\mathcal{K}}}) =F$ since
$\mathcal{N}_1 \subseteq \hat{\mathcal{K}}$. Therefore, the key question is whether $\mu_0 \in \mathcal{M}_{\hat{\mathcal{J}}}$.

\textbf{Remark 2}\phantomsection\label{R2} Suppose that for $\mathcal{K} = \mathcal{N}_1$ and $\mathcal{J} = \mathcal{N}_0$ with $N_1$ fixed and $N_0,T_0 \rightarrow \infty$, the conditions of Lemma \hyperref[L1]{1} are satisfied so $\hat{\mu}_0 \overset{p}{\rightarrow} \mu_0$. Let $\gamma_{N_0} = \min_{\Tilde{\mathcal{J}} \subset \mathcal{N}_0 : \mu_0 \notin \mathcal{M}_{\Tilde{\mathcal{J}}}} \min_{W \in \Delta^J} ||\mu_0 - \mu_{\mathcal{J}}W||_2$. Suppose that $\inf_{N_0}\gamma_{N_0} = \gamma^* >0$. Since $\hat{\mu}_0 \overset{p}{\rightarrow} \mu_0$, $||\hat{\mu}_0 - \mu_0||_2 < \gamma^*$ with probability approaching one. Therefore, $\mathcal{P} := \{i \in \mathcal{N}_0: \hat{W}_i > 0\} \not\subseteq \Tilde{\mathcal{J}}$ for all $\Tilde{\mathcal{J}} \subset \mathcal{N}_0$ such that $\mu_0 \notin \mathcal{M}_{\Tilde{\mathcal{J}}}$ with probability approaching one. Then for $\hat{\mathcal{J}} = \mathcal{P}$ and $\hat{\mathcal{K}} = \mathcal{N}_1 \cup \mathcal{N}_0 \setminus \mathcal{P}$, $(\hat{\mathcal{J}}, \hat{\mathcal{K}}) \in \mathcal{L}^0$ with probability approaching one.
\par
Remark \hyperref[R2]{2} makes a fairly simple argument that if there is a lower bound for all the incorrect choices of controls (ones where $\mu_0 \notin \mathcal{M}_{\mathcal{J}}$) on how close the SC can come to reconstructing the factor loadings of the unit of interest, then the probability of selecting any of these incorrect choices of controls must converge to zero since $\hat{\mu}_0 \overset{p}{\rightarrow} \mu_0$ by Lemma \hyperref[L1]{1}. Both of these model selection procedures have their own limitations, so it is quite possible that superior procedures exist. Further exploration of model selection methods is a promising area for future research, as the results may be applicable to other SCEs in addition to the one presented here. I compare the performance of the two-step procedure with the sequential version of \cite{MomentSelection}'s approach in the simulations below.

\section{Simulations} \label{Simulations}

The asymptotic results motivate some interest in this method, particularly when both the number of pre-treatment and post-treatment time periods are large. However, they do not guarantee that this estimator will have better finite sample performance than existing panel methods. I therefore also conduct simulations to compare the finite sample performance of this GMM-SCE to existing SCE and a method that estimates the factor structure. I compare them in terms of bias, MSE for time period specific treatment effects, and MSE for the equally weighted average treatment effect, which I denote as $\Bar{\alpha}$. For similar reasons as with the formal results, when conducting simulations I focus on the case where the data follow a linear factor model. However, I would also like to conduct the analysis using data that accurately reflects the properties of real data that researchers using the method are working with, like in the Placebo Studies of \cite{BertrandDufloMullainathan} and \cite{SDID}. Therefore, I conduct a Placebo Study by following the approach of \cite{SDID} by first fitting a linear factor model on GDP data from the \cite{WB-GDP-Growth}. This data set contains the annual growth rates of real GDP from 86 countries for 60 consecutive years from 1961 to 2020. This fits with many SC applications since there are many time periods, the outcome is observed for most units in the population of interest, and it is plausible that many policy interventions we would like to know the effect of would happen to either a single unit or a small number of units. To estimate the linear factor model, I estimate the number of factors using the Singular Value Thresholding method of \cite{OptimalSVThresholding}.\footnote{Here, the number of factors is chosen to be equal to the number of singular values greater than the median singular value times 2.858.} Using this method, I estimate that there are four factors. I then estimate the factor loadings and factor realizations using Principal Components Analysis (see e.g., \cite{Bai2009_ife} for more details):
\begin{equation*}
    (\hat{\lambda}, \hat{\mu}) = \argmin_{\lambda, \mu} \sum_{i} \sum_{t} (Y_{it} - \sum_{f =1}^4 \lambda_{tf}\mu_{fi})^2 \; s.t. \; \lambda'\lambda/60 = I_4, \; \mu \mu' = \text{diagonal}.
\end{equation*}
\par
In order to allow the number of time periods to vary, I fit the estimated values of the factors $\hat{\lambda}$ and the residuals $\hat{\epsilon}_{it} = Y_{it} - \sum_{f =1}^4 \hat{\lambda}_{tf}\hat{\mu}_{fi}$ to models and then re-sample them. For the factors, I fit each factor to an ARIMA model, using AIC for model selection and QMLE to estimate the parameters.\footnote{This is done using the auto.arima() function in the forecast package in R.} For the idiosyncratic shocks, I have them be mean zero, normally distributed, and independent across both unit and time. I also have each unit have the same variance over time but allow for heteroscedasticity across units by using the sample variance of $\{\hat{\epsilon}_{it}\}_{t=1961}^{2020}$ for each $i$, which is consistent with my formal results. I have the factor loadings remain fixed but now re-sample which unit is the unit of interest as well as which units are in $\mathcal{N}_0$ and $\mathcal{N}_1$. I focus on presenting results for the case where treatment assignment is done uniformly at random, but in Appendix \hyperref[ApD]{D} I also follow \cite{SDID}'s example by considering a case where the probability of being treated is estimated by fitting on the World Bank's indicator for country income level. When I do this, I obtain results very similar to those presented below, indicating that these results are still relevant when treatment assignment is correlated with the factor loadings. 
\par 
I compare variations of my GMM-SCE to an SCE with uniform control weights ($W_j = \frac{1}{J} \; \text{for all } j \in \mathcal{J}$), an approach that estimates the factor model, the method of \cite{Powell2021}, and the OLS-SCE discussed above. While placing uniform weights on the control units is not used in practice often for SCEs, it provides a benchmark to see how much improvement is being achieved by fitting on different sets of predictor variables relative to a simpler matching approach. For the method that explicitly estimates the factor structure, it follows an approach similar to \cite{IFE_with_SC}. The never-treated units outcomes in both $\mathcal{T}_0$ and $\mathcal{T}_1$ are used to estimate the number of factors and realizations of the factors. This is done using the same Singular Value Thresholding and Principal Components Analysis methods mentioned above. The factor loadings of the unit of interest are then estimated by regressing the unit of interest on the estimated factors in the pre-treatment time periods so $\hat{\mu}_0 = (\hat{\lambda}^{pre'}\hat{\lambda}^{pre})^{-1}\hat{\lambda}^{pre'}Y_0^{pre'}$. Then the treatment effects are estimated as $\hat{\alpha}_{0t} = \hat{\lambda}_t \hat{\mu}_0$. For the OLS-SCE, since it uses pre-treatment outcomes as predictors and weights each of these predictors evenly, it chooses the control weights to minimize pre-treatment MSE. This is the SCE analyzed by \cite{LargeSampleProperties}, \cite{ImperfectFit}, \cite{Li2020}, and others. As noted by \cite{IncludingAllLags}, for this choice of predictors, this is also an optimal solution to \cite{Abadie2010}'s algorithm for weighting predictors, which is one of the most commonly used methods in practice. It involves choosing a vector of weights $V$ for the predictors using the bi-level optimization problem below\footnote{Here there are $N$ predictors $\{X_1,...,X_N\}$. The reasoning for why choosing to weight the predictors evenly is an optimal solution is based on the fact that when the predictors are equal to the outcomes in each pre-treatment time period, by making $V_n = \frac{1}{T_0} \; \text{for all } n$, the objective functions of the inner and outer optimization problem become the same function of $W$.}:
\begin{equation}  \label{eq: AndG 2003 SC}
	\min_{V \in \Delta^N} \sum_{t \in \mathcal{T}_0} \frac{1}{T_0} (Y_{0t} - \sum_{j=1}^J W_j(V) Y_{jt})^2 
\end{equation}
where
\begin{equation*}
	W(V) = \argmin_{W \in \Delta^J} \sum_{n=1}^N V_n (X_{0n} - \sum_{j=1}^J W_{j} X_{jn})^2.
\end{equation*}
For both my GMM-SCE and the estimator of \cite{Powell2021}, I estimate with the moments weighted evenly ($A_{T_0} = I_{K+1}$). In Appendix \hyperref[ApD]{D}, I show that using $A_{T_0} = \hat{\mathcal{V}}(\hat{W})^{-1}$  produces similar results. For the approach of \cite{Powell2021}, in addition to estimating a synthetic unit for the unit of interest $i=0$, one is estimated for every never-treated unit $i \in \mathcal{N}_0 = \{1,...,N_0\}$ using the other never-treated units and the unit of interest as the set of controls. The weights for the synthetic $i$-th unit, denoted as $W^i$, are estimated using the moment conditions $\frac{1}{T_0}\sum_{t \in \mathcal{T}_0}Y_{jt}(Y_{it} - \sum_{ 0 \leq h \ne i,j} W_h^i Y_{ht} - W_j^i \hat{Y}_{jt})$ for each $0 \leq j \ne i$ where $\hat{Y}_{jt}$ is an estimated value of the $j$-th unit. These estimated values of the units are calculated using all the synthetic units. Specifically,
$$\hat{Y}_{jt} = \frac{\sum_{0 \leq i \ne j}a_i [(Y_{it} - \sum_{0 \leq h \ne i,j} \hat{W}_h^i Y_{ht})\hat{W}_j^i] + a_j \sum_{0 \leq h \ne j} \hat{W}_h^j Y_{ht}}{a_j + \sum_{0 \leq i \ne j} a_i (\hat{W}_j^i)^2},$$
where $a_0, ..., a_{N_0}$ are weights capturing the goodness-of-fit for each synthetic unit with $a_i$ equal to the minimized value of the objective function for the $i$-th synthetic unit. Since the estimated units and the weights depend on one another, they are both repeatedly re-estimated.\footnote{For my simulations, I use the uniform control weights as the initial values of both the control weights $W$ and the goodness-of-fit weights $a$. I repeatedly solve for the weights and estimated units 10 times.} Once the synthetic units have been found, they are used to estimate treatment effects by regressing $(Y_{it} - \sum_{0 \leq j \ne i} \hat{W}_j^i Y_{jt})$ on $(D_{it} - \sum_{0 \leq j \ne i} \hat{W}_j^i D_{jt})$. He suggests doing this using a weighted least squares regression where $a_0, ..., a_{N_0}$ are the weights.
 \par
 As discussed in section \ref{Moment and Model Selection}, in some contexts there will be units which can be included as instruments but not as controls, while in other contexts there may be only a set of never-treated units that must be divided into instruments and controls. Because the relative performance of the GMM-SCE may depend on this, I consider two different scenarios for the simulations: one where the unit of interest is the only treated unit ($N_1 = 0$) and one where there are 20 other treated units which can be only be used as instruments ($N_1 = 20$). In the second case, the unit of interest is selected uniformly at random from the treated units. In both cases, $N_0$ never-treated units are sampled uniformly at random from the remaining units. The OLS-SCE and the SCE with equal weights use all of these units as controls so $J = N_0$. For the GMM-SCE in the first scenario, the $N_0$ units are divided into $\mathcal{J}$ and $\mathcal{K}$ using the sequential version of \cite{MomentSelection}'s method where the pvalue is set to .05.\footnote{Similar results are obtained for other choices of the pvalue such as .1 and .01.} For the second scenario, I additionally include a GMM-SCE that uses the two-step model selection procedure from section \ref{Moment and Model Selection}. I also add a version of the GMM-SCE that is unconstrained in this second scenario. The unconstrained GMM-SCE uses $\mathcal{K} = \mathcal{N}_1$ and selects the unconstrained minimum to the objective function in equation \eqref{eq: Time-Series Averaged SC} with the minimum L2 norm. This allows for a comparison of the performance of the GMM-SCE with the simplex constraints to one using an alternative regularization method.
\par
For all simulations, I have the number of post-treatment time periods $T_1$ equal to 50. In Table \hyperref[Table1]{1}, there is the absolute value of the bias of the estimated average treatment effect $\hat{\Bar{\alpha}}$ (Bias Magnitude)\footnote{Note that by the linearity of expectations, the bias of the $\hat{\Bar{\alpha}}$ is equal to the average bias of the time period specific treatment effects $\hat{\alpha}_{0t}$. Therefore, these results are relevant both if someone is interested in individual treatment effects and if they are only interested in the $\Bar{\alpha}$.}, the MSE for time period specific treatment effects ($\hat{\alpha}_{0t}$ MSE), and the MSE for average treatment effects ($\hat{\Bar{\alpha}}$ MSE) under each of these conditions for the OLS-SCE, the Sequential GMM-SCE (the GMM-SCE using the sequential version of \cite{MomentSelection}'s model selection method), the uniform SCE, the factor method, and \cite{Powell2021}'s method.

\begin{table}[t]\centering\caption{Main Results with One Treated Unit \label{Table1}}\scalebox{.95}{
\begin{threeparttable}
\begin{tabular}{l c*{5}{c}} \hline
          &\multicolumn{1}{c}{OLS-SCE}         &\multicolumn{1}{c}{Sequential} &\multicolumn{1}{c}{Uniform SCE} &\multicolumn{1}{c}{Factor Model}
          &\multicolumn{1}{c}{Powell (2021)}
          \\
          &\multicolumn{1}{c}{}                &\multicolumn{1}{c}{GMM-SCE} &\multicolumn{1}{c}{} &\multicolumn{1}{c}{Estimate}
          &\multicolumn{1}{c}{}\\

\toprule        
\multicolumn{5}{l}{\textbf{Bias Magnitude}} \\ 
\midrule
$T_0 = 25$,$N_0 = 10$ & 0.530 & 0.303 & 1.342 & 0.268 & 1.342\\
$T_0 = 50$,$N_0 = 10$  & 0.482 & 0.202 & 1.342 & 0.251 & 1.502\\
$T_0 = 100$,$N_0 = 10$  & 0.447 & 0.163 & 1.342 & 0.227 & 1.624\\

$T_0 = 25$,$N_0 = 50$  & 0.124 & 0.078 & 1.351 & 0.030 & 1.558\\
$T_0 = 50$,$N_0 = 50$  & 0.102 & 0.055 & 1.351 & 0.025 & 1.873\\
$T_0 = 100$,$N_0 = 50$  & 0.081 & 0.038 & 1.351 & 0.006 & 2.112\\

\hline 
\multicolumn{5}{l}{\textbf{$\hat{\alpha}_{0t}$ MSE}} \\                                 \midrule
$T_0 = 25$,$N_0 = 10$  & 4.754 & 5.875 & 6.171 & 4.791 & 14.022\\
$T_0 = 50$,$N_0 = 10$  & 4.523 & 5.464 & 6.171 & 4.365 & 18.348\\
$T_0 = 100$,$N_0 = 10$  & 4.347 & 5.322 & 6.171 & 4.198 & 22.138\\

$T_0 = 25$,$N_0 = 50$  & 1.927 & 2.419 & 5.728 & 4.287 & 13.217\\
$T_0 = 50$,$N_0 = 50$ & 1.841 & 2.212 & 5.728 & 3.944 & 17.928\\
$T_0 = 100$,$N_0 = 50$ & 1.765 & 2.074 & 5.728 & 3.828 & 22.508\\

\hline 
\multicolumn{5}{l}{\textbf{$\hat{\Bar{\alpha}}$ MSE}} \\                                 \midrule
$T_0 = 25$,$N_0 = 10$ & 0.536 & 0.487 & 2.083 & 0.348 & 4.126\\
$T_0 = 50$,$N_0 = 10$  & 0.445 & 0.326 & 2.083 & 0.252 & 5.550\\
$T_0 = 100$,$N_0 = 10$  & 0.380 & 0.235 & 2.083 & 0.199 & 6.260\\

$T_0 = 25$,$N_0 = 50$ & 0.124 & 0.167 & 1.928 & 0.231 & 6.278\\
$T_0 = 50$,$N_0 = 50$ & 0.092 & 0.107 & 1.928 & 0.152 & 8.887\\
$T_0 = 100$,$N_0 = 50$ & 0.067 & 0.080 & 1.928 & 0.116 & 11.225\\

\hline
\end{tabular}
\begin{tablenotes}
      \small
      \item Notes: All simulations are done with a thousand replications.
\end{tablenotes}

\end{threeparttable}}
\end{table}

\par 
Looking at the results in Table \hyperref[Table1]{1} for the case where $N_1 = 0$, we can see that the factor method, the OLS-SCE, and the GMM-SCE all outperform the SCE with equal weights and the method of \cite{Powell2021} across the various specifications in terms of bias and MSE for both individual and average treatment effects. In the case of \cite{Powell2021}'s method, it is quite possible that there are alternative ways to implement the method that would improve its performance. While increasing the number of times the weights and estimated units are solved for from 10 to 100 does not seem to noticeably improve its performance, exploring alternative algorithms for solving for the equilibrium values of the weights and estimated units may improve its performance and computational cost. In terms of MSE for time period specific treatment effects, we can see the OLS-SCE performs best with $N_0 = 50$ but the factor method largely performs best with $N_0 = 10$. The GMM-SCE underperforms the OLS-SCE, although the difference becomes more modest when the number of units is large. While the OLS-SCE provides the lowest MSE out of the SCEs in these simulations, there is no guarantee that this will be true in general, especially if the factors are not stationary.\footnote{In this case, the estimated ARIMA models for two out of the four factors happen to be stationary.} The bias of the GMM-SCE is smaller than that of the OLS-SCE, even when the number of pre-treatment time periods is small. However, the factor method provides the lowest bias is several cases. For the MSE of $\hat{\Bar{\alpha}}$, the results are more mixed. When $N_0 = 10$, the GMM-SCE outperforms the OLS-SCE but is outperformed by the factor method. While for $N_0 = 50$, the OLS-SCE performs best but the GMM-SCE outperforms the factor method. It may seem surprising that the factor method would be outperformed in some cases, given that it is taking advantage of the true data generating process. One possible explanation is that the factor method tends to underestimate the number of factors, even when using the same method used to construct the simulations. This is likely due to the fact that the first factor explains a large amount of the variation relative to the remaining three. Consistent estimation of the number of factors requires not only both the number of time periods and units be large, but that they are large relative to the amount of signal provided by the latent factors. These results raise the possibility that SCEs may be more robust to problems of some latent factors providing weak signals. I also check in each simulation whether $\mu_0 \in \mathcal{M}_{\mathcal{N}_0}$ to verify that it is feasible to give the SC the same loadings as the unit of interest, since this was assumed in the formal results. When $N_0 = 50$, I find that it is true in 100\% of the simulations, which is unsurprising given that the number of units is substantially larger than the number of factors. When $N_0 = 10$, it is still the case that $\mu_0 \in \mathcal{M}_{\mathcal{N}_0}$ in approximately 56\% of the simulations. 

\begin{table}[t]\centering\caption{Main Results with $N_1$ Treated Units \label{Table2}}\scalebox{.95}{
\begin{threeparttable}
\begin{tabular}{l c*{5}{c}} \hline
          &\multicolumn{1}{c}{OLS-SCE}                &\multicolumn{1}{c}{Sequential}&\multicolumn{1}{c}{Two-Step MS} &\multicolumn{1}{c}{Factor Model}&\multicolumn{1}{c}{Unconstrained} \\
          &\multicolumn{1}{c}{}                &\multicolumn{1}{c}{GMM-SCE}&\multicolumn{1}{c}{GMM-SCE} &\multicolumn{1}{c}{Estimate}&\multicolumn{1}{c}{GMM-SCE}\\

\toprule        
\multicolumn{5}{l}{\textbf{Bias Magnitude}} \\ 
\midrule
$T_0 = 25$,$N_0 = 10$ & 0.530 & 0.259 & 0.240 & 0.268 & 0.121\\
$T_0 = 50$,$N_0 = 10$  & 0.482 & 0.158 & 0.160 & 0.251 & 0.036\\
$T_0 = 100$,$N_0 = 10$ & 0.447 & 0.116 & 0.104 &  0.227 & 0.040\\

$T_0 = 25$,$N_0 = 50$  & 0.124 & 0.076 & 0.045 &  0.030 & 0.071\\
$T_0 = 50$,$N_0 = 50$ & 0.102 & 0.050 & 0.015 & 0.025 & 0.000\\
$T_0 = 100$,$N_0 = 50$ & 0.081 & 0.007 & 0.001 &  0.006 &  0.037\\

\hline 
\multicolumn{5}{l}{\textbf{$\hat{\alpha}_{0t}$ MSE}} \\                                 \midrule
$T_0 = 25$,$N_0 = 10$  & 4.754 & 5.694 & 5.029 & 4.791 & 13.444\\
$T_0 = 50$,$N_0 = 10$  & 4.523 & 5.333 & 4.964 &  4.365 & 12.848\\
$T_0 = 100$,$N_0 = 10$ & 4.347 & 5.209 & 5.002 &  4.198 & 15.235\\

$T_0 = 25$,$N_0 = 50$  & 1.927 & 2.411 & 2.053 &  4.287 & 4.356\\
$T_0 = 50$,$N_0 = 50$ & 1.841 & 2.198 & 2.027 & 3.944 & 5.059\\
$T_0 = 100$,$N_0 = 50$ & 1.765 & 2.071 & 2.015 & 3.828 & 6.705\\

\hline 
\multicolumn{5}{l}{\textbf{$\hat{\Bar{\alpha}}$ MSE}} \\                                 \midrule
$T_0 = 25$,$N_0 = 10$  & 0.536 & 0.422 & 0.316 &  0.348 & 0.803\\
$T_0 = 50$,$N_0 = 10$  & 0.445 & 0.278 & 0.227 &  0.252 & 0.515\\
$T_0 = 100$,$N_0 = 10$ & 0.380 & 0.202 & 0.180 &  0.199 & 0.480\\

$T_0 = 25$,$N_0 = 50$  & 0.124 & 0.164 & 0.109 &  0.231 & 0.230\\
$T_0 = 50$,$N_0 = 50$ & 0.092 & 0.106 & 0.084 & 0.152 & 0.205\\
$T_0 = 100$,$N_0 = 50$ & 0.067 & 0.075 & 0.062 & 0.116 & 0.216\\

\hline
\end{tabular}
\begin{tablenotes}
      \small
      \item Notes: All simulations are done with a thousand replications.
\end{tablenotes}
    \end{threeparttable}}
\end{table}

In Table \ref{Table2}, the results are presented for the case with $N_1 = 20$. Because the conditions are the same for the OLS-SCE, the factor method, \cite{Powell2021}'s method, and the SCE with equal weights, the performance of these estimators is the same as before. Therefore, for conciseness I don't include the results for the SCE with equal weights and \cite{Powell2021}'s method in the table. For the Sequential GMM-SCE, its MSE for individual time period effects is slightly higher but its bias and MSE for $\hat{\Bar{\alpha}}$ are lower. Its performance is relatively similar to the Two-Step GMM-SCE, but the Two-Step GMM-SCE consistently performs slightly better. When comparing the OLS-SCE to the GMM-SCEs, similarly to before, the GMM-SCEs tend to achieve lower bias but the OLS-SCE has lower MSE for estimates of effects in individual time periods. In this case, however, the Two-Step GMM-SCE has lower $\hat{\Bar{\alpha}}$ MSE than both the OLS-SCE and the factor method. The unconstrained GMM-SCE tends to also have low bias, and it is significantly lower than the constrained versions when $N_0 = 10$. This may be due to the unconstrained version still being able to reconstruct the factor loadings of the treated unit even when $\mu_0 \notin \mathcal{M}_{\mathcal{J}}$. However, its MSE is much higher, especially for $\hat{\alpha}_{0t}$. This means it has a much higher variance, and therefore the simplex constraints are providing a stronger form of regularization. 
\par 
We should expect the contribution to the MSE of $\hat{\Bar{\alpha}}$ coming from the variances of individual treatment effect estimates to become smaller and smaller as $T_1$ becomes larger and larger. Therefore, the OLS-SCE's advantage of having lower variance for $\hat{\alpha}_{0t}$ becomes less and less relevant as $T_1$ becomes large. While \cite{LargeSampleProperties} shows that the OLS-SCE can also be asymptotically unbiased when both $J$ and $T_0$ go to infinity, additional assumptions are needed and here we see the OLS-SCE having higher bias even when $J = 50$. On the other hand, if $T_1 = 1$, the $\hat{\alpha}_{0t}$ MSE and $\hat{\Bar{\alpha}}$ MSE would become the same, so the case for using OLS-SCE would be stronger when $T_1$ is very small. Therefore, if bias is our primary criterion for which estimator to use, then the GMM-SCE is likely a better choice. If MSE of $\hat{\Bar{\alpha}}$ is our primary criterion, then these simulation results suggest that the GMM-SCE may be the best choice even when $T_0$ is relatively small, provided that $T_1$ isn't too small.

\section{Empirical Application} \label{Empirical Application}

While the simulations conducted can give some evidence about the finite-sample performance of this SCE and are designed to be similar to real data, it is still worth analyzing how it can be used in an actual empirical application. Therefore, I replicate the work of \cite{Abadie2015} and examine how their estimates change when a GMM-SCE is used. \cite{Abadie2015} evaluate the impact of the German Reunification in 1990 on GDP per capita in West Germany. The pre-treatment time periods are 1960 through 1989 and the post-treatment periods are 1990 through 2003. The set of units they use as controls are the 16 OECD countries: Australia, Austria, Belgium, Denmark, France, Greece, Italy, Japan, the Netherlands, New Zealand, Norway, Portugal, Spain, Switzerland, the United Kingdom, and the United States. The way that they arrived at these 16 countries as their set of controls was by starting with all the countries which were in the OECD in 1990 besides West Germany. This gives them 23 countries, but they excluded 7 of these countries: Luxembourg, Iceland, Canada, Finland, Sweden, Ireland, and Turkey. In the case of Canada, Finland, Sweden, and Ireland, their justification for excluding them is based on economic shocks that happened in the post-treatment time periods which didn't also occur in West Germany.\footnote{More specifically, they cite the Celtic Tiger expansion period in Ireland and the financial and fiscal crises that occurred in Canada, Finland, and Sweden in the 1990s.} This is a reasonable explanation for not including these countries as controls, but because these shocks happened in the post-treatment time periods, it is still reasonable to use them as instrument units when using a GMM-SCE since only the pre-treatment values of instrument units are used. Therefore, this is an example of how even when there is only a single treated unit (only one country is unifying with East Germany), there can still be units which can function as instrument units but not as control units. In the case of Iceland, Luxembourg, and Turkey, they exclude them because their economies are different from the other OECD countries.\footnote{For Iceland and Luxembourg, they cite their small size and "peculiarities of their economies" and for Turkey, they cite its GDP per capita being significantly lower than the other OECD member countries in 1990.} It is less clear to what extent this justification extends to also not using these countries as instrument units. The key assumption for a country to be a valid instrument unit is that the factors playing a role in determining both its GDP per capita and West Germany's GDP per capita should be the same as the factors playing a role in determining both its GDP per capita and GDP per capita in the control countries, in the pre-treatment time periods. This could be true even if there are many factors influencing GDP per capita in these 3 countries that aren't having an effect in West Germany, as long as they also aren't having an effect in the control countries. I present results for the case where for this case with $N_1 = 7$ and $N_0 = 16$, although I find similar results when only Canada, Finland, Sweden, and Ireland are included in $\mathcal{N}_1$.\footnote{GDP per capita data for West Germany and the 16 countries in $\mathcal{N}_0$ is provided by \cite{SynthPackage} and the GDP per capita data for the 7 countries in $\mathcal{N}_1$ is from \cite{pwt-GDP}.}
\par
\cite{Abadie2015} include several specifications for their SCE which vary based on which combination of covariates and pre-treatment values of GDP per capita were used as predictors, as well as which of the 16 control units were included. They report similar estimates across these specifications. In their main specification they include average GDP per capita, investment rate, trade openness, schooling, inflation rate, and industry share as their predictors and all 16 countries are used as control units. Unlike previous work, they use a cross-validation approach to weighing these predictors similar to the approach of \cite{Abadie2010} in equation \eqref{eq: AndG 2003 SC} but using a two-step procedure. They use a training period from 1971 to 1980 and a validation period from 1981 to 1990. In the first step, the nested optimization problem is solved using only training period data in the inner optimization problem and only validation period data in the outer optimization problem. In the second step, the predictor weighting vector $V$ from the first step is used along with data from the validation period in the inner optimization problem to produce the control weights $W$.\footnote{Using this procedure, the weights that they give to each predictor are: average GDP per capita (0.442), investment rate (0.245), trade openness (0.134), schooling (0.107), inflation rate (0.072), and industry share (0.001).} As is common when the Synthetic Control is constrained to be a convex combination of the control units, the weights they find end up being rather sparse. Specifically, only 5 of the 16 control countries included receive positive weight: Austria (0.42), United States (0.22), Japan (0.16), Switzerland (0.11), and the Netherlands (0.09). Using this SC as their counterfactual, they find that on average over the post-treatment time periods, GDP per capita was reduced by about 1600 USD (2002) per year in West Germany due to German Reunification. 

\begin{figure}[h!]  \caption{Gap between West
Germany and Synthetic West
Germany \label{F1}} 
        \centering
        \includegraphics[scale=0.8]{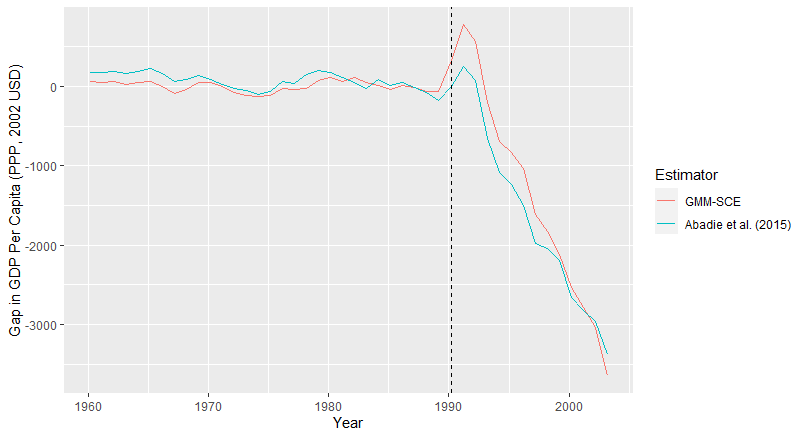}
\end{figure}

\par
When re-estimating the Synthetic West Germany, I use the same set of pre-treatment time periods. Because there are countries in $\mathcal{N}_1$, either model selection procedure could be used. Since the Two-step model selection method is simpler and performs better in the simulations, I focus on using that version of the GMM-SCE here. Using this method, all 16 countries in $\mathcal{N}_0$ are selected to be included as controls, although the weights are concentrated in just a few of them. For the GMM-SCE using $A_{T_0} = I_{K+1}$, the weights are concentrated on the United States (.31), Austria (.18), Switzerland (.14), Greece (.10), and Portugal (.06). This is relative similar to the SC of \cite{Abadie2015} with the U.S., Austria, and Switzerland still receiving a significant amount of weight. I find similar estimates of the average treatment effect using $A_{T_0} = I_{K+1}$ and $A_{T_0} = \hat{\mathcal{V}}(\hat{W})$ for weighting the moment equations, with GDP per capita being reduced by approximately 2000 USD (2002) on average from 1990 to 2003 in West Germany due to German Reunification.\footnote{The exact estimate for the GMM-SCE with $A_{T_0} = \hat{\mathcal{V}}(\hat{W})^{-1}$ is a 1906 average decrease and a 2024 average decrease for the GMM-SCE with $A_{T_0} = I_{K+1}$.}
\par 
The fact that this method gives a similar estimate of the average effect on GDP per capita provides evidence of the robustness of the original finding. The gaps between the Synthetic West Germany and the actual West Germany for both the original SCE and the GMM-SCE are given in Figure \ref{F1}. As illustrated in the figure, the main reason why the estimated magnitude of the average treatment effect is larger using the GMM-SCE is because the estimates begin to diverge around 1996 despite produces quite similar estimates in the initial years following reunification. In fact, by 2003, the estimated decrease in GDP per Capita is more than \$1200 larger using the GMM-SCE.

\section{Discussion and Inference Options} \label{Discussion}

How to conduct inference for SCEs is an active area of research in the literature. Here, I discuss some recently proposed inference methods as well as remaining challenges. In the permutation method of \cite{Abadie2010}, synthetic units are also estimated for each control unit and then these are used to calculate the Root Mean Squared Error (RMSE) for each unit in the post-treatment time periods. This is then used as the test statistic for each unit, possibly after adjusting for pre-treatment RMSE or excluding some units based on pre-treatment RMSE, and then the p-value is calculated using the quantile that the unit of interest's test statistic falls in. As previously noted (see \cite{Abadie2010} and \cite{Abadie2015}), this approach reduces to a traditional Fisher Randomization Test in the case where treatment is randomly assigned. While this would mean that these tests have exact size from a design-based perspective, this assumption of random treatment assignment is unrealistic in almost all current SC applications. Also, \cite{SCandInference} show that when a repeated sampling perspective is taken, strong functional form assumptions must be placed on the distributions of the idiosyncratic shocks to guarantee that there won't be size distortions. Additionally, in cases where the number of controls is very small, the sizes for the test that can be chosen will be restricted (e.g., a test with a size of .05 requires that there be at least 19 control units and a test with a size of .01 requires that there be at least 99 control units). Based on this, especially when $J$ is small and information about the treatment assignment process is lacking, exploring alternative approaches to inference is desirable. When the data show evidence of stationarity, two plausible choices are the conformal inference method of \cite{ConformalInference} and the end-of-sample instability test originally introduced by \cite{Andrews2003} and suggested for SCE by \cite{SCandInference}. Also, \cite{t-test2022} propose a t-test inference method based on a K-fold cross-fitting procedure. Their method will have asymptotically correct size when both $T_0$ and $T_1$ are large. \cite{Li2020} proposed a subsampling method that they show will have asymptotically correct size when both $T_0$ and $T_1$ are large without needing to impose stationarity conditions. While using a block subsampling approach has the advantage of allowing for more general forms of nonstationarity, \cite{ProblemWithSubampling} show that subsampling and m out of n bootstrap methods can have incorrect asymptotic size when the parameter is close to the boundary of the parameter space. \cite{ConformalInference}'s and \cite{Andrews2003}'s methods require the idiosyncratic shocks to be strictly stationary, but they both have the potential advantage that the sizes of their tests will be asymptotically correct when $T_1$ is fixed and only $T_0 \rightarrow \infty$. This means that they may be preferable when $T_1$ is small. The methods of \cite{Abadie2010}, \cite{ConformalInference}, and \cite{Andrews2003} are designed to test the sharp null of there being no effect in every post-treatment time period, rather than testing a null hypothesis about the average treatment effect. While it depends on context, usually testing the sharp null hypothesis is of less interest. However, as pointed out by \cite{ConformalInference}, they can also be used to test the weak null hypothesis by collapsing the data by non-overlapping blocks with length $T_1$. Other approaches have also been suggested, such as the Bayesian methods of \cite{amjad2017robust}, \cite{KimEtAl2020}, and \cite{Tuomaala2019TheBS}. 
\par 
Another issue to considered with inference is the fact that many SCE, including the one presented here, involve a model selection step. Therefore, the researcher must decide whether to conduct inference conditional on the chosen model being selected or unconditionally. Even if the model selection method is consistent, ignoring this step when conducting inference may cause size distortions. Some inference methods, such as the resampling and randomization based approaches, can incorporate the model selection procedure. For example, when using the method of \cite{Li2020}, the sets $\mathcal{J}$ and $\mathcal{K}$ could be chosen for each subsample or when using the method of \cite{Abadie2010}, the choice of $\mathcal{J}$ and $\mathcal{K}$ could be made separately for each synthetic unit constructed. This shows how, if it is desirable, the post-treatment data for never-treated units used as instruments can still be used in inference even if it wasn't used in estimation of treatment effects. However, such approaches can have their own limitations. For example, the finite sample probabilities of choosing different models may be significantly different in each subsample than when the whole sample is used. In addition to the model selection issue, the fact that $W$ is high dimensional and possibly partially identified also present challenges for characterizing the asymptotic distribution of the average treatment effect and performing inference. I show in Appendix \hyperref[ApB]{B} how under stronger conditions, the asymptotic distribution of the average treatment effect can be characterized and how this can be used to justify a block subsampling method along the lines of \cite{Li2020}. However, these results have similar limitations to most other formal inference results in the literature, such as treating the number of units as fixed in the asymptotics. All of this suggests that the optimal way to conduct inference for SCEs is still an open question, and further proposals and comparisons of possible methods is a promising area for future work.

\section*{Appendix A. Proofs of the Main Results}\phantomsection\label{ApA}


\textbf{Lemma 1} Suppose that Assumptions \hyperref[A1]{1} and \hyperref[A2]{2} hold. Additionally, $A_{T_0} \overset{p}{\rightarrow} A$ where $A$ is positive definite, as $T_0 \rightarrow \infty$ and $K$ is fixed while either
\begin{enumerate}[label=(\roman*)]
    \item $J$ is fixed and $\mu_0 \in \mathcal{M}_{\mathcal{J}}$ or
    \item $J \rightarrow \infty$ and $\mu_0 \in \mathcal{M}$.
\end{enumerate}
For $\hat{\mu}_0 \coloneqq \mu_{\mathcal{J}}\hat{W}$ where $\hat{W}$ is defined by equation \eqref{eq: Time-Series Averaged SC}, $\hat{\mu}_0 \overset{p}{\rightarrow} \mu_0$.

\textbf{Proof:} 
 I first consider the case where $J$ is fixed and $\mu_0 \in \mathcal{M}_{\mathcal{J}}$.  Let $$H (W) = (\mu_0 - \mu_{\mathcal{J}}W)'\Lambda_{\mathcal{K}}'A\Lambda_{\mathcal{K}}(\mu_0 - \mu_{\mathcal{J}}W).$$ Let $\mathcal{W}^* = \argmin_{W \in \Delta^J} H (W)$. Note that since $\mu_0 \in \mathcal{M}_{\mathcal{J}}$, $A$ is positive definite, and $rank(\Lambda_{\mathcal{K}}) = F$, $\mu_0 = \mu_{\mathcal{J}}W^* \; \text{for all } W^* \in \mathcal{W}^*$.

By Assumption \hyperref[A2]{2}, 
\begin{equation*}
    \frac{1}{T_0} \Tilde{Y}_{\mathcal{K}}^{pre}Y_{i}^{pre'}  \overset{p}{\rightarrow} \Lambda_{\mathcal{K}}\mu_i \; \text{for all } i \in \{0,1,...,J\}.
\end{equation*}

Let $H_{T_0}(W)$ be equal to the objective function from equation \eqref{eq: Time-Series Averaged SC}. Note that by the triangle inequality, $\max_{W \in \Delta^J} ||\Lambda_{\mathcal{K}}(\mu_0 - \mu_{\mathcal{J}}W) - \frac{1}{T_0} \Tilde{Y}_{\mathcal{K}}^{pre}(Y_{0}^{pre'} - Y_{\mathcal{J}}^{pre'}W) ||_2 \leq 2 \max_{i \in \{0,1,...,J\}} ||\Lambda_{\mathcal{K}}\mu_i - \frac{1}{T_0} \Tilde{Y}_{\mathcal{K}}^{pre}Y_{i}^{pre'}||_2 = o_p(1)$. This implies that $\max_{W \in \Delta^J} ||\frac{1}{T_0} \Tilde{Y}_{\mathcal{K}}^{pre}(Y_{0}^{pre'} - Y_{\mathcal{J}}^{pre'}W)||_2 = O_p(1)$. Therefore, using the triangle inequality and the fact that the Frobenius norm is submultiplicative, $\max_{W \in \Delta^J} |H (W) - H_{T_0}(W)| \leq \max_{W \in \Delta^J} 2 ||\frac{1}{T_0} \Tilde{Y}_{\mathcal{K}}^{pre}(Y_{0}^{pre'} - Y_{\mathcal{J}}^{pre'}W)||_2 ||A - A_{T_0}||_2 + ||\frac{1}{T_0} \Tilde{Y}_{\mathcal{K}}^{pre}(Y_{0}^{pre'} - Y_{\mathcal{J}}^{pre'}W) ||_2 ||A||_2 ||\Lambda_{\mathcal{K}}(\mu_0 - \mu_{\mathcal{J}}W) - \frac{1}{T_0} \Tilde{Y}_{\mathcal{K}}^{pre}(Y_{0}^{pre'} - Y_{\mathcal{J}}^{pre'}W) ||_2 + ||\Lambda_{\mathcal{K}}(\mu_0 - \mu_{\mathcal{J}}W)||_2 ||A||_2 ||\Lambda_{\mathcal{K}}(\mu_0 - \mu_{\mathcal{J}}W) - \frac{1}{T_0} \Tilde{Y}_{\mathcal{K}}^{pre}(Y_{0}^{pre'} - Y_{\mathcal{J}}^{pre'}W) ||_2  = o_p(1)$ because $||A - A_{T_0}||_2 = o_p(1)$. Therefore, $H_{T_0}(W)$ converges uniformly in probability to $H(W)$ for $W \in 
\Delta^J$.
\par 
Since $H(W)$ is continuous and $\Delta^J$ is compact, I follow Theorem 2.1 from \cite{Newey-McFadden} to show that this implies $||\hat{\mu}_{0} - \mu_0||_2 = o_p(1)$. Let $\eta > 0$ and $W^* \in \mathcal{W}^*$. With probability approaching one (wpa1), $H_{T_0}(\hat{W}) < H_{T_0}(W^*) + \frac{\eta}{3}$ since $\hat{W} = \argmin_{W \in \Delta^J} H_{T_0}(W)$. From uniform convergence in probability, $H(\hat{W}) < H_{T_0}(\hat{W}) + \frac{\eta}{3} $ and $H_{T_0}(W^*) < H(W^*) + \frac{\eta}{3}$ wpa1. Combining these three inequalities gives $H(\hat{W}) < H(W^*) + \eta$ wpa1. Now let $\delta > 0$ and define the open set $\mathcal{V} = \{ W  : \exists W^* \in \mathcal{W}^* \text{ s.t. } ||W^* - W||_2 < \delta \}$. Since $\Delta^J$ is compact, so is $\Delta^J \cap \mathcal{V}^c$. Then since  $H(W)$ is continuous, I can define a vector $\Tilde{W}$ taking values in $\Delta^J \cap \mathcal{V}^c$ such that $\inf_{W \in \Delta^J\cap\mathcal{V}^c} H(W) = H(\Tilde{W}) > H(W^*)$. Letting $\eta = H(\Tilde{W}) - H(W^*) \; \text{for all } W^* \in \mathcal{W}^*$, then we have that wpa1 $H(\hat{W}) < H(W^*) + \eta = H(\Tilde{W})$ which implies that $\hat{W} \in \mathcal{V}$ wpa1. Then $ \inf_{W^* \in \mathcal{W}^*} ||\hat{W} - W^*||_2 \overset{p}{\rightarrow} 0$. Therefore, $||\hat{\mu}_0 - \mu_0||_2 = \inf_{W^* \in \mathcal{W}^*}||\mu_{\mathcal{J}}\hat{W} - \mu_{\mathcal{J}}W^*||_2 \leq ||\mu_{\mathcal{J}}||_2 \inf_{W^* \in \mathcal{W}^*} ||\hat{W} - W^*||_2 \overset{p}{\rightarrow} 0$.
\par
For the case where $J \rightarrow \infty$ and $\mu_0 \in \mathcal{M}$, I follow the approach used in Proposition 3.1 of \cite{LargeSampleProperties} by reformulating the optimization problem to choose the control weights in equation \eqref{eq: Time-Series Averaged SC} as one to choose the factor loadings of the Synthetic Control. I make the feasible set equal to $\mathcal{M}$ instead of $\Delta^J$ so it no longer depends on $J$ and note that $\mathcal{M}$ is compact by Assumption \hyperref[A2]{2.3} and $||W||_1 = 1$. Then instead of using the objective function from equation \eqref{eq: Time-Series Averaged SC}, I use the alternative objective function given below: 
\begin{equation*}
    H_J(\mu) =  \min_{\Tilde{\mu} \in \mathcal{M}_{\mathcal{J}}} \{ \min_{W \in \Delta^J: \mu_{\mathcal{J}} W = \Tilde{\mu}} \{  (\frac{1}{T_0} \Tilde{Y}_{\mathcal{K}}^{pre}(Y_{0}^{pre'} - Y_{\mathcal{J}}^{pre'}W))'A_{T_0}(\frac{1}{T_0} \Tilde{Y}_{\mathcal{K}}^{pre}(Y_{0}^{pre'} - Y_{\mathcal{J}}^{pre'}W))  \} + \eta ||\mu - \Tilde{\mu}||_2 \}.
\end{equation*}

$\eta$ will be defined later in the proof but for now it suffices to note that $\eta > 0$ and $\eta = O_p(1)$. Then it will be the case that for $\hat{W}$ defined in equation \eqref{eq: Time-Series Averaged SC}, $\hat{\mu}_0 \coloneqq \mu_{\mathcal{J}} \hat{W} = \argmin_{\mu \in \mathcal{M}} H_J(\mu)$. Let $H (\mu) = (\mu_0 - \mu)'\Lambda_{\mathcal{K}}' A\Lambda_{\mathcal{K}}(\mu_0 - \mu)$ and $\{W^J\}_{J \in \mathbb{N}}$ with $W^J \in \Delta^J \; \text{for all } J$ be a sequence of control weights such that for $\mu^J \coloneqq \mu_{\mathcal{J}} W^J$, $\lim_{J \rightarrow \infty} ||\mu_0 - \mu^J||_2 = 0$. Such a sequence of weights will exist because $\mu_0 \in \mathcal{M}$. I first show that a sequence of functions which upper bounds $H_J(\mu)$ converges in probability to $H(\mu)$ at $\mu = \mu_0$. I define this upper bound as: 
\begin{equation*}
    H^{UB}_J(\mu) =  (\frac{1}{T_0} \Tilde{Y}_{\mathcal{K}}^{pre}(Y_{0}^{pre'} - Y_{\mathcal{J}}^{pre'}W^J))'A_{T_0}(\frac{1}{T_0} \Tilde{Y}_{\mathcal{K}}^{pre}(Y_{0}^{pre'} - Y_{\mathcal{J}}^{pre'}W^J)) + \eta||\mu - \mu^J||_2.
\end{equation*}
Since $W^J$ and $\mu^J$ form a pair of feasible solutions to the inner and outer minimization problems in $H_J(\mu)$ respectively, it must be the case that $H_J(\mu) \leq H_J^{UB}(\mu) \; \text{for all } \mu \in \mathcal{M}$. Since $\lim_{J \rightarrow \infty} ||\mu_0 - \mu^J||_2 = 0$ and $\eta = O_p(1)$, $\eta||\mu_0 - \mu^J||_2 = o_p(1)$.
Note that
$$||\frac{1}{T_0} \Tilde{Y}_{\mathcal{K}}^{pre}(\lambda^{pre}(\mu_0 - \mu_{\mathcal{J}}W^J) + \epsilon_{0}^{pre'} - \epsilon_{\mathcal{J}}^{pre'}W^J)||_2 \leq ||\frac{1}{T_0} \Tilde{Y}_{\mathcal{K}}^{pre}\lambda^{pre}||_2 ||\mu_0 - \mu_{\mathcal{J}}W^J||_2 + ||\frac{1}{T_0} \epsilon_{0}^{pre'}||_2 +$$
$$\max_{j \in \mathcal{J}}||\frac{1}{T_0} \Tilde{Y}_{\mathcal{K}}^{pre}\epsilon_{j}^{pre'}||_2 = O_p(1) ||\mu_0 - \mu_{\mathcal{J}}W||_2 + o_p(1)$$
by Assumption \hyperref[A2]{2}. Then since
$ ||\mu_0 - \mu_{\mathcal{J}}W^J||_2 = o(1)$, $|H_J^{UB}(\mu_0)| = o_p(1)$. Therefore, because $H (\mu_0) = 0$, $|H_J^{UB}(\mu_0) - H (\mu_0)| = o_p(1)$.
\par 
Next, I show that a sequence of functions which lower bounds $H_J(\mu)$ converges uniformly in probability to $H(\mu)$. I define this sequence of functions as
\begin{equation*}
    H^{LB}_J(\mu) = \min_{\Tilde{\mu} \in \mathcal{M}} \{ Q_J(\Tilde{\mu}) + \eta ||\mu - \Tilde{\mu}||_2 \}
\end{equation*}
where $Q_J(\Tilde{\mu}) = \min_{b \in \mathcal{W}} \{  (\frac{1}{T_0} \Tilde{Y}_{\mathcal{K}}^{pre}(Y_{0}^{pre'} -\lambda^{pre}\Tilde{\mu} - \epsilon_{\mathcal{J}}^{pre'}b))'A_{T_0}(\frac{1}{T_0} \Tilde{Y}_{\mathcal{K}}^{pre}(Y_{0}^{pre'} - \lambda^{pre}\Tilde{\mu} - \epsilon_{\mathcal{J}}^{pre'}b)) \}$ and $\mathcal{W} = \{b : ||b||_1 \leq 1\}$. Note that this is the same way that $H_J(\mu)$ is defined except for the feasible set for the outer minimization problem is now $\mathcal{M} \supseteq \mathcal{M}_{\mathcal{J}}$ and the feasible set for the inner minimization problem has also been expanded since $\mathcal{W} =  \{W : ||W||_1 \leq 1\} \supset \{W\in \Delta^J: \mu_{\mathcal{J}} W = \Tilde{\mu}\}$ which implies that $H_J(\mu) \geq H^{LB}_J(\mu) \; \text{for all } \mu \in \mathcal{M}$. In order to show uniform convergence in probability, I first show that the function $Q_J(\Tilde{\mu})$ is Lipschitz with a constant that is $O_p(1)$. Since $\mathcal{M}$ is convex and $Q_J(\Tilde{\mu})$ is differentiable by the Envelope Theorem, by the Mean Value Theorem $\text{for all } \mu, \mu' \in \mathcal{M}$, $\exists \Bar{\mu} \in \mathcal{M}$ such that 
\begin{equation*}
    |Q_J(\mu) - Q_J(\mu')| = |Q_J'(\Bar{\mu})(\mu - \mu')| \leq ||Q_J'(\Bar{\mu})||_2 ||\mu - \mu'||_2
\end{equation*}
where the inequality follows from the Cauchy-Schwartz Inequality. Let $\hat{b}(\Tilde{\mu}) = \argmin_{b \in \mathcal{W}}  Q_J(\Tilde{\mu})$. Then using the Envelope Theorem, 
\begin{equation*}
    ||Q_J'(\Bar{\mu})||_2  ||\mu - \mu'||_2 = ||(\frac{1}{T_0}\Tilde{Y}_{\mathcal{K}}^{pre}(Y_{0}^{pre'} -\lambda^{pre}\Bar{\mu} - \epsilon_{\mathcal{J}}^{pre'}\hat{b}(\Bar{\mu}))(A_{T_0} + A_{T_0}')(-\frac{1}{T_0}\Tilde{Y}_{\mathcal{K}}^{pre}\lambda^{pre}) ||_2 | ||\mu - \mu'||_2   
\end{equation*}
\begin{equation*}
    \leq ||\frac{1}{T_0}\Tilde{Y}_{\mathcal{K}}^{pre}(Y_{0}^{pre'} -\lambda^{pre}\Bar{\mu} - \epsilon_{\mathcal{J}}^{pre'}\hat{b}(\Bar{\mu})||_2 ||A_{T_0} + A_{T_0}'||_2 ||\frac{1}{T_0}\Tilde{Y}_{\mathcal{K}}^{pre}\lambda^{pre}||_2 ||\mu - \mu'||_2
\end{equation*}
where $||\frac{1}{T_0}\Tilde{Y}_{\mathcal{K}}^{pre}\lambda^{pre}||_2 \overset{p}{\rightarrow} ||\Lambda_{\mathcal{K}}||_2$ by Assumption \hyperref[A2]{2}, $\max_{\mu,\mu' \in \mathcal{M}} ||\mu-\mu'||_2 = O(1)$ since $\mathcal{M}$ is compact, and $||A_{T_0} + A_{T_0}'||_2 = O_p(1)$ since $A_{T_0} \overset{p}{\rightarrow} A$. Using the Triangle Inequality and $||\hat{b}(\Tilde{\mu})||_1 \leq 1$,
\begin{equation*}
    ||\frac{1}{T_0}\Tilde{Y}_{\mathcal{K}}^{pre}(Y_{0}^{pre'} -\lambda^{pre}\Bar{\mu} - \epsilon_{\mathcal{J}}^{pre'}\hat{b}(\Bar{\mu})||_2  \leq ||\frac{1}{T_0}\Tilde{Y}_{\mathcal{K}}^{pre}Y_{0}^{pre'}||_2 + ||\frac{1}{T_0}\Tilde{Y}_{\mathcal{K}}^{pre}\lambda^{pre}\Bar{\mu}||_2 +  \max_{j \in \mathcal{J}}||\frac{1}{T_0}\Tilde{Y}_{\mathcal{K}}^{pre}\epsilon_{j}^{pre'}||_2.
\end{equation*}
Using Assumption \hyperref[A2]{2}, $ ||\frac{1}{T_0}\Tilde{Y}_{\mathcal{K}}^{pre}Y_{0}^{pre'}||_2 \overset{p}{\rightarrow} ||\Lambda_{\mathcal{K}} \mu_0||_2$, $\max_{\Bar{\mu} \in \mathcal{M}}||\frac{1}{T_0}\Tilde{Y}_{\mathcal{K}}^{pre}\lambda^{pre}\Bar{\mu}||_2 \leq ||\frac{1}{T_0}\Tilde{Y}_{\mathcal{K}}^{pre}\lambda^{pre}||_2 \max_{\Bar{\mu} \in \mathcal{M}} ||\Bar{\mu}||_2 \overset{p}{\rightarrow} ||\Lambda_{\mathcal{K}}||_2 \max_{\Bar{\mu} \in \mathcal{M}} ||\Bar{\mu}||_2$, and $\max_{j \in \mathcal{J}}||\frac{1}{T_0}\Tilde{Y}_{\mathcal{K}}^{pre}\epsilon_{j}^{pre'}|||_2 \overset{p}{\rightarrow} 0$. Therefore, I can define $\Tilde{\eta}$ as
\begin{equation*}
    \Tilde{\eta} = (||\frac{1}{T_0}\Tilde{Y}_{\mathcal{K}}^{pre}Y_{0}^{pre'}||_2 + ||\frac{1}{T_0}\Tilde{Y}_{\mathcal{K}}^{pre}\lambda^{pre}||_2 \max_{\Bar{\mu} \in \mathcal{M}} ||\Bar{\mu}||_2 +  \max_{j \in \mathcal{J}}||\frac{1}{T_0}\Tilde{Y}_{\mathcal{K}}^{pre}\epsilon_{j}^{pre'}||_2)||A_{T_0} + A_{T_0}'||_2 ||\frac{1}{T_0}\Tilde{Y}_{\mathcal{K}}^{pre} \lambda^{pre}||_2.
\end{equation*}
Then there exists $\Tilde{\eta}$ such that  $|Q_J(\mu) - Q_J(\mu')| \leq ||Q_J'(\Bar{\mu})||_2 ||\mu - \mu'||_2 \leq \Tilde{K}||\mu - \mu'||_2$, $\Tilde{\eta} = O_p(1)$, and it does not depend on $\mu$ and $\mu'$. I now define the $K$ in $H_J(\mu)$ as $\eta = \Tilde{\eta} + 1$ so it is in fact the case that $\eta >0$ and $\eta = O_p(1)$. Because $\eta$ is greater than the Lipschitz constant of $Q_J(\mu)$, it will be the case that $H^{LB}_J(\mu) = Q_J(\mu) =  (\frac{1}{T_0} \Tilde{Y}_{\mathcal{K}}^{pre}(Y_{0}^{pre'} -\lambda^{pre}\mu - \epsilon_{\mathcal{J}}^{pre'}\hat{b}(\mu)))'A_{T_0}(\frac{1}{T_0} \Tilde{Y}_{\mathcal{K}}^{pre}(Y_{0}^{pre'} - \lambda^{pre}\mu - \epsilon_{\mathcal{J}}^{pre'}\hat{b}(\mu))) \; \text{for all } \mu \in \mathcal{M}$. Therefore $\sup_{\mu \in \mathcal{M}} |H^{LB}_J(\mu) - H(\mu)| = \sup_{\mu \in \mathcal{M}} |(\frac{1}{T_0} \Tilde{Y}_{\mathcal{K}}^{pre}(Y_{0}^{pre'} -\lambda^{pre}\mu - \epsilon_{\mathcal{J}}^{pre'}\hat{b}(\mu)))'A_{T_0}(\frac{1}{T_0} \Tilde{Y}_{\mathcal{K}}^{pre}(Y_{0}^{pre'} - \lambda^{pre}\mu - \epsilon_{\mathcal{J}}^{pre'}\hat{b}(\mu))) - (\mu_0 -\mu)'\Lambda_{\mathcal{K}}' A \Lambda_{\mathcal{K}}(\mu_0 - \mu)| \leq \sup_{\mu \in \mathcal{M}} 2||\frac{1}{T_0} \Tilde{Y}_{\mathcal{K}}^{pre}(Y_{0}^{pre'} - \lambda^{pre}\mu - \epsilon_{\mathcal{J}}^{pre'}\hat{b}(\mu))||_2 ||A - A_{T_0}||_2 + ||\frac{1}{T_0} \Tilde{Y}_{\mathcal{K}}^{pre}(Y_{0}^{pre'} - \lambda^{pre}\mu - \epsilon_{\mathcal{J}}^{pre'}\hat{b}(\mu)) ||_2 ||A||_2 ||\Lambda_{\mathcal{K}}(\mu_0 - \mu) - \frac{1}{T_0} \Tilde{Y}_{\mathcal{K}}^{pre}(Y_{0}^{pre'} -  \lambda^{pre}\mu - \epsilon_{\mathcal{J}}^{pre'}\hat{b}(\mu)) ||_2 + ||\Lambda_{\mathcal{K}}(\mu_0 - \mu)||_2 ||A||_2 ||\Lambda_{\mathcal{K}}(\mu_0 - \mu) - \frac{1}{T_0} \Tilde{Y}_{\mathcal{K}}^{pre}(Y_{0}^{pre'} -  \lambda^{pre}\mu - \epsilon_{\mathcal{J}}^{pre'}\hat{b}(\mu)) ||_2$. Again using Assumption \hyperref[A2]{2} and $||\hat{b}(\mu)||_1 \leq 1$,

$$\sup_{\mu \in \mathcal{M}} ||\Lambda_{\mathcal{K}}(\mu_0 - \mu) -\frac{1}{T_0} \Tilde{Y}_{\mathcal{K}}^{pre}(\lambda^{pre}(\mu_0 - \mu) + \epsilon_{0}^{pre'} - \epsilon_{\mathcal{J}}^{pre'}\hat{b}(\mu))||_2 \leq  $$
$$\sup_{\mu \in \mathcal{M}} ||\Lambda_{\mathcal{K}} - \Tilde{Y}_{\mathcal{K}}^{pre}\lambda^{pre}||_2 || \mu_0 - \mu||_2 + ||\Tilde{Y}_{\mathcal{K}}^{pre}\epsilon_{0}'||_2 + \max_{j \in \mathcal{J}}||\Tilde{Y}_{\mathcal{K}}^{pre}\epsilon_{j}'||_2 = o_p(1).$$
This also implies that $\sup_{\mu \in \mathcal{M}} ||\frac{1}{T_0} \Tilde{Y}_{\mathcal{K}}^{pre}(Y_{0}^{pre'} -  \lambda^{pre}\mu - \epsilon_{\mathcal{J}}^{pre'}\hat{b}(\mu)) ||_2 = O_p(1)$. Combining this with $||A_{T_0} - A||_2 = o_p(1)$ gives us that $\sup_{\mu \in \mathcal{M}}|H_J^{LB}(\mu) - H(\mu)| = o_p(1)$.
\par 
Lastly, I again use an extension of Theorem 2.1 from \cite{Newey-McFadden} to show that $(i) H_J(\mu_0) \leq H^{UB}_J(\mu_0)$ with $|H^{UB}_J(\mu_0) - H(\mu_0)| \overset{p}{\rightarrow} 0$ and $(ii) H_J(\mu) \geq H^{LB}_J(\mu)$ with $\sup_{\mu \in \mathcal{M}}| H^{LB}_J(\mu) - H(\mu)| \overset{p}{\rightarrow} 0$ together imply $||\hat{\mu}_0 - \mu_0||_2 = o_p(1)$. Let $\eta > 0$. $H_J(\hat{\mu}_0) < H_J(\mu_0) + \frac{\eta}{3}$ wpa1 since $\hat{\mu}_0 = \argmin_{\mu \in \mathcal{M}} H_J(\mu)$. From (ii), $H(\hat{\mu}_0) < H^{LB}_J(\hat{\mu}_0) + \frac{\eta}{3} \leq H_J(\hat{\mu}_0) + \frac{\eta}{3}$ wpa1. Together (i) and (ii) imply that $H_J(\mu_0) \overset{p}{\rightarrow} H(\mu_0)$ which means that $H_J(\mu_0) < H(\mu_0) + \frac{\eta}{3}$ wpa1. Combining these three inequalities gives $H(\hat{\mu}_0) < H(\mu_0) + \eta$ wpa1. Now let $\delta > 0$ and let $\mathcal{V} = \{ \mu : ||\mu - \mu_0||_2 < \delta \}$. Since $\mathcal{M}$ is compact, so is $\mathcal{M} \cap \mathcal{V}^c$. Note that $\mu_0$ is the unique minimum of $H(\mu)$ as a result of $rank(\Lambda_{\mathcal{K}}) = F$, $A$ being positive definite, and $\mu_0 \in \mathcal{M}$. Because $\mu_0$ is the unique minimum of $H(\mu)$ and $H(\mu)$ is continuous, I can define a vector $\Tilde{\mu}$ taking values in $\mathcal{M} \cap \mathcal{V}^c$ such that $\inf_{\mu \in \mathcal{M}\cap\mathcal{V}^c} H(\mu) = H(\Tilde{\mu}) > H(\mu_0)$. Letting $\eta = H(\Tilde{\mu}) - H(\mu_0)$, then we have that wpa1 $H(\hat{\mu}_0) < H(\mu_0) + \eta = H(\Tilde{\mu})$ which implies that $\hat{\mu}_0 \in \mathcal{V}$ wpa1. Therefore $||\hat{\mu}_0 - \mu_0||_2 \overset{p}{\rightarrow} 0$.

\textbf{Proposition 1} Suppose that Assumptions \hyperref[A1]{1} and \hyperref[A2]{2} are satisfied and $A_{T_0} \overset{p}{\rightarrow} A$ where $A$ is positive definite, as $T_0 \rightarrow \infty$ and $K$ is fixed while either 
\begin{enumerate}[label=(\roman*)]
    \item $J$ is fixed and $\mu_0 \in \mathcal{M}_{\mathcal{J}}$ or
    \item $J \rightarrow \infty$, $\mu_0 \in \mathcal{M}$, and $\epsilon_{\mathcal{J}}^{post} \indep Y^{pre}$.
\end{enumerate}
Then  $\text{for all } t \in \mathcal{T}_1, \; \lim_{T_0 \rightarrow \infty}E[\hat{\alpha}_{0t} - \alpha_{0t}] = 0$.

\textbf{Proof}: We can first note that by $\hat{W} \in \Delta^J$ and Assumption \hyperref[A2]{2.3},  $||\hat{\mu}_0 - \mu_0||_2$ is bounded. Therefore, since $||\hat{\mu}_0 - \mu_0||_2 = o_p(1)$ by Lemma \hyperref[L1]{1}, $\lim_{T_0 \rightarrow \infty} E[||\hat{\mu}_0 - \mu_0||_2^2] = 0$. $E[\hat{\alpha}_{0t} - \alpha_{0t}] = E[\lambda_t (\mu_0 -  \hat{\mu}_0) + \epsilon_{0t} - \sum_{j \in \mathcal{J}} \hat{W}_j \epsilon_{jt}]$.  $|E[\lambda_t (\mu_0 - \hat{\mu}_0)]| \leq E[|\lambda_t (\mu_0 - \hat{\mu}_0)|] \leq E[||\lambda_t||_2 ||\mu_0 - \hat{\mu}_0||_2|] \leq E[||\lambda_t||_2^2]^{\frac{1}{2}}  E[||\mu_0 - \hat{\mu}_0||_2^2]^{\frac{1}{2}} \rightarrow 0$ as $T_0 \rightarrow \infty$ using the Cauchy-Schwartz Inequality. For case (i), let $\mathcal{W}^* \coloneqq \{W : \mu_{\mathcal{J}}W = \mu_0\}$. Then as shown in the proof of Lemma \hyperref[L1]{1}, $\inf_{W \in \mathcal{W}^*} ||\hat{W} - W||_2 = o_p(1)$ and because this is bounded $\inf_{W \in \mathcal{W}^*} E||\hat{W} - W||_2^2] \rightarrow 0$ as $T_0 \rightarrow \infty$. Note that for each $W \in \mathcal{W}^*$,
\begin{equation*}
   | E[\sum_{j \in \mathcal{J}} \hat{W}_j \epsilon_{jt}]| = |E[\sum_{j \in \mathcal{J}} (\hat{W}_j - W_j)\epsilon_{jt}] + E[\sum_{j \in \mathcal{J}} W_j \epsilon_{jt}]| = |E[\sum_{j \in \mathcal{J}} (\hat{W}_j - W_j)\epsilon_{jt}]|
\end{equation*}
\begin{equation*}
    \leq E||\hat{W} - W||_2^2]^{\frac{1}{2}} E[||\sum_{j \in \mathcal{J}} \epsilon_{jt}||_2^2]^{\frac{1}{2}}.
\end{equation*}
Therefore $| E[\sum_{j \in \mathcal{J}} \hat{W}_j \epsilon_{jt}]| \leq \inf_{W \in \mathcal{W^*}} E||\hat{W} - W||_2^2]^{\frac{1}{2}} E[||\sum_{j \in \mathcal{J}} \epsilon_{jt}||_2^2]^{\frac{1}{2}} \rightarrow 0$ as $T_0 \rightarrow \infty$ while $J$ is fixed. For case (ii), because $\hat{W}_j \; \text{for all } j \in \mathcal{J}$ is a bounded measurable function of $Y^{pre}$ which is independent of $\epsilon_{\mathcal{J}}^{post}$, $E[\sum_{j \in \mathcal{J}} \hat{W}_j \epsilon_{jt}] = 0$. Therefore, in both cases we have 
\begin{equation*}
    |E[\hat{\alpha}_{0t} - \alpha_{0t}]| = |E[\lambda_t (\hat{\mu}_0 - \mu_0) + \epsilon_{0t} - \sum_{j \in \mathcal{J}} \hat{W}_j \epsilon_{jt}]| \leq |E[\lambda_t (\hat{\mu}_0 - \mu_0)]| + |E[\sum_{j \in \mathcal{J}} \hat{W}_j \epsilon_{jt}]| \rightarrow 0 \text{ as } T_0 \rightarrow \infty.
\end{equation*}

\textbf{Proposition 2} Suppose that Assumptions \hyperref[A1]{1}, \hyperref[A2]{2}, and \hyperref[A3]{3} are satisfied and $A_{T_0} \overset{p}{\rightarrow} A$ where $A$ is positive definite, as $T_0 , T_1 \rightarrow \infty$ and $K$ is fixed and while either 
\begin{enumerate}[label=(\roman*)]
    \item $J$ is fixed and $\mu_0 \in \mathcal{M}_{\mathcal{J}}$ or
    \item $J \rightarrow \infty$ and $\mu_0 \in \mathcal{M}$.
\end{enumerate}
Then $\sum_{t \in \mathcal{T}_1} v_t (\hat{\alpha}_{0t} - \alpha_{0t}) \overset{p}{\rightarrow} 0$.

\textbf{Proof:} First note that since $\sup_{t \in \mathcal{T}_1} E[||\lambda_t||_2^2] < \infty$ and $v \in \Delta^{T_1}$, then $\text{for all } f \in \{1,...,F\}$, $\text{for all } \eta > 0$,
\begin{equation*}
    P(|\sum_{t \in \mathcal{T}_1} v_t \lambda_{tf}| > \eta) \leq \frac{E[(\sum_{t \in \mathcal{T}_1} v_t \lambda_{tf})^2]}{\eta^2} \leq \frac{\sum_{t \in \mathcal{T}_1} \sum_{s \in \mathcal{T}_1} v_t v_s E[\lambda_{tf}\lambda_{sf}] }{\eta^2} \leq \frac{\sup_{t \in \mathcal{T}_1} E[\lambda_{tf}^2]}{\eta^2}.
\end{equation*}
Therefore, $\sum_{t \in \mathcal{T}_1} v_t \lambda_{tf} = O_p(1) \; \text{for all } \{1,...,F\}$. $\sum_{t \in \mathcal{T}_1} v_t ( \hat{\alpha}_{0t} - \alpha_{0t})$ is equal to
$$\sum_{t \in \mathcal{T}_1} v_t (\lambda_t (\mu_0 - \hat{\mu}_0) + \epsilon_{0t} - \sum_{j \in \mathcal{J}} \hat{W}_j \epsilon_{jt}) = ( \sum_{t \in \mathcal{T}_1}v_t \lambda_t) (\mu_0 - \hat{\mu}_0) + \sum_{t \in \mathcal{T}_1}  v_t \epsilon_{0t} - \sum_{j \in \mathcal{J}} \hat{W}_j  \sum_{t \in \mathcal{T}_1} v_t \epsilon_{jt}.$$ Using Assumption \hyperref[A3]{3}, $\sum_{t \in \mathcal{T}_1} v_t \epsilon_{0t} = o_p(1)$, and $|\sum_{j \in \mathcal{J}} \hat{W}_j \sum_{t \in \mathcal{T}_1} v_t \epsilon_{jt}| \leq \max_{j \in \mathcal{J}} |\sum_{t \in \mathcal{T}_1} v_t\epsilon_{jt}| = o_p(1)$ where the inequality comes from the fact that $||\hat{W}||_1 = 1$. Because $||\mu_0 - \hat{\mu}_0||_2 = o_p(1)$ by Lemma \hyperref[L1]{1}, $| (\sum_{t \in \mathcal{T}_1}v_t \lambda_t) (\mu_0 - \hat{\mu}_0)| \leq ||\sum_{t \in \mathcal{T}_1}v_t \lambda_t||_2 ||\mu_0 - \hat{\mu}_0||_2 = O_p(1) o_p(1) = o_p(1)$. Therefore, we have that $\sum_{t \in \mathcal{T}_1} v_t( \alpha_{0t} - \hat{\alpha}_{0t}) = o_p(1)$.

\textbf{Proposition 3} Suppose Assumptions \hyperref[A1*]{1*}, \hyperref[A2*]{2*}, and \hyperref[A3*]{3*} hold and $\mathcal{BC} \cap \mathcal{L}^0 \ne \emptyset$. Furthermore, suppose that for all $(\mathcal{J},\mathcal{K}) \in \mathcal{BC}$, if $|\mathcal{K}| \geq |\mathcal{K}^*|$ for some $(\mathcal{J}^*,\mathcal{K}^*) \in \mathcal{MBCL}^0$, then $rank(\Lambda_{\mathcal{K}}) = F$. Then $P((\hat{\mathcal{J}},\hat{\mathcal{K}}) \in \mathcal{MBCL}^0) \rightarrow 1$ as $T_0 \rightarrow \infty$ and $N_0$ and $N_1$ are fixed. If additionally, $\mathcal{MBCL}^0$ contains a single element $(\mathcal{J}^0,\mathcal{K}^0)$ and there is a unique $W^0 \in \Delta^{|\mathcal{J}^0|}$ such that $\mu_0 = \mu_{\mathcal{J}^0} W^0$, then $\hat{W} \overset{p}{\rightarrow} W^0$. 

\textbf{Proof:} The result almost follows directly from Theorem 2 of \cite{MomentSelection}, with a slight exception.  First, let $\mathcal{N} = \mathcal{N}_0 \cup \mathcal{N}_1$ and let 
$$
g_{T_0}(W) = \Tilde{Y}_{\mathcal{N}}^{pre}(Y_0^{pre'} - Y_{\mathcal{N}_0}^{pre'}W)/T_0,$$
$$g(W) = \Lambda_{\mathcal{N}}(\mu_0 - \mu_{\mathcal{N}_0}W) + \begin{pmatrix}
    0_{N_1} \\
    W_1 \sigma_1^2 \\
    ... \\
    W_{N_0} \sigma_{N_0}^2
\end{pmatrix},$$ 
$$g_{T_0,\mathcal{J},\mathcal{K}}(W) = \Tilde{Y}_{\mathcal{K}}^{pre}(Y_0^{pre'} - Y_{\mathcal{J}}^{pre'}W)/T_0 \text{ , and}$$
$$g_{\mathcal{J},\mathcal{K}}(W) = \Lambda_{\mathcal{K}}(\mu_0 - \mu_{\mathcal{J}}W) \text{ for each }(\mathcal{J},\mathcal{K}) \in \mathcal{L}.$$
\cite{MomentSelection} define $\mathcal{L}^0$ so that $\mathcal{L}^0 = \{(\mathcal{J},\mathcal{K}) \in \mathcal{L} : g_{\mathcal{J},\mathcal{K}}(W) = 0 \text{ for some } W \in \Delta^J\}$. I defined it be $\mathcal{L}^0 = \{(\mathcal{J},\mathcal{K}) \in \mathcal{L} : \mu_{\mathcal{J}}W = \mu_0 \text{ for some } W \in \Delta^J \text{ and } rank(\Lambda_{\mathcal{K}}) = F \}$. However, I impose that for all $(\mathcal{J},\mathcal{K}) \in \mathcal{BC}$, if $|\mathcal{K}| \geq |\mathcal{K}^*|$ for some $(\mathcal{J}^*,\mathcal{K}^*) \in \mathcal{MBCL}^0$, then $rank(\Lambda_{\mathcal{K}}) = F$. This ensures that the set $\mathcal{MBCL}^0$ will be the same on both definitions, since for each $(\mathcal{J},\mathcal{K}) \in \mathcal{MBCL}^0$, $rank(\Lambda_{\mathcal{K}}) = F$ so $g_{\mathcal{J},\mathcal{K}}(W) = 0$ for some $W \in \Delta^J$ if and only if $\mu_0 = \mu_{\mathcal{J}}W$ for some $W \in \Delta^J$. Therefore, the results of Theorem 2 using their definition of $\mathcal{MBCL}^0$ hold for my definition of $\mathcal{MBCL}^0$ as well. 
\par 
In addition to the Assumption \hyperref[A3*]{3*} stated in section \ref{Moment and Model Selection}, \cite{MomentSelection} impose the following conditions for their Theorem 2:

\textbf{Assumption 1 of \cite{MomentSelection}}
\begin{enumerate}[label=\alph*]
    \item $g_{T_0}(W) = g(W) +O_p(T_0^{-1/2})$ for any $W \in \Delta^{N_0}$.
    \item For each, $(\mathcal{\mathcal{J}},\mathcal{K}) \in \mathcal{L}$, there exists a non-stochastic matrix $A_{\mathcal{J},\mathcal{K}}$ such that $A_{T_0,\mathcal{J},\mathcal{K}} \overset{p}{\rightarrow} A_{\mathcal{J},\mathcal{K}}$.
    \item For all $(\mathcal{J},\mathcal{K}) \in \mathcal{L}$, 
    
    $\inf_{W \in \Delta^J} g_{T_0,\mathcal{J},\mathcal{K}}(W)A_{T_0,\mathcal{J},\mathcal{K}}g_{T_0,\mathcal{J},\mathcal{K}}(W) \overset{p}{\rightarrow} \inf_{W \in \Delta^J} g_{\mathcal{J},\mathcal{K}}(W)A_{\mathcal{J},\mathcal{K}} g_{\mathcal{J},\mathcal{K}} (W) = g_{\mathcal{J},\mathcal{K}}(W^*)A_{\mathcal{J},\mathcal{K}} g_{\mathcal{J},\mathcal{K}}(W^*)$ for some $W^* \in \Delta^J$ which may depend on $(\mathcal{J},\mathcal{K})$.
\end{enumerate}

Assumption 1.b is implied directly by Assumption \hyperref[A2*]{2.6*}. For Assumption 1.a, note that for each $i \in \mathcal{N}$ and any $W \in \Delta^{N_0}$,

$$Y_{i}^{pre}(Y_0^{pre'} - Y_{\mathcal{N}_0}^{pre'}W)/T_0 =
\mu_{i}'\lambda^{pre'}\lambda^{pre}/T_0(\mu_0 - \mu_{\mathcal{N}_0}W) + \mu_{i}'\lambda^{pre'}(\epsilon_0^{pre'} - \epsilon_{\mathcal{N}_0}^{pre'}W)/T_0
$$

$$ + \epsilon_{i}^{pre}(\epsilon_0^{pre'} - \epsilon_{\mathcal{N}_0}^{pre'}W)/T_0  + \epsilon_i^{pre}\lambda^{pre}/T_0(\mu_0 - \mu_{\mathcal{N}_0}W) .$$
By Assumption \hyperref[A2*]{2*}, this will be equal to $ \mu_i' \Omega_0 (\mu_0 - \mu_{\mathcal{N}_0}W) + W_i \sigma_i^2 + O_p(T_0^{-\frac{1}{2}})$ for $i \in \mathcal{N}_0$ and $\mu_i' \Omega_0 (\mu_0 - \mu_{\mathcal{N}_0}W) + O_p(T_0^{-\frac{1}{2}})$ for $i \in \mathcal{N}_1$. Similarly, any $W \in \Delta^{N_0}$,
$$\sum_{t \in \mathcal{T}_0}(Y_{0t} - Y_{\mathcal{N}_0,t}W)/T_0 = \sum_{t \in \mathcal{T}_0} \lambda_t /T_0 (\mu_0 - \mu_{\mathcal{N}_0}W) + \sum_{t \in \mathcal{T}_0}(\epsilon_{0t} - \epsilon_{\mathcal{N}_0,t}W)/T_0$$
$$= \Bar{\lambda}_0 (\mu_0 - \mu_{\mathcal{N}_0}W) + O_p(T_0^{-\frac{1}{2}}).$$
So $g_{T_0}(W) - g(W) = O_p(T_0^{-\frac{1}{2}})$.

For Assumption 1.c, the reasoning is similar to in Lemma \hyperref[L1]{1}. By Assumption \hyperref[A2*]{2*}, 
\begin{equation*}
    \frac{1}{T_0} \Tilde{Y}_{\mathcal{K}}^{pre}Y_{i}^{pre'}  \overset{p}{\rightarrow} \Lambda_{\mathcal{K}} \mu_i \; \text{for all } i \in \{0\} \cup \mathcal{J}.
\end{equation*}

Let $H_{T_0}(W)$ be equal to the objective function from equation \eqref{eq: Time-Series Averaged SC}. Note that by the triangle inequality, $\max_{W \in \Delta^J} ||\Lambda_{\mathcal{K}}(\mu_0 - \mu_{\mathcal{J}}W) - \frac{1}{T_0} \Tilde{Y}_{\mathcal{K}}^{pre}(Y_{0}^{pre'} - Y_{\mathcal{J}}^{pre'}W) ||_2 \leq 2 \max_{i \in \{0\} \cup \mathcal{J}} ||\Lambda_{\mathcal{K}}\mu_i - \frac{1}{T_0} \Tilde{Y}_{\mathcal{K}}^{pre}Y_{i}^{pre'}||_2 = o_p(1)$. This implies that $\max_{W \in \Delta^J} ||\frac{1}{T_0} \Tilde{Y}_{\mathcal{K}}^{pre}(Y_{0}^{pre'} - Y_{\mathcal{J}}^{pre'}W)||_2 = O_p(1)$. Therefore, using the triangle inequality and the fact that the Frobenius norm is submultiplicative, $\max_{W \in \Delta^J} |H (W) - H_{T_0}(W)| \leq \max_{W \in \Delta^J} 2 ||\frac{1}{T_0} \Tilde{Y}_{\mathcal{K}}^{pre}(Y_{0}^{pre'} - Y_{\mathcal{J}}^{pre'}W)||_2 ||A_{\mathcal{J},\mathcal{K}} - A_{T_0,\mathcal{J},\mathcal{K}}||_2 + ||\frac{1}{T_0} \Tilde{Y}_{\mathcal{K}}^{pre}(Y_{0}^{pre'} - Y_{\mathcal{J}}^{pre'}W) ||_2 ||A_{\mathcal{J},\mathcal{K}}||_2 ||\Lambda_{\mathcal{K}}(\mu_0 - \mu_{\mathcal{J}}W) - \frac{1}{T_0} \Tilde{Y}_{\mathcal{K}}^{pre}(Y_{0}^{pre'} - Y_{\mathcal{J}}^{pre'}W) ||_2 + ||\Lambda_{\mathcal{K}}(\mu_0 - \mu_{\mathcal{J}}W)||_2 ||A_{\mathcal{J},\mathcal{K}}||_2 ||\Lambda_{\mathcal{K}}(\mu_0 - \mu_{\mathcal{J}}W) - \frac{1}{T_0} \Tilde{Y}_{\mathcal{K}}^{pre}(Y_{0}^{pre'} - Y_{\mathcal{J}}^{pre'}W) ||_2  = o_p(1)$ because $||A_{\mathcal{J},\mathcal{K}} - A_{T_0,\mathcal{J},\mathcal{K}}||_2 = o_p(1)$. Therefore, $H_{T_0}(W)$ converges uniformly in probability to $H(W)$ for $W \in 
\Delta^J$. Therefore, by the Continuous Mapping Theorem, we have that for each
$(\mathcal{J},\mathcal{K}) \in \mathcal{L}$, 
$$\inf_{W \in \Delta^J} g_{T_0,\mathcal{J},\mathcal{K}}(W)A_{T_0,\mathcal{J},\mathcal{K}}g_{T_0,\mathcal{J},\mathcal{K}}(W) \overset{p}{\rightarrow} \inf_{W \in \Delta^J} g_{\mathcal{J},\mathcal{K}}(W)A_{\mathcal{J},\mathcal{K}} g_{\mathcal{J},\mathcal{K}}(W).$$ Furthermore, since $\Delta^J$ is compact and $H_J$ is continuous, there is guaranteed to exist some $W^* \in \Delta^J$ such that $\inf_{W \in \Delta^J} g_{\mathcal{J},\mathcal{K}}(W)A_{\mathcal{J},\mathcal{K}} g_{\mathcal{J},\mathcal{K}}(W) = g_{\mathcal{J},\mathcal{K}}(W^*)A_{\mathcal{J},\mathcal{K}} g_{\mathcal{J},\mathcal{K}}(W^*)$.

\section*{Appendix B. Additional Formal Results}\phantomsection\label{ApB}
\textbf{Example for Assumption 2:} 
Suppose $(\lambda_t, \epsilon_t)_{t \in \mathbb{Z}}$ is $\alpha$-mixing with exponentially decaying mixing coefficients, $E[\epsilon_{it}|\epsilon_{kt}] = 0$ for all $k < 0$, $i \geq 0$, $E[\epsilon_{it}|\lambda_{tf}] = 0$ for all $i$ and all $f \in \{1,...,F\}$, $\sup_{i,t} E[\epsilon_{it}^4] < \infty$, and
$$\sup_{i,t} P(|\epsilon_{it}| > a) \leq c_1 \exp(-c_2 a^{q_1}) \text{ and } \sup_{f,t} P(|\lambda_{tf}| > a) \leq c_1 \exp(-c_2 a^{q_2})$$
for all $a > 0$ for some $q_1,q_2 > 0$ and $c_1,c_2 > 0$ which do not depend on $i,t,$ and $f$.

By Lemma A4 in \cite{Dendramis2021}, an exponential tail bound will also hold for products of the idiosyncratic shocks and factors so, for example, 
$\sup_{i,t,f} P(|\lambda_{ft}\epsilon_{it}| > a) \leq c_1 exp(-c_2 a^{q})$ for all $a > 0$ where $q = q_1 q_2/(q_1 + q_2)$. Then by Lemma 1 of \cite{Dendramis2021},

$$P(|\frac{1}{\sqrt{T_0}} \sum_{t \in \mathcal{T}_0} \lambda_{ft}\epsilon_{it}| > a) \leq c_3 [\exp(-c_4 a^2) + \exp(-c_5 (\frac{a\sqrt{T_0}}{\log^2 T_0})^{q/(q+1)})]$$
for all $a >0$ where $c_3$, $c_4$, and $c_5$ don't depend on $i, f,$ and $t$. Then let $a = \kappa \log J/2$ for some $\kappa$ so that,

$$P(\max_{i \in \{0,...,J\}} |\frac{1}{T_0} \sum_{t \in \mathcal{T}_0} \lambda_{ft}\epsilon_{it}| > \kappa \log J/2) \leq $$
$$\sum_{i=0}^J P(|\frac{1}{\sqrt{T_0}} \sum_{t \in \mathcal{T}_0} \lambda_{ft}\epsilon_{it}| > \kappa \log J/2 \sqrt{T_0}) \leq $$
$$(J+1) c_3 \exp(-c_4 (\kappa \log J /2)^2) + (J+1)c_3 \exp(-c_5 (\frac{\kappa \log J /2\sqrt{T_0}}{\log^2 T_0})^{q/(q+1)}) := r_J + r_{J,T_0}.$$

Let $\gamma >0$. Then we can choose $\kappa$ such that $c_4 (\kappa/2)^2 > 1 + \gamma$. Then
$r_J \leq (J+1)\exp(-(1+\gamma) \log J) = c_3 J^{-\gamma} \rightarrow 0$ as $J \rightarrow \infty$.

Furthermore, if $J = o(T_0^\delta)$ for some $\delta > 0$. Then we have $T_0^{\frac{1}{4}} \geq (J+1)^{\frac{1}{4 \delta}}$ and $T_0^{\frac{1}{4}} > 2\log^2(T_0)$ as $T_0 \rightarrow \infty$. Then 
$$c_5 (\frac{\kappa \log J /2\sqrt{T_0}}{\log^2 T_0})^{q/(q+1)} \geq c_5 (\kappa(J^{\frac{1}{4\delta}} \sqrt{\log J})^{q/(q+1)} > (1 + \gamma) \log J$$
as $J \rightarrow \infty$. Therefore, $ 0 \leq r_{J,T_0} \leq r_J \rightarrow 0$ as $T_0,J \rightarrow \infty$. Hence $\max_{i \geq 0} |\frac{1}{T_0} \sum_{t \in \mathcal{T}_0} \lambda_{ft}\epsilon_{it}| = o_p(1)$ for all $f \in \{1,...,F\}$. Analogous arguments can be made to show that  $\max_{i \geq 0, k< 0} |\frac{1}{T_0} \sum_{t \in \mathcal{T}_0} \epsilon_{kt}\epsilon_{it}| = o_p(1)$ and $\max_{i \geq 0}|\frac{1}{T_0} \sum_{t \in \mathcal{T}_0} \epsilon_{it}| = o_p(1)$. One limitation of applying these conditions for Assumption \hyperref[A2]{2.2} is that they require that the factors first and second moments are invariant (i.e. imposing $E[\lambda_t] = \Bar{\lambda}_0$ and $E[\lambda_t'\lambda_t] = \Omega_0$ for all $t \in \mathcal{T}_0$ to show that $\frac{1}{T_0}\sum_{t \in \mathcal{T}_0}\lambda_t \overset{p}{\rightarrow} \Bar{\lambda}_0$ and $\frac{1}{T_0}\sum_{t \in \mathcal{T}_0}\lambda_t'\lambda_t \overset{p}{\rightarrow} \Omega_0$. However, the conditions can be altered to allow for diverging factors. 
\par 
Suppose there is an unrestricted factor $\phi_t$ that all units have equal exposure to so 
$$Y_{it} = \alpha_{it}D_{it} + \phi_t + \lambda_t \mu_i + \epsilon_{it},$$
for all $i,t$. Instead of using $(1, Y_{-Kt}, ..., Y_{-1t})$ as instruments, we could instead use $(1,Y_{-Kt}-Y_{-1t},...,Y_{-2t}-Y_{-1t})$ as instruments. Then for each $k \in \{-K,...,-2\}$, our sample moments conditions would become
$$\frac{1}{T_0}\sum_{t \in \mathcal{T}_0} (Y_{kt}-Y_{-1t})(Y_{0t} - Y_{\mathcal{J},t}'W) =$$
$$\frac{1}{T_0}\sum_{t \in \mathcal{T}_0}((\mu_k - \mu_{-1})'\lambda_t' + \epsilon_{kt} - \epsilon_{-1t})(\lambda_t(\mu_0 - \mu_{\mathcal{J}}W) + \epsilon_{0t} + \epsilon_{\mathcal{J},t}'W).$$
Therefore $\phi_t$ does not effect the estimation of $\hat{W}$ at all. 
\par 
It is also possible to allow for multiple diverging factors. Suppose that Assumptions \hyperref[A2]{2.1}, \hyperref[A2]{2.3}, \hyperref[A2]{2.4}, and \hyperref[A2]{2.5} hold and $\sum_{t \in \mathcal{T}_1} \lambda^{pre'}\lambda^{pre}/T_0^\delta \overset{p}{\rightarrow} \Omega_0$ and $\sum_{t \in \mathcal{T}_1} \lambda_t /T_0^\delta \overset{p}{\rightarrow} \Bar{\lambda}_0$ for some $\delta \geq 1$. Under these conditions, we can instead consider the normalized version of the same moment conditions
$$\max_{W \in \Delta^J}|| \Tilde{Y}_{\mathcal{K}}^{pre}(Y_0^{pre'} - Y_{\mathcal{J}}^{pre'} W)/T_0^\delta - \Lambda_{\mathcal{K}} (\mu_0 - \mu_{\mathcal{J}}W)||_2 \leq 2 \max_{i \geq 0} ||\Tilde{Y}_{\mathcal{K}}^{pre}Y_i^{pre'}/T_0^\delta - \Lambda_{\mathcal{K}} \mu_i ||_2 = o_p(1).$$
Therefore, by the same reasoning as in the proof of Lemma \hyperref[L1]{1}, the of objective function in equation \eqref{eq: Time-Series Averaged SC} divided by $T_0^{\delta -1}$ is converging uniformly in probability to $(\mu_0 - \mu_{\mathcal{J}}W)'\Lambda_{\mathcal{K}}'A \Lambda_{\mathcal{K}}(\mu_0 - \mu_{\mathcal{J}}W)$ when $A_{T_0} \overset{p}{\rightarrow} A$. Because of this, the formal results in section \ref{Model} continue to hold. This allows for cases where the factors are diverging, as long as they are diverging at the same rate. For example, consider the case of a polynomial time trend where $\lambda_{tf} = \sum_{m=1}^M c_m t^m$ with constants $c_1,...,c_M$ where $M \geq 1$ is fixed. Then for $\delta = M+1$, $\sum_{t \in \mathcal{T}_0} \lambda_{tf}/T_0^\delta \rightarrow c_M/2$.

\textbf{Examples for Assumption 3:} For Assumption \hyperref[A3]{3.2}, consider the case where all idiosyncratic shocks are independent across unit and time, $E[\epsilon_{it}] = 0$, and $\sup_{i,t} E[\epsilon_{it}^4] < \infty$. Then $\text{for all } \eta > 0$,
\begin{equation*}
    P(\max_{0 \leq i \leq J} \{ |\sum_{t \in \mathcal{T}_1} v_t \epsilon_{it}|\} > \eta ) \leq \sum_{i=0}^J P(|\sum_{t \in \mathcal{T}_1} v_t \epsilon_{it}| > \eta) 
\end{equation*}
\begin{equation*}
    \leq \sum_{i=0}^J \frac{E[(\sum_{t \in \mathcal{T}_1} v_t \epsilon_{it})^4]}{\eta^4} \leq (J+1) C ||v||_2^4 
\end{equation*}
for some constant $C$. Therefore, if $J^{\frac{1}{4}}||v||_2 \rightarrow 0$, then $\max_{0 \leq i \leq J} \{| \sum_{t \in \mathcal{T}_1} v_t \epsilon_{it} | \} \overset{p}{\rightarrow} 0$ as $T_1 \rightarrow \infty$. 

\section*{Appendix C. Asymptotic Distribution and Inference Results}\phantomsection\label{ApC}

Here, I show how it is possible to derive the asymptotic distribution of the average treatment effect and how this can be used to justify a subsampling approach to inference. The parameter of interest will be the limit of a sequence of weighted averages of treatment effects $\alpha_v$. 

\textbf{Assumption C1}\label{AC1} Suppose that $\sum_{t \in \mathcal{T}_1} v_t \alpha_{0t} \overset{p}{\rightarrow} \alpha_v$ for some constant $\alpha_v$ as $T_1 \rightarrow \infty$.

I focus exclusively on the case where $J$ is fixed and there is a unique set of “true” control weights $W^* \in \Delta^J$ because what $\hat{W}$ is converging to will influence the asymptotic distribution of $\sqrt{T_1}(\hat{\Bar{\alpha}}_v - \Bar{\alpha}_v)$. While the asymptotic normality of standard GMM estimators is well established, here there are some complications to the analysis of the asymptotic behavior of $\hat{W}$. First, the presence of the constraints in equation \eqref{eq: Time-Series Averaged SC} imply that results for the asymptotic normality of unconstrained GMM estimators cannot be employed. However, \cite{Andrews2002} provides the asymptotic distribution for GMM estimators when the parameter is on the boundary of the parameter space. Using his approach, one can first consider the asymptotic behavior of the unconstrained version of the GMM-SCE and then consider how this unconstrained GMM-SCE is related to the constrained version. I define the unconstrained version of the GMM-SCE as 

\begin{equation} \label{eq: Unconstrained}
    \hat{W}^{GMM} \in \argmin_{W} (\frac{1}{T_0} \Tilde{Y}_{\mathcal{K}}^{pre}
     (Y_{0}^{pre'} - Y_{\mathcal{J}}^{pre'} W))'A_{T_0} (\frac{1}{T_0} \Tilde{Y}_{\mathcal{K}}^{pre}(Y_{0}^{pre'} - Y_{\mathcal{J}}^{pre'} W)).
\end{equation}

Because of the quadratic nature of the objective function in equations \eqref{eq: Time-Series Averaged SC} and \eqref{eq: Unconstrained} and the fact that $\Delta^J$ is a convex set, there is a direct connection between the unconstrained solution $\hat{W}^{GMM}$ and the constrained solution $\hat{W}$. More precisely, we can think of the constrained solution as the unconstrained solution projected onto the set $\Delta^J$ using some projection function $\Pi_{\Delta^J,T_0}$. I show below that 

\begin{equation} \label{eq: Projection}
    \hat{W} = \Pi_{\Delta^J,T_0 } (\hat{W}^{GMM}) \coloneqq \argmin_{W \in \Delta^J} (\hat{W}^{GMM} - W)' \frac{1}{T_0} Y_{\mathcal{J}}^{pre} \Tilde{Y}_{\mathcal{K}}^{pre'} A_{T_0} \frac{1}{T_0}\Tilde{Y}_{\mathcal{K}}^{pre} Y_{\mathcal{J}}^{pre'} (\hat{W}^{GMM} - W).
\end{equation}

Furthermore, it can be shown that $\Pi_{\Delta^J,T_{0}}(W)$ will be asymptotically equivalent to $\Pi_{\Delta^J} (W)$ where $\Pi_{\Delta^J}$ is defined the same except its objective function is  equal to what the objective function in equation \eqref{eq: Projection} converges to in probability point-wise w.r.t. $W$ as $T_0 \rightarrow \infty$. This means that $\Pi_{\Delta^J}(\hat{W}^{GMM})$ will be given by

\begin{equation*} \label{eq: Asymptotic Projection}
  \hat{W} \overset{a}{\sim}  \Pi_{\Delta^J} (\hat{W}^{GMM}) \coloneqq \argmin_{W \in \Delta^J} (\hat{W}^{GMM} - W)' \mu_{\mathcal{J}}' \Lambda_{\mathcal{K}}' A \Lambda_{\mathcal{K}} \mu_{\mathcal{J}} (\hat{W}^{GMM} - W).
\end{equation*}

From here, one may wish to show the asymptotic normality of $\sqrt{T_0}(\hat{W}^{GMM} - W^*)$ and use the fact that $\hat{W}$ is a function of $\hat{W}^{GMM}$ to characterize its asymptotic distribution. However, even if $W^*$ is the unique set of control weights in $\Delta^J$ such that $\mu_0 = \mu_{\mathcal{J}} W^*$, when $rank(\mu_{\mathcal{J}}) < J$ the “true” control weights for the unconstrained estimator will still only be partially identified. To combat this problem, a similar trick as in Lemma \hyperref[L1]{1} is used, where the problem is reformulated in terms of the estimated factor loadings of the treated unit instead of the control weights. So letting $\hat{\mu}_0^{GMM} \coloneqq \mu_{\mathcal{J}}\hat{W}^{GMM}$, it will still be the case that $\sqrt{T_0}(\hat{\mu}_0^{GMM} - \mu_0)$ is asymptotically normal. Then using a change of variable we can instead consider the projection
\begin{equation*}
  \sqrt{T_0}(\hat{\mu}_0 - \mu_0) \overset{a}{\sim}  \Pi_{T_{\mathcal{M}_{\mathcal{J}}}(\mu_0)}(\sqrt{T_0}(\hat{\mu}_0^{GMM} - \mu_0)) \coloneqq \argmin_{\delta \in T_{\mathcal{M}_{\mathcal{J}}}(\mu_0)} (\sqrt{T_0}(\hat{\mu}_0^{GMM} - \mu_0) - \delta)' \Lambda_{\mathcal{K}}' A \Lambda_{\mathcal{K}}(\sqrt{T_0}(\hat{\mu}_0^{GMM} - \mu_0) - \delta)
\end{equation*}
where $T_{\mathcal{M}_{\mathcal{J}}} (\mu_0)$ is the tangent cone of $\mathcal{M}_{\mathcal{J}}$ at $\mu_0$.\footnote{This means that $T_{\mathcal{M}_{\mathcal{J}}} (\mu_0)$ is equal to the cone formed by the closure of all the rays in $\mathbb{R}^F$ that emanate from $\mu_0$ and intersect $\mathcal{M}_{\mathcal{J}}$ in at least one point distinct from $\mu_0$.} Therefore, because $\Pi_{T_{\mathcal{M}_{\mathcal{J}}}(\mu_0)}$ is a continuous function, the Continuous Mapping Theorem can be applied to show that the asymptotic distribution of  $\sqrt{T_0}(\hat{\mu}_0 - \mu_0)$  will be equal to the asymptotic distribution of  $\sqrt{T_0}(\hat{\mu}_0^{GMM} - \mu_0)$  projected onto the set $T_{\mathcal{M}_{\mathcal{J}}} (\mu_0)$ using the function $\Pi_{T_{\mathcal{M}_{\mathcal{J}}}(\mu_0)}$. In order to show this, additional assumptions are required.  

\textbf{Assumption C2} \phantomsection\label{AC2} 
\begin{enumerate}
   \item There exists a unique $W^* \in \Delta^J$ such that $\mu_0 = \mu_{\mathcal{J}} W^*$.
    \item $\{(\lambda_t, \epsilon_t',\alpha_{0t})\}_{t \in \mathbb{Z}}$ is $\alpha$-mixing.
    \item $(\frac{1}{\sqrt{T_0}} \Tilde{Y}_{\mathcal{K}}^{pre} (Y_{0}^{pre'} - Y_{\mathcal{J}}^{pre'} W^*)) \overset{d}{\rightarrow} N(0, \mathcal{V})$ as $T_0 \rightarrow \infty$ for some nonrandom matrix $\mathcal{V}$.
    \item $\sum_{t \in \mathcal{T}_1} v_t\lambda_t \overset{p}{\rightarrow} \Bar{\lambda}_v$ as $T_1 \rightarrow \infty$ for some constant $\Bar{\lambda}_v$.
    \item $\sqrt{T_1} \sum_{t \in \mathcal{T}_1} v_t(\alpha_{0t} - \Bar{\alpha}_v + \epsilon_{0t} - \sum_{j \in \mathcal{J}} W_j^* \epsilon_{jt} ) \overset{d}{\rightarrow} N(0, \Sigma_v)$ as $T_1 \rightarrow \infty$ for some non-random matrix $\Sigma_v$.

\end{enumerate}

Assumption \hyperref[AC2]{C2.1} imposes that the control weights $W^* \in \Delta^J$ which allow us to reconstruct the unit of interest's factor loadings are unique. This helps to guarantee identification of these “true” control weights when the parameter space is being restricted to be the unit simplex. The mixing condition in Assumption \hyperref[AC2]{C2.2} will be used for guarantying the validity of the subsampling procedure. Assumption \hyperref[AC2]{C2.3} ensures the asymptotic normality of the sample moment equations. This is the typical condition imposed to show the asymptotic normality of unconstrained GMM estimators. Assumption \hyperref[AC2]{C2.5} helps ensure that the variability in $\hat{\Bar{\alpha}}_v$ due to the variation in treatment effects and due to the idiosyncratic shocks in the post-treatment time periods can be estimated by estimating the long-run variance $\Sigma_v$. The asymptotic normality conditions in Assumptions \hyperref[AC2]{C2.3} and \hyperref[AC2]{C2.5} can be satisfied even when the factors, treatment effects, and idiosyncratic shocks are not stationary and even when $v \ne (\frac{1}{T_1},...,\frac{1}{T_1})$. 

\textbf{Lemma C.1} \phantomsection\label{LC1} Suppose that the conditions of Assumptions \hyperref[A1]{1}, \hyperref[A2]{2}, and \hyperref[AC2]{C2} are satisfied, $A_{T_0} \overset{p}{\rightarrow} A$ where $A$ is positive definite, and $T_0\rightarrow \infty$ while $J$ is fixed. Then,
$$\sqrt{T_0}(\hat{\mu}_0 - \mu_0) \overset{d}{\rightarrow} \Pi_{T_{\mathcal{M}_{\mathcal{J}}}(\mu_0)} (Z_{1}) \text{ where }$$
$$Z_1 \sim  N(0, (\Lambda_{\mathcal{K}}' A \Lambda_{\mathcal{K}} \mu_{\mathcal{J}} \mu_{\mathcal{J}}' \Lambda_{\mathcal{K}}' A \Lambda_{\mathcal{K}})^{-1} \Lambda_{\mathcal{K}}' A \Lambda_{\mathcal{K}} \mu_{\mathcal{J}} \mu_{\mathcal{J}}' \Lambda_{\mathcal{K}}' A \mathcal{V} A \Lambda_{\mathcal{K}} \mu_{\mathcal{J}} \mu_{\mathcal{J}}' \Lambda_{\mathcal{K}}' A \Lambda_{\mathcal{K}} (\Lambda_{\mathcal{K}}' A \Lambda_{\mathcal{K}} \mu_{\mathcal{J}} \mu_{\mathcal{J}}' \Lambda_{\mathcal{K}}' A \Lambda_{\mathcal{K}})^{-1}).$$ 
\par
Using the asymptotic distribution of $\hat{\mu}_0$, we can then determine the asymptotic distribution of $\hat{\Bar{\alpha}}_v$ in the case where both the number of pre-treatment time periods and post-treatment time periods go to infinity. Specifically, using the fact that $\mu_0 = \mu_{\mathcal{J}}W^*$, we can rewrite $\sqrt{T_1}(\hat{\Bar{\alpha}}_v - \Bar{\alpha}_v)$ in three terms:

$$-\sqrt{\frac{T_1}{T_0}}(\sum_{t \in \mathcal{T}_1} v_t \lambda_t)(\sqrt{T_0}(\hat{\mu}_0 - \mu_0)) + \sqrt{T_1}\sum_{t \in \mathcal{T}_1} v_t (\alpha_{0t} - \Bar{\alpha}_v + \epsilon_{0t} - \sum_{j \in \mathcal{J}} W_j^* \epsilon_{jt}) + \sqrt{T_1}\sum_{t \in \mathcal{T}_1}v_t \sum_{j \in \mathcal{J}} (W_j^* - \hat{W}_j)\epsilon_{jt}.$$
The asymptotic distribution of the first term can be found using Lemma \hyperref[LC1]{C.1}, the second term will be asymptotically normal from Assumption \hyperref[AC2]{C2.5}, and the last term can be shown to converge in probability to zero. This decomposition of the error is similar to other decomposition that use a set of "oracle weights" (in this case $W^*$) such as \cite{SDID}. Also, similar to those cases, the term based on the error of estimating with the oracle weights will be smaller when $||W^*||_2$ is small.

\textbf{Proposition C.1} \phantomsection\label{PC1} Suppose that the conditions of Assumptions \hyperref[A1]{1}, \hyperref[A2]{2}, \hyperref[A3]{3}, \hyperref[AC1]{C1}, and \hyperref[AC2]{C2} are satisfied, $A_{T_0} \overset{p}{\rightarrow} A$ where $A$ is positive definite, and $T_0, T_1\rightarrow \infty$ such that $\lim_{T_0,T_1 \rightarrow \infty} \sqrt{\frac{T_1}{T_0}} = \phi$ while $J$ and $K$ are fixed. Then
\begin{equation*} 
    \sqrt{T_1}(\hat{\Bar{\alpha}}_v - \Bar{\alpha}_v) \overset{d}{\rightarrow} - \phi \Bar{\lambda}_v \Pi_{T_{\mathcal{M}_{\mathcal{J}}}(\mu_0)}(Z_1) + Z_2
\end{equation*}
where $Z_2 \sim N(0, \Sigma_v)$ and $Z_2$ is independent of $ \phi \Bar{\lambda}_v \Pi_{T_{\mathcal{M}_{\mathcal{J}}}(\mu_0)}(Z_1)$.
\par 
Note that Proposition \hyperref[PC1]{C.1} allows for the case where $\phi = 0$, in which case the asymptotic distribution simplifies to $Z_2$. The fact that the two terms in the asymptotic distribution are independent is because the variation in the first term is asymptotically only being driven by the variation in $\hat{W}$ which is estimated using only pre-treatment outcomes, while the variation in the second term is only being driven by post-treatment outcomes. I do not assume that pre-treatment outcomes and post-treatment outcomes are independent, but as $T_0$ and $T_1$ grow, time periods are added that are farther and farther apart from each other. Assumption \hyperref[AC2]{C2.2} helps to ensure that as this happens, the degree of dependence between these terms will gradually decrease because as the outcomes become farther apart, their degree of dependence converges to zero. Because $\hat{\alpha}_{0t} - \hat{\Bar{\alpha}}_v$ is converging to $\alpha_{0t} - \Bar{\alpha}_v + \epsilon_{0t} - \sum_{j \in \mathcal{J}}W^*_j \epsilon_{jt}$ at a sufficiently fast rate, $\Sigma_v$ can be estimated using the estimated deviations from the weighted average effect of treatment $\{\hat{\alpha}_{0t} - \hat{\Bar{\alpha}}_v\}_{t \in \mathcal{T}_1}$ using a variety of long-run variance estimators. 
\par
Unfortunately, $-\phi \lambda_v \Pi_{T_{\mathcal{M}_{\mathcal{J}}}(\mu_0)}(Z_1)$ is non-standard and so numerical methods will be required to calculate the asymptotic distribution. \cite{BootstrapInconsistency} shows that the standard bootstrap is inconsistent when the parameter is on the boundary of the parameter space. Therefore, I instead follow \cite{Li2020} by focusing on using a subsampling approach for estimating this term. Specifically, I can take advantage of the fact that this first term in finite samples can also be written as $-\sqrt{\frac{T_1}{T_0}}(\sum_{t \in \mathcal{T}_1} v_t Y_{\mathcal{J}t})(\sqrt{T_0}(\hat{W} - W^*))$ and use a subsampling approach based on resampling pre-treatment time periods and re-estimating the control weights. I let $m$ denote the subsample size. In order to employ standard asymptotic results for subsampling methods, we will want the size of our subsamples to grow as our sample size grows while also having the subsample size become a smaller and smaller fraction of the sample size, so in this context $m \rightarrow \infty$ and $m/T_0 \rightarrow 0$ as $T_0 \rightarrow \infty$. For collecting the subsamples, I focus on a block subsampling approach because of its robustness to temporal dependence in the outcomes. If we index time periods so that $\mathcal{T}_0 = \{1,...,T_0\}$, then our subsamples will be $\{(Y_{-Kt}^*,...,Y_{0t}^*,...,Y_{Jt}^*)\}_{t = 1}^m = \{(Y_{-Kt},...,Y_{0t},...,Y_{Jt})\}_{t = b}^{b+m-1}$, for $b = 1,...,T_0 - m +1$. This means we will have a total of $T_0 - m + 1$ subsamples. If we were instead willing to impose that $\{Y_t\}_{t \in \mathcal{T}_0}$ is $i.i.d.$, then the subsampling could done $i.i.d.$ as well. So for each $t = 1,...,m$, we draw $(Y_{-Kt}^*,...,Y_{0t}^*,...,Y_{Jt}^*)$ uniformly at random from $\{(Y_{-Kt},...,Y_{0t},...,Y_{Jt})\}_{t \in \mathcal{T}_0}$. While this allows for more subsamples, it is an unrealistic assumption in a setting with panel data. For each subsample, we can then calculate an estimate of the control weights $\hat{W}_b^*$ the same as in equation \eqref{eq: Time-Series Averaged SC} but replacing $\{(Y_{-Kt},...,Y_{0t},...,Y_{Jt})\}_{t \in \mathcal{T}_0}$ with $\{(Y_{-Kt}^*,...,Y_{0t}^*,...,Y_{Jt}^*)\}_{t = 1}^m$. Since $\phi \lambda_v \Pi_{T_{\mathcal{M}_{\mathcal{J}}}(\mu_0)}(Z_1)$ and $Z_2$ are independent, we can independently sample $s^*$ from $N(0,\hat{\Sigma}_v)$ and pair it with one of these subsampled control weights $\hat{W}_b^*$ selected uniformly at random. A statistic for the weighted average of treatment effects is then given by
\begin{equation*}
    \hat{\alpha}^* = - \frac{1}{\sqrt{T_0}}(\sum_{t \in \mathcal{T}_1} v_t Y_{\mathcal{J} t}') \sqrt{m} (\hat{W}_b^* - \hat{W}) + \frac{1}{\sqrt{T_1}} s^*.
\end{equation*}
We repeat this process a large number of times which I denote as $N$. We can then use these statistics to construct confidence intervals for $\Bar{\alpha}_v$ by subtracting them from our estimate $\hat{\Bar{\alpha}}_v$. We sort the statistics such that $\hat{\alpha}^*_1 \leq \hat{\alpha}^*_2 \leq ... \leq \hat{\alpha}^*_N$
and then for a $1 - \delta$ confidence interval for $\Bar{\alpha}_v$, we use 
\begin{equation*}
    [\hat{\Bar{\alpha}}_v - \hat{\alpha}^*_{(1-\delta/2)N},  \hat{\Bar{\alpha}}_v - \hat{\alpha}^*_{(\delta/2)N}].
\end{equation*}
I show that under the same conditions as in Proposition \hyperref[PC1]{C.1}, this method for constructing confidence intervals will provide asymptotically correct coverage.
\par

\textbf{Proposition C.2}\phantomsection\label{PC2}
Suppose the same conditions as in Proposition \hyperref[PC1]{C.1}. Additionally, suppose that $N,m \rightarrow \infty$, $m/T_0 \rightarrow \infty$, and $\hat{\Sigma}_v \overset{p}{\rightarrow} \Sigma_v$ as $T_0, T_1 \rightarrow \infty$.  
Then $\text{for all } \delta \in (0,1)$,
\begin{equation*}
    \lim_{T_0,T_1 \rightarrow \infty} P(\hat{\Bar{\alpha}}_v - \hat{\alpha}^*_{(1-\delta/2)N} \leq \Bar{\alpha}_v \leq \hat{\Bar{\alpha}}_v - \hat{\alpha}^*_{(\delta/2)N}) = 1 - \delta.
\end{equation*}

\subsection*{Proofs of Inference Results}

\textbf{Lemma C.1 Proof:} Let the set of optimal solutions for the unconstrained estimator in equation \eqref{eq: Unconstrained} be denoted by
\begin{equation*}
    \mathcal{W}^{GMM} = \argmin_{W} 
    (\frac{1}{T_0}\Tilde{Y}_{\mathcal{K}}^{pre}(Y_{0}^{pre'} - Y_{\mathcal{J}}^{pre'}\hat{W}^{GMM}))'A_{T_0}(\frac{1}{T_0}\Tilde{Y}_{\mathcal{K}}^{pre}(Y_{0}^{pre'} - Y_{\mathcal{J}}^{pre'}\hat{W}^{GMM}))
\end{equation*}
and because $\mathcal{W}^{GMM}$ is closed and convex, let $\hat{W}^{GMM} \in \argmin_{W \in \mathcal{W}^{GMM}} ||W - W^*||_2$. Note that $\hat{W}^{GMM}$ must satisfy the first-order condition, 
$$\frac{1}{T_0}Y_{\mathcal{J}}^{pre} \Tilde{Y}_{\mathcal{K}}^{pre'}A_{T_0} (\frac{1}{T_0}\Tilde{Y}_{\mathcal{K}}^{pre}(Y_{0}^{pre'} - Y_{\mathcal{J}}^{pre'}\hat{W}^{GMM})) = 0.$$
Using the fact that
$$ \frac{1}{T_0}\Tilde{Y}_{\mathcal{K}}^{pre}(Y_{0}^{pre'} - Y_{\mathcal{J}}^{pre'}\hat{W}^{GMM}) =  \frac{1}{T_0}\Tilde{Y}_{\mathcal{K}}^{pre}(Y_{0}^{pre'} - Y_{\mathcal{J}}^{pre'}W^*) - \frac{1}{T_0}\Tilde{Y}_{\mathcal{K}}^{pre} Y_{\mathcal{J}}^{pre'}(\hat{W}^{GMM} - W^*),$$
we can rewrite the first-order condition as 
$$\frac{1}{T_0}Y_{\mathcal{J}}^{pre} \Tilde{Y}_{\mathcal{K}}^{pre'}A_{T_0} \frac{1}{T_0}\Tilde{Y}_{\mathcal{K}}^{pre} Y_{\mathcal{J}}^{pre'} \sqrt{T_0}(\hat{W}^{GMM} - W^*) = - \frac{1}{T_0}Y_{\mathcal{J}}^{pre} \Tilde{Y}_{\mathcal{K}}^{pre} A_{T_0} (\frac{1}{\sqrt{T_0}}\Tilde{Y}_{\mathcal{K}}^{pre}(Y_{0}^{pre'} - Y_{\mathcal{J}}^{pre'}W^*)). $$
Expanding the left side of the equation,
$$\frac{1}{T_0}Y_{\mathcal{J}}^{pre} \Tilde{Y}_{\mathcal{K}}^{pre'}A_{T_0} (\frac{1}{T_0}\Tilde{Y}_{\mathcal{K}}^{pre} (\lambda^{pre}\mu_{\mathcal{J}} + \epsilon_{\mathcal{J}}^{pre'}) \sqrt{T_0}(\hat{W}^{GMM} - W^*) = - \frac{1}{T_0}Y_{\mathcal{J}}^{pre} \Tilde{Y}_{\mathcal{K}}^{pre} A_{T_0} (\frac{1}{\sqrt{T_0}}\Tilde{Y}_{\mathcal{K}}^{pre}(Y_{0}^{pre'} - Y_{\mathcal{J}}^{pre'}W^*)) .$$
By Assumption \hyperref[A2]{2} and $A_{T_0} = O_p(1)$, $\max_{j \in \mathcal{J}} \frac{1}{T_0}Y_{\mathcal{J}}^{pre} \Tilde{Y}_{\mathcal{K}}^{pre'}A_{T_0}  \frac{1}{T_0} \Tilde{Y}_{\mathcal{K}}^{pre} \epsilon_{j}^{pre'} = o_p(1)$ and I show in Lemma \hyperref[LC3]{C.3} that $\sqrt{T_0}||\hat{W}^{GMM} - W^*||_2 = O_p(1)$, so $\frac{1}{T_0}Y_{\mathcal{J}}^{pre} \Tilde{Y}_{\mathcal{K}}^{pre'}A_{T_0} \frac{1}{T_0} \Tilde{Y}_{\mathcal{K}}^{pre} \epsilon_{\mathcal{J}}^{pre'}\sqrt{T_0}(\hat{W}^{GMM} - W^*) = o_p(1)$. Therefore, $\sqrt{T_0}\mu_{\mathcal{J}}(\hat{W}^{GMM} - W^*)$ is equal to
$$(\lambda^{pre'}\Tilde{Y}_{\mathcal{K}}^{pre'}A_{T_0} \Tilde{Y}_{\mathcal{K}}^{pre}Y_{\mathcal{J}}^{pre'} Y_{\mathcal{J}}^{pre} \Tilde{Y}_{\mathcal{K}}^{pre'}A_{T_0} \Tilde{Y}_{\mathcal{K}}^{pre} \lambda^{pre}/T_0^4)^{-1} \frac{1}{T_0} Y_{\mathcal{J}}^{pre} \Tilde{Y}_{\mathcal{K}}^{pre'} A_{T_0}(\frac{1}{\sqrt{T_0}}\Tilde{Y}_{\mathcal{K}}^{pre}(Y_{0}^{pre'} - Y_{\mathcal{J}}^{pre'}W^*)) + o_p(1).$$
Then letting $\hat{\mu}_0^{GMM} = \mu_{\mathcal{J}}\hat{W}^{GMM}$ and using Assumptions \hyperref[A2]{2} and \hyperref[AC2]{C2} we have that $ \sqrt{T_0}(\hat{\mu}_0^{GMM} - \mu_0) \overset{d}{\rightarrow} Z_1 \sim$
$$N(0, (\Lambda_{\mathcal{K}}' A \Lambda_{\mathcal{K}} \mu_{\mathcal{J}} \mu_{\mathcal{J}}' \Lambda_{\mathcal{K}}' A \Lambda_{\mathcal{K}})^{-1} \Lambda_{\mathcal{K}}' A \Lambda_{\mathcal{K}} \mu_{\mathcal{J}} \mu_{\mathcal{J}}' \Lambda_{\mathcal{K}}' A \mathcal{V} A \Lambda_{\mathcal{K}} \mu_{\mathcal{J}} \mu_{\mathcal{J}}' \Lambda_{\mathcal{K}}' A \Lambda_{\mathcal{K}} (\Lambda_{\mathcal{K}}' A \Lambda_{\mathcal{K}} \mu_{\mathcal{J}} \mu_{\mathcal{J}}' \Lambda_{\mathcal{K}}' A \Lambda_{\mathcal{K}})^{-1}).$$
I show in Lemma \hyperref[LC2]{C.2} that $\hat{\mu}_0 = \mu_{\mathcal{J}} \Pi_{\Delta^J}(\hat{W}^{GMM}) + o_p(T_0^{-\frac{1}{2}})$. Therefore, 
$$\sqrt{T_0}(\hat{\mu}_0 - \mu_0) = \sqrt{T_0}\mu_{\mathcal{J}}(\Pi_{\Delta^J} (\hat{W}^{GMM}) - W^*) + o_p(1) $$
$$= \Pi_{T_{\mathcal{M}_{\mathcal{J}}}(\mu_0)}(\sqrt{T_0}(\hat{\mu}_0^{GMM} - \mu_0)) + o_p(1),$$
where the third equality follows from performing a change of variable in the optimization problem defined by equation \eqref{eq: Asymptotic Projection}. By Berge’s Maximum Theorem, the function 
 $\Pi_{T_{\mathcal{M}_{\mathcal{J}}}(\mu_0)}$ must be continuous. Hence, by the Continuous Mapping Theorem,
$$\sqrt{T_0}(\hat{\mu}_0 - \mu_0) \overset{d}{\rightarrow} 
 \Pi_{T_{\mathcal{M}_{\mathcal{J}}}(\mu_0)}(Z_1).$$

\textbf{Proposition C.1 Proof:} First $\sqrt{T_1}(\hat{\Bar{\alpha}}_v - \Bar{\alpha}_v)$ can be rewritten as $\sqrt{T_1}(\hat{\Bar{\alpha}}_v - \Bar{\alpha}_v) = \hat{\Bar{\alpha}}_1 + \hat{\Bar{\alpha}}_2 + \hat{\Bar{\alpha}}_3$ where
\begin{equation*} 
    \hat{\Bar{\alpha}}_1 = - (\sum_{t \in \mathcal{T}_1} v_t \lambda_t)\sqrt{\frac{T_1}{T_0}}\sqrt{T_0}(\hat{\mu}_0 - \mu_0), \quad
    \hat{\Bar{\alpha}}_2 = \sqrt{T_1}\sum_{t \in \mathcal{T}_1}v_t (\epsilon_{0t} - \sum_{j \in \mathcal{J}} W_j^* \epsilon_{jt} + \alpha_{0t} - \Bar{\alpha}_v), 
\end{equation*}
\begin{equation*}
    \hat{\Bar{\alpha}}_3 = \sqrt{T_1}\sum_{t \in \mathcal{T}_1} v_t \sum_{j \in \mathcal{J}} (W_j^* - \hat{W}_j)\epsilon_{jt}.
\end{equation*}
By Assumption \hyperref[A3]{3}, $\sum_{t \in \mathcal{T}_1} v_t \epsilon_{jt} = o_p(1)$ for each $j \in \mathcal{J}$. I show in Lemma \hyperref[LC3]{C.3} that $\sqrt{T_0}||W^* - \hat{W}||_2 = O_p(1)$. Since $\lim_{T_0,T_1 \rightarrow \infty} \sqrt{\frac{T_1}{T_0}} = \phi < \infty$, this implies that  $\sqrt{T_1}||W^* - \hat{W}||_2 = O_p(1)$. Therefore, $\hat{\Bar{\alpha}}_3 = o_p(1)$.
By Assumption \hyperref[AC2]{C2.5}, we have that $\hat{\Bar{\alpha}}_2 \overset{d}{\rightarrow} Z_2 \sim N(0, \Sigma_v)$. Also by Assumption \hyperref[AC2]{C2}, $\sum_{t \in \mathcal{T}_1} v_t \lambda_t \overset{p}{\rightarrow} \Bar{\lambda}_v$ and by Lemma \hyperref[LC1]{C.1} $\sqrt{T_0}(\hat{\mu}_0 -\mu_0) \overset{d}{\rightarrow} \Pi_{T_{\mathcal{M}_{\mathcal{J}}}(\mu_0)} (Z_1)$. Therefore, because  $\lim_{T_0,T_1 \rightarrow \infty} \sqrt{\frac{T_1}{T_0}} = \phi$, $\hat{\Bar{\alpha}}_1 \overset{d}{\rightarrow} -\phi \lambda_v \Pi_{T_{\mathcal{M}_{\mathcal{J}}}(\mu_0)}(Z_1)$. Let $s$ be equal to the first post-treatment time period, so that $\mathcal{T}_1 = \{s,...,s+T_1 - 1\}$. Then it will also be the case that $\hat{\Bar{\alpha}}_2^p = \sum_{t = p}^{s+T_1 -1} v_t (\epsilon_{0t} - \sum_{j \in \mathcal{J}} W_j^* \epsilon_{jt} + \alpha_{0t} - \Bar{\alpha}_v) = \sum_{t = p}^{s+T_1 -1} v_t (Y_{0t} - \sum_{j \in \mathcal{J}} W_j^* Y_{jt} - \Bar{\alpha}_v)  \overset{d}{\rightarrow} Z_1$ as $p, T_1 \rightarrow \infty$ if $\frac{p}{T_1} \rightarrow 0$. However, because $\{(\lambda_t, \epsilon_t' , \alpha_{0t})\}_{t \in \mathbb{Z}}$ is $\alpha$-mixing and $\hat{\Bar{\alpha}}_1$ is some measurable function of $\{Y_t\}_{t \in \mathcal{T}_0}$ and $\hat{\Bar{\alpha}}_2^p$ is some measurable function of $\{Y_t\}_{t \geq p}$, it must be the case that $$\sup_{A ,B \in \mathcal{B}} |P(\hat{\Bar{\alpha}}_1 \in A, \hat{\Bar{\alpha}}_2^p \in B) - P(\hat{\Bar{\alpha}}_1 \in A)P( \hat{\Bar{\alpha}}_2^p \in B)| \rightarrow 0$$
where $\mathcal{B}$ is the Borel sets. Combining this with the convergence in distribution results, we have for all $A,B \in \mathcal{B}$, $P(\hat{\Bar{\alpha}}_1 \in A, \hat{\Bar{\alpha}}_2^p \in B) \rightarrow P(-\phi \Bar{\lambda}_v \Pi_{T_{\mathcal{M}_{\mathcal{J}}}(\mu_0)}(Z_1) \in A)P(Z_2 \in B)$ as $T_0, T_1, p \rightarrow \infty$ with $\frac{p}{T_1} \rightarrow 0$. Then $\hat{\Bar{\alpha}}_1 + \hat{\Bar{\alpha}}_2^p \overset{d}{\rightarrow} -\phi \Bar{\lambda}_v \Pi_{T_{\mathcal{M}_{\mathcal{J}}}(\mu_0)} (Z_1) + Z_2$ where $\phi \Bar{\lambda}_v \Pi_{T_{\mathcal{M}_{\mathcal{J}}}(\mu_0)} (Z_1)$ and $Z_2$ are independent. Then because $ \hat{\Bar{\alpha}}_2 = \hat{\Bar{\alpha}}_2^p + o_p(1)$ by Assumption \hyperref[AC2]{C2}, it must also be the case that $\sqrt{T_1}(\hat{\Bar{\alpha}}_v - \Bar{\alpha}_v) \overset{d}{\rightarrow} -\phi \Bar{\lambda}_v \Pi_{T_{\mathcal{M}_{\mathcal{J}}}(\mu_0)} \Pi (Z_1) + Z_2$ where $ -\phi \Bar{\lambda}_v \Pi_{T_{\mathcal{M}_{\mathcal{J}}}(\mu_0)} \Pi (Z_1)$ and $Z_2$ are independent.

\textbf{Proposition C.2 Proof:} Let $A \overset{d}{\sim} B$ denote that $A$ and $B$ converge in distribution to the same distribution. Then $v^* \overset{d}{\sim} \sum_{t \in \mathcal{T}_1} v_t (\alpha_{0t} - \Bar{\alpha}_0 + \epsilon_{0t} - \epsilon_{\mathcal{J}t}'W^*) \overset{d}{\rightarrow} Z_2$ by Assumption \hyperref[AC2]{C2.5} and Proposition \hyperref[PC1]{C.1}. For each subsample, let $\hat{\mu}_b^* = \mu_{\mathcal{J}} \hat{W}_b^*$. Because $m \rightarrow \infty$ and $m/T_0 \rightarrow 0$ as $T_0 \rightarrow \infty$, $\sqrt{T_0}(\hat{\mu}_0 - \mu_0)$ converges in distribution by Proposition \hyperref[PC1]{C.1}, and $\{Y_t\}_{t \in \mathcal{T}_0}$ is $\alpha$-mixing by Assumption \hyperref[AC2]{C2.2}, it follows from Theorem 4.3.1 of \cite{SubsamplingBook} that $\sqrt{T_0}\mu_{\mathcal{J}}(\hat{W}_b^* - \hat{W}) = \sqrt{T_0}(\hat{\mu}_b^* - \hat{\mu}_0) \overset{d}{\sim} \sqrt{T_0}(\hat{\mu}_0 - \mu_0)$. Therefore, $\hat{\Bar{\alpha}}_v - \hat{\alpha}^* \overset{d}{\sim} \sqrt{T_1}(\hat{\Bar{\alpha}}_v - \Bar{\alpha}_v)$.

\textbf{Lemma C.2} \phantomsection\label{LC2} Under the same conditions as in Lemma \hyperref[LC1]{C.1}, we have
\begin{equation*}
    \hat{\mu}_0 = \mu_{\mathcal{J}} \Pi_{\Delta^J} (\hat{W}^{GMM}) + o_p(T_0^{-\frac{1}{2}})
\end{equation*}
where $\hat{W}^{GMM} = \argmin_{W \in \mathcal{W}^{GMM}} ||W - W^*||_2$ and $\mathcal{W}^{GMM}$ is the set of optimal solutions for the estimator defined by equation \eqref{eq: Unconstrained}.

\textbf{Proof:} The proof follows Lemma C.1 from \cite{Li2020}. Let $\Tilde{W} = \Pi_{\Delta^J}(\hat{W}^{GMM})$ and $\Tilde{\mu}_0 = \mu_{\mathcal{J}}\Tilde{W}$. Then let $\epsilon > 0$ and suppose $\sqrt{T_0}||\hat{\mu}_0 - \Tilde{\mu}_0||_2 > \epsilon$. Because $\sqrt{T_0}||\mu_{\mathcal{J}}||_2 ||\hat{W} - \Tilde{W}||_2 \geq \sqrt{T_0}||\mu_{\mathcal{J}}(\hat{W} - \Tilde{W})||_2 > \epsilon$, $\sqrt{T_0}||\hat{W} - \Tilde{W}||_2 > \epsilon/||\mu_{\mathcal{J}}||_2$. Let $\pi_{\Delta^J,T_0} (\theta) = \theta' (Y_{\mathcal{J}}^{pre} \Tilde{Y}_{\mathcal{K}}^{pre'} A_{T_{0}} \Tilde{Y}_{\mathcal{K}}^{pre} Y_{\mathcal{J}}^{pre'}/T_0) \theta = \sqrt{T_0}\theta' (Y_{\mathcal{J}}^{pre} \Tilde{Y}_{\mathcal{K}}^{pre'} A_{T_{0}} \Tilde{Y}_{\mathcal{K}}^{pre} Y_{\mathcal{J}}^{pre'}/T_0^2) \sqrt{T_0}\theta$. Then because $\hat{W} \ne \Tilde{W}$, $\Tilde{W} \in \Delta^J$, and $\hat{W} = \Pi_{\Delta^J, T_0} (\hat{W}^{GMM})$, $$\pi_{\Delta^J,T_0} (\hat{W}^{GMM} - \hat{W}) < \pi_{\Delta^J,T_0} (\hat{W}^{GMM} - \Tilde{W}). \quad (B.1)$$ Then note that because of the quadratic form of $\pi_{\Delta^J,T_0}(\theta)$, 
\begin{equation*}
    \pi_{\Delta^J,T_0}(\hat{W}^{GMM} - \hat{W}) = \pi_{\Delta^J,T_0}(\hat{W}^{GMM} - \Tilde{W} + \Tilde{W} - \hat{W}) = \pi_{\Delta^J,T_0}(\hat{W}^{GMM} - \Tilde{W}) + \pi_{\Delta^J,T_0}(\Tilde{W} - \hat{W}) + 
\end{equation*}
\begin{equation*}
    2\sqrt{T_0}(\Tilde{W} - \hat{W}^{GMM})'(Y_{\mathcal{J}}^{pre} \Tilde{Y}_{\mathcal{K}}^{pre'} A_{T_{0}} \Tilde{Y}_{\mathcal{K}}^{pre} Y_{\mathcal{J}}^{pre'}/T_0^2) \sqrt{T_0}(\hat{W} - \Tilde{W}). \quad (B.2)
\end{equation*}
Combining B.1 and B.2, the last two terms in B.2 are negative so $D_{1,T_0} + D_{2,T_0} < 0$ where $D_{1,T_0} \coloneqq \pi_{\Delta^J,T_0}(\Tilde{W} - \hat{W})$ and $D_{2,T_0} \coloneqq 2\sqrt{T_0}(\Tilde{W} - \hat{W}^{GMM})'(Y_{\mathcal{J}}^{pre} \Tilde{Y}_{\mathcal{K}}^{pre'} A_{T_{0}} \Tilde{Y}_{\mathcal{K}}^{pre} Y_{\mathcal{J}}^{pre'}/T_0^2) \sqrt{T_0}(\hat{W} - \Tilde{W})$.
\par 
Let $\mathcal{S}^J = \{a \in \mathbb{R}^J: ||a||_2 = 1\}$ denote the unit sphere in $\mathbb{R}^J$. Note that 
\begin{equation*}
    D_{1,T_0} = \sqrt{T_0}(\Tilde{W} - \hat{W})' (Y_{\mathcal{J}}^{pre} \Tilde{Y}_{\mathcal{K}}^{pre'} A_{T_{0}} \Tilde{Y}_{\mathcal{K}}^{pre} Y_{\mathcal{J}}^{pre'}/T_0^2) \sqrt{T_0}(\Tilde{W} - \hat{W})
\end{equation*}
\begin{equation*}
    = \sqrt{T_0}(\Tilde{W} - \hat{W})' (\mu_{\mathcal{J}}'\lambda^{pre'} \Tilde{Y}_{\mathcal{K}}^{pre'} A_{T_{0}} \Tilde{Y}_{\mathcal{K}}^{pre} \lambda^{pre} \mu_{\mathcal{J}} /T_0) \sqrt{T_0}(\Tilde{W} - \hat{W}) + o_p(1)
\end{equation*}
\begin{equation*}
    = ||\sqrt{T_0}(\Tilde{\mu}_0 - \hat{\mu}_0)||_2^2 [\frac{\sqrt{T_0}(\Tilde{\mu}_0 - \hat{\mu}_0)'}{||\sqrt{T_0}(\Tilde{\mu}_0 - \hat{\mu}_0)||_2} (\lambda^{pre'} \Tilde{Y}_{\mathcal{K}}^{pre'} A_{T_{0}} \Tilde{Y}_{\mathcal{K}}^{pre} \lambda^{pre'}/T_0) \frac{\sqrt{T_0}(\Tilde{\mu}_0 - \hat{\mu}_0)}{||\sqrt{T_0}(\Tilde{\mu}_0 - \hat{\mu}_0)||_2}] + o_p(1)
\end{equation*}
\begin{equation*}
    \geq T_0 ||\Tilde{\mu}_0 - \hat{\mu}_0||_2^2 \inf_{a \in \mathcal{S}^J} a'(\lambda^{pre'} \Tilde{Y}_{\mathcal{K}}^{pre'} A_{T_{0}} \Tilde{Y}_{\mathcal{K}}^{pre} \lambda^{pre}/T_0^2)a = T_0 ||\Tilde{\mu}_0 - \hat{\mu}_0||_2^2 \lambda_{min} (\lambda^{pre}\Tilde{Y}_{\mathcal{K}}^{pre'} A_{T_{0}} \Tilde{Y}_{\mathcal{K}}^{pre} \lambda^{pre}/T_0^2) + o_p(1)
\end{equation*}
\begin{equation*}
    \geq \epsilon^2 \lambda_{min} (\lambda^{pre'}\Tilde{Y}_{\mathcal{K}}^{pre'} A_{T_{0}} \Tilde{Y}_{\mathcal{K}}^{pre} \lambda^{pre}/T_0^2) + o_p(1) \overset{p}{\rightarrow} \epsilon^2 \lambda_{min}( \Lambda_{\mathcal{K}}'  A \Lambda_{\mathcal{K}} ) > 0,
\end{equation*}
where the minimum eigenvalue of a square matrix $M$ is denoted by $\lambda_{min} (M)$. The second equality come from Assumptions \hyperref[A2]{2.1}, \hyperref[A2]{2.4}, and \hyperref[A2]{2.5}. The third equality follows from Lemma C.2 of \cite{Li2020}. The second inequality follows because $\sqrt{T_0}||\Tilde{\mu}_0 - \hat{\mu}_0||_2 \geq \epsilon$ and the final term is strictly positive because $\Lambda_{\mathcal{K}}'  A \Lambda_{\mathcal{K}}$ is nonsingular. By rewriting $Y_{\mathcal{J}}^{pre} \Tilde{Y}_{\mathcal{K}}^{pre'} A_{T_{0}} \Tilde{Y}_{\mathcal{K}}^{pre} Y_{\mathcal{J}}^{pre'}/T_0^2 = \mu_{\mathcal{J}}'\Lambda_{\mathcal{K}}' A \Lambda_{\mathcal{K}} \mu_{\mathcal{J}} + Y_{\mathcal{J}}^{pre} \Tilde{Y}_{\mathcal{K}}^{pre'} A_{T_{0}} \Tilde{Y}_{\mathcal{K}}^{pre} Y_{\mathcal{J}}^{pre'}/T_0^2 - \mu_{\mathcal{J}}' \Lambda_{\mathcal{K}}' A \Lambda_{\mathcal{K}} \mu_{\mathcal{J}}$, $D_{2,T_0}$ can be rewritten as $D_{2,1,T_0} + D_{2,2,T_0}$ where
\begin{equation*}
    D_{2,1,T_0} \coloneqq 2 \sqrt{T_0}(\Tilde{W} - \hat{W}^{GMM})' \mu_{\mathcal{J}}' \Lambda_{\mathcal{K}}' A \Lambda_{\mathcal{K}} \Omega_0 \mu_{\mathcal{J}} \sqrt{T_0} (\hat{W} - \Tilde{W}) \text{ and }
\end{equation*}
\begin{equation*}
    D_{2,2,T_0} \coloneqq 2 \sqrt{T_0}(\Tilde{W} - \hat{W}^{GMM})'(Y_{\mathcal{J}}^{pre} \Tilde{Y}_{\mathcal{K}}^{pre'} A_{T_{0}} \Tilde{Y}_{\mathcal{K}}^{pre} Y_{\mathcal{J}}^{pre'}/T_0^2 - \mu_{\mathcal{J}}' \Lambda_{\mathcal{K}}' A \Lambda_{\mathcal{K}} \mu_{\mathcal{J}}) \sqrt{T_0} (\hat{W} - \Tilde{W}).
\end{equation*}
By definition of $\Tilde{W}$ and Lemma 1.1 in \cite{Zarantonello1971}, $D_{2,1,T_0} \geq 0$. By Assumption \hyperref[A2]{2}, 
$$||Y_{\mathcal{J}}^{pre} \Tilde{Y}_{\mathcal{K}}^{pre'} A_{T_{0}} \Tilde{Y}_{\mathcal{K}}^{pre} Y_{\mathcal{J}}^{pre'}/T_0^2 - \mu_{\mathcal{J}}' \Lambda_{\mathcal{K}}' A \Lambda_{\mathcal{K}} \mu_{\mathcal{J}}||_2 = o_p(1).$$ Additionally, $\sqrt{T_0}(\hat{W} - \Tilde{W}) = O_p(1)$ because
\begin{equation*}
    ||\sqrt{T_0}(\hat{W} - \Tilde{W})||_2 \leq ||\sqrt{T_0}(\hat{W} - W^*)||_2 + ||\sqrt{T_0}(\Tilde{W} - W^*)||_2
\end{equation*}
\begin{equation*}
    =\sqrt{T_0}||\hat{W} - W^*||_2 + \sqrt{T_0}||\Pi_{\Delta^J} (\hat{W}^{GMM})-\Pi_{\Delta^J}(W^*)||_2
\end{equation*}
\begin{equation*}
    \leq \sqrt{T_0}||\hat{W} - W^*||_2 + \sqrt{T_0}C||\hat{W}^{GMM} - W^*||_2 = O_p(1),
\end{equation*}
where C is some constant. The second inequality follows from the Lipschitz continuity of projection operators shown by \cite{Zarantonello1971} and the final equality follows from Lemma \hyperref[LC3]{C.3}. Therefore, $D_{2,2,T_0} = o_p(1)$. Then since $D_{2,1,T_0} \geq 0$ implies that $D_{2,T_0} \geq o_p(1)$.
\par  
Hence, we have that $\sqrt{T_0}||\hat{\mu}_0 - \Tilde{\mu}_0||_2 > \epsilon$ implies that $D_{T_0} < 0$ and that $D_{T_0} = D_{1,T_0} + D_{2,T_0} \geq \epsilon^2 \lambda_{min}( \lambda^{pre'} \Tilde{Y}_{\mathcal{K}}^{pre'} A_{T_{0}} \Tilde{Y}_{\mathcal{K}}^{pre} \lambda^{pre}/T_0^2) + o_p(1)$. Using this, we have that
\begin{equation*}
    P(\sqrt{T_0}||\hat{\mu}_0 - \Tilde{\mu}_0||_2 > \epsilon) \leq P(D_{T_0} < 0) \leq 
\end{equation*}
\begin{equation*}
    P(o_p(1) + \epsilon^2 \lambda_{min}( Y_{\mathcal{J}}^{pre} \Tilde{Y}_{\mathcal{K}}^{pre'} A_{T_{0}} \Tilde{Y}_{\mathcal{K}}^{pre} Y_{\mathcal{J}}^{pre'}/T_0) < 0) \rightarrow P(\epsilon^2 \lambda_{min}(\mu_{\mathcal{J}}' \Omega_0 \mu_{\mathcal{K}} A \mu_{\mathcal{K}}' \Omega_0 \mu_{\mathcal{J}}) \leq 0) = 0
\end{equation*}
Therefore, $T_0||\hat{\mu}_0 - \Tilde{\mu}_0||_2 = o_p(1)$, or equivalently $\hat{\mu}_0 = \mu_{\mathcal{J}}\Pi_{\Delta^J}(\hat{W}^{GMM}) + o_p(T_0^{-\frac{1}{2}})$.

\textbf{Lemma C.3}\phantomsection\label{LC3} Suppose the same conditions as Lemma \hyperref[LC1]{C.1}. Let $\mathcal{W}^{GMM}$ denote the set of optimal solutions for the unconstrained estimator in equation \eqref{eq: Unconstrained} and let $\hat{W}^{GMM} = \argmin_{W \in \mathcal{W}^{GMM}} ||W - W^*||_2$. Then $\sqrt{T_0}||\hat{W}^{GMM} - W^*||_2 = O_p(1)$ and $\sqrt{T_0}||\hat{W} - W^*||_2 = O_p(1)$.

\textbf{Proof:}
 Let $H_{T_0}(W)$ equal the objective function in equation \eqref{eq: Unconstrained} and 
\begin{equation*}
    H(W) = (\mu_0 - \mu_{\mathcal{J}}W)'\Lambda_{\mathcal{K}}' A \Lambda_{\mathcal{K}} (\mu_0 - \mu_{\mathcal{J}}W).
\end{equation*}
As shown in Lemma \hyperref[L1]{1} for the case where $J$ is fixed, $H_{T_0}(W)$ converges uniformly in probability to $H(W)$ on $\Delta^J$. Then the fact that $W^*$ is the unique solution to $\argmin_{W \in \Delta^J} H(W)$ and Theorem 3.2 of \cite{PartialGMM} imply that $\sqrt{T_0}||\hat{W} - W^*||_2 = O_p(1)$ for each $\hat{W} \in \argmin_{W \in \Delta^J}  H_{T_0}(W)$. Let $\delta > 0$ and $\mathcal{B} = \{W : ||W - W^*||_2 \leq 2\delta \}$. By the same reasoning as in Lemma \hyperref[L1]{1}, because $\mathcal{B}$ is compact, $H_{T_0}(W)$ converges uniformly in probability to $H(W)$ on $\mathcal{B}$. Therefore, for
the solution set $\Tilde{W}^{GMM} = \argmin_{W \in \mathcal{B}} H_{T_0}(W)$, Theorem 3.2 of \cite{PartialGMM} also implies that $\inf_{W \in \Tilde{W}^{GMM}} ||W - W^*||_2 = O_p(\frac{1}{\sqrt{T_0}})$. So the event that there exists $\Tilde{W} = \argmin_{W \in \Tilde{\mathcal{W}}^{GMM}} ||W - W^*||_2$ such that $||\Tilde{W} - W^*||_2 < \delta$ occurs with probability approaching one (wpa1). In this event, for $W \notin \mathcal{B}$, there exists $0 < \lambda < 1$ such that $\lambda \Tilde{W} + (1 - \lambda)W \in \mathcal{B}$ so that
\begin{equation*}
    H_{T_0}(\Tilde{W}) \leq H_{T_0}(\lambda \Tilde{W} + (1-\lambda)W) \leq  \lambda H_{T_0}(\Tilde{W}) + 
    (1-\lambda)H_{T_0}  W)
\end{equation*}
where the second inequality follows from the convexity of $H_{T_0}(W)$. Therefore, $H_{T_0}(\Tilde{W}) \leq H_{T_0}(W)$ for all $W$ wpa1 so $\Tilde{W} \in \mathcal{W}^{GMM}$ wpa1. Because $\hat{W}^{GMM} = \argmin_{W \in \mathcal{W}^{GMM}} ||W - W^*||_2$,  $||\hat{W}^{GMM} - W^*||_2 \leq ||\Tilde{W} - W^*||_2$ wpa1. Hence, it also the case that $\sqrt{T_0}||\hat{W}^{GMM} - W^*||_2 = O_p(1)$.

\textbf{Proof that $\hat{W} = \Pi_{\Delta^J,T_0}(\hat{W}^{GMM})$:} Let $\hat{W}^{GMM}$ be defined as in equation \eqref{eq: Unconstrained}. In order to see how the objective function defining $\hat{W}$ in equation \eqref{eq: Time-Series Averaged SC} can be rewritten in terms of $\hat{W}^{GMM}$, let $\hat{u} = Y_{0}^{pre'} - Y_{\mathcal{J}}^{pre'} \hat{W}^{GMM}$ and note that 
\begin{equation*}
    (Y_{0}^{pre'} - Y_{\mathcal{J}}^{pre'} W)'\Tilde{Y}_{\mathcal{K}}^{pre'} A_{T_0} \Tilde{Y}_{\mathcal{K}}^{pre} (Y_{0}^{pre'} - Y_{\mathcal{J}}^{pre'} W) = (\hat{u} + Y_{\mathcal{J}}^{pre'}(\hat{W}^{GMM} - W))' \Tilde{Y}_{\mathcal{K}}^{pre'} A_{T_0} \Tilde{Y}_{\mathcal{K}}^{pre} (\hat{u} + Y_{\mathcal{J}}^{pre'} (\hat{W} - W)) = 
\end{equation*}
\begin{equation*}
    \hat{u}' \Tilde{Y}_{\mathcal{K}}^{pre'} A_{T_0} \Tilde{Y}_{\mathcal{K}}^{pre} \hat{u} + 2\hat{u}' \Tilde{Y}_{\mathcal{K}}^{pre'} A_{T_0} \Tilde{Y}_{\mathcal{K}}^{pre} Y_{\mathcal{J}}^{pre'} (\hat{W}^{GMM} - W) + (\hat{W}^{GMM} - W)' Y_{\mathcal{J}}^{pre} \Tilde{Y}_{\mathcal{K}}^{pre'} A_{T_0} \Tilde{Y}_{\mathcal{K}}^{pre} Y_{\mathcal{J}}^{pre'} (\hat{W}^{GMM} - W)
\end{equation*}
\begin{equation*}
    \hat{u}' \Tilde{Y}_{\mathcal{K}}^{pre'} A_{T_0} \Tilde{Y}_{\mathcal{K}}^{pre} \hat{u} + (\hat{W}^{GMM} - W)' Y_{\mathcal{J}}^{pre} \Tilde{Y}_{\mathcal{K}}^{pre'} A_{T_0} \Tilde{Y}_{\mathcal{K}}^{pre} Y_{\mathcal{J}}^{pre'} (\hat{W}^{GMM} - W)
\end{equation*}
where the last equality follows from the first order condition for $\hat{W}^{GMM}$ which is that $\hat{u}' \Tilde{Y}_{\mathcal{K}}^{pre'} A_{T_0} \Tilde{Y}_{\mathcal{K}}^{pre} Y_{\mathcal{J}}^{pre'} = 0$. Since $\hat{u}' \Tilde{Y}_{\mathcal{K}}^{pre'} A_{T_0} \Tilde{Y}_{\mathcal{K}}^{pre} \hat{u}$ is not dependent on $W$, the set of minimizers of the objective function in equation \eqref{eq: Time-Series Averaged SC} is equal to the set of minimizers of $(\hat{W}^{GMM} - W)' Y_{\mathcal{J}}^{pre} \Tilde{Y}_{\mathcal{K}}^{pre'} A_{T_0} \Tilde{Y}_{\mathcal{K}}^{pre} Y_{\mathcal{J}}^{pre'} (\hat{W}^{GMM} - W)$.

\section*{Appendix D. Additional Simulation Results}\phantomsection\label{ApD}

\setcounter{table}{0}
\renewcommand{\thetable}{A\arabic{table}}

Tables \ref{Table1A} and \ref{Table2A} contain the bias and MSE results for the same two scenarios described in section \ref{Simulations}, except now with the probability of treatment being estimated rather than being uniform. Similarly to \cite{SDID}, I estimate the probability of each country becoming treated by predicting the World Bank's indicators for the country's income level. I do this by performing logistic regression of an indicator for whether a country is low-income on the units' factor loadings. I then let the probabilities of being treated be proportional to their estimated probability of being low-income. This helps to capture the fact that in SC applications, we should expect treatment assignment to be correlated with unobserved variables related to the outcome variable. The results also differ from those in section \ref{Simulations} in that the GMM-SCEs use $A_{T_0} = \hat{\mathcal{V}}(\hat{W})$ rather than $A_{T_0} = I_{K+1}$.

\newpage
\begin{table}[t]\centering\caption{Main Results with One Treated Unit \label{Table1A}}\scalebox{.95}{
\begin{threeparttable}
\begin{tabular}{l c*{4}{c}} \hline
          &\multicolumn{1}{c}{OLS-SCE}                &\multicolumn{1}{c}{Sequential} &\multicolumn{1}{c}{Uniform SCE} &\multicolumn{1}{c}{Factor Model}&\multicolumn{1}{c}{Powell (2021)}
          
          \\
          &\multicolumn{1}{c}{}                &\multicolumn{1}{c}{GMM-SCE} &\multicolumn{1}{c}{} &\multicolumn{1}{c}{Estimate}&\multicolumn{1}{c}{}\\

\toprule        
\multicolumn{4}{l}{\textbf{Bias Magnitude}} \\ 
\midrule
$T_0 = 25$,$N_0 = 10$ & 0.525 & 0.350 & 1.356 & 0.242 & 1.353\\
$T_0 = 50$,$N_0 = 10$  & 0.492 & 0.305 & 1.356 & 0.245 & 1.482\\
$T_0 = 100$,$N_0 = 10$  & 0.459 & 0.234 & 1.356 & 0.225 & 1.672\\

$T_0 = 25$,$N_0 = 50$ & 0.129 & 0.083 & 1.353 & 0.032 & 1.528 \\
$T_0 = 50$,$N_0 = 50$  & 0.104 & 0.066 & 1.353 & 0.023 & 1.888\\
$T_0 = 100$,$N_0 = 50$  & 0.079 & 0.062 & 1.353 & 0.010 & 2.227\\

\hline 
\multicolumn{4}{l}{\textbf{$\hat{\alpha}_{0t}$ MSE}} \\                                 \midrule
$T_0 = 25$,$N_0 = 10$ & 4.700 & 5.725 & 6.176 & 4.737 & 13.879\\
$T_0 = 50$,$N_0 = 10$  & 4.479 & 5.414 & 6.176 & 4.342 & 17.983\\
$T_0 = 100$,$N_0 = 10$  & 4.328 & 5.245 & 6.176 & 4.167 &  23.243\\

$T_0 = 25$,$N_0 = 50$   & 1.879 &  2.310 & 5.723 & 4.278 &  13.188\\
$T_0 = 50$,$N_0 = 50$  & 1.790 & 2.148 & 5.723 & 3.958 & 18.251\\
$T_0 = 100$,$N_0 = 50$  & 1.711 & 1.965 & 5.723 & 3.830 & 23.121\\

\hline 
\multicolumn{4}{l}{\textbf{$\hat{\Bar{\alpha}}$ MSE}} \\                                 \midrule
$T_0 = 25$,$N_0 = 10$ & 0.533 & 0.503 & 2.104 & 0.316 & 4.259\\
$T_0 = 50$,$N_0 = 10$  & 0.449 & 0.396 & 2.104 & 0.244 & 5.257\\
$T_0 = 100$,$N_0 = 10$  & 0.389 & 0.288 & 2.104 &  0.200 & 7.252\\

$T_0 = 25$,$N_0 = 50$   & 0.124 & 0.150  & 1.929 & 0.233 & 6.226\\
$T_0 = 50$,$N_0 = 50$  & 0.094 &  0.105 & 1.929 & 0.150 & 9.150\\
$T_0 = 100$,$N_0 = 50$  & 0.066 & 0.073 & 1.929 & 0.115 & 11.722\\

\hline
\end{tabular}
\begin{tablenotes}
      \small
      \item Notes: All simulations are done with a thousand replications.
\end{tablenotes}
    \end{threeparttable}}
\end{table}

\begin{table}[t]\centering\caption{Main Results with $N_1$ Treated Units \label{Table2A}}\scalebox{.95}{
\begin{threeparttable}
\begin{tabular}{l c*{6}{c}} \hline
          &\multicolumn{1}{c}{OLS-SCE}                &\multicolumn{1}{c}{Sequential}&\multicolumn{1}{c}{Two-Step MS} &\multicolumn{1}{c}{Uniform SCE} &\multicolumn{1}{c}{Factor Model}&\multicolumn{1}{c}{Unconstrained}\\
          &\multicolumn{1}{c}{}                &\multicolumn{1}{c}{GMM-SCE}&\multicolumn{1}{c}{GMM-SCE} &\multicolumn{1}{c}{} &\multicolumn{1}{c}{Estimate}&\multicolumn{1}{c}{GMM-SCE}\\

\toprule        
\multicolumn{5}{l}{\textbf{Bias Magnitude}} \\ 
\midrule
$T_0 = 25$,$N_0 = 10$ & 0.525 & 0.254 & 0.462 & 1.356 & 0.242 & 0.039\\
$T_0 = 50$,$N_0 = 10$  & 0.492 & 0.291 & 0.211 & 1.356 & 0.245 & 0.030\\
$T_0 = 100$,$N_0 = 10$  & 0.459 & 0.270 & 0.174 & 1.356 & 0.225 & 0.068\\

$T_0 = 25$,$N_0 = 50$   & 0.129 & 0.084 & 0.122 & 1.353 & 0.032 & 0.068\\
$T_0 = 50$,$N_0 = 50$ & 0.104 & 0.037 & 0.079 & 1.353 & 0.023 & 0.012\\
$T_0 = 100$,$N_0 = 50$  & 0.079 & 0.060 & 0.013 & 1.353 & 0.010 & 0.030\\

\hline 
\multicolumn{5}{l}{\textbf{$\hat{\alpha}_{0t}$ MSE}} \\                                 \midrule
$T_0 = 25$,$N_0 = 10$ & 4.700 & 5.536 & 5.277 & 6.176 &  4.737 & 13.556\\
$T_0 = 50$,$N_0 = 10$  & 4.479 & 5.337 & 4.788 & 6.176 & 4.342 & 12.773\\
$T_0 = 100$,$N_0 = 10$  & 4.328 & 5.090 & 4.720 & 6.176 & 4.167 & 14.542\\

$T_0 = 25$,$N_0 = 50$   & 1.879 & 2.290 & 2.126 & 5.723 & 4.278 & 4.275\\
$T_0 = 50$,$N_0 = 50$  & 1.790 & 2.058 & 2.151 & 5.723 & 3.958 & 5.014\\
$T_0 = 100$,$N_0 = 50$  & 1.711 & 1.962 & 1.905 & 5.723 & 3.830 & 6.690\\

\hline 
\multicolumn{5}{l}{\textbf{$\hat{\Bar{\alpha}}$ MSE}} \\                                 \midrule
$T_0 = 25$,$N_0 = 10$ & 0.533 & 0.384  & 0.601  &  2.104 & 0.316 & 0.760 \\
$T_0 = 50$,$N_0 = 10$  & 0.449 & 0.370 & 0.242 &  2.104 & 0.244 & 0.519 \\
$T_0 = 100$,$N_0 = 10$  & 0.389 & 0.324 & 0.193 & 2.104 & 0.200 & 0.463 \\

$T_0 = 25$,$N_0 = 50$ & 0.124 & 0.149 & 0.188 & 1.929 & 0.233 & 0.226\\
$T_0 = 50$,$N_0 = 50$  & 0.094 & 0.086  & 0.163  & 1.929 & 0.150 & 0.194\\
$T_0 = 100$,$N_0 = 50$  & 0.066 & 0.073  & 0.056 & 1.929 & 0.115 & 0.215\\

\hline
\end{tabular}
\begin{tablenotes}
      \small
      \item Notes: All simulations are done with a thousand replications.
\end{tablenotes}
    \end{threeparttable}}
\end{table}

\FloatBarrier
\noindent
\textbf{Acknowledgements}: I am grateful for helpful comments and feedback from Adam Mccloskey, Carlos Martins-Filho, Daniel Kaffine, and Taylor Jaworski.

\footnotesize
\bibliographystyle{econ-aea}
\bibliography{references}

\end{document}